# Condensation sequence of circumstellar cluster seeds (CSCCS)

David Gobrecht

## Abstract

Traditionally, the condensation sequence of circumstellar dust is predicted based on the thermodynamic stabilities of specific condensates in the macroscopic bulk phase. However, at the (sub-)nanometer scale clusters with non-crystalline structures and significantly different properties are energetically favoured.

For this reason, we study the thermodynamic stabilities of metal oxide clusters with generic stoichiometries of $M_2O_3$ and $M_3O_4$, where M represents a metal atom. With an upper size limit of 50 atoms, we consider clusters with sizes n=1-10 for $(M_2O_3)_n$, and n=1-7 for $(M_3O_4)_n$. The $M_2O_3$ clusters comprise alumina ($Al_2O_3$), Mg-rich pyroxene ($MgSiO_3$) and a size-limited sample of titanates ($CaTiO_3$), whereas the $M_3O_4$ clusters include spinel ($MgAl_2O_4$), Mg-rich olivine ($Mg_2SiO_4$) and calcium aluminates ($CaAl_2O_4$).

We find that, apart from the alumina monomer, the aluminum-bearing clusters $(Al_2O_3)_n$, n=1-10, and $(MgAl_2O_4)_n$, n=1-7, are favoured over their silicate counterparts $(MgSiO_3)_n$, n=1-10 and $(Mg_2SiO_4)_n$, n=1-7. Also, we find that calcium aluminate clusters, $CaAl_2O_4$, are energetically more favourable than magnesium aluminate clusters, $MgAl_2O_4$. Furthermore, for a limited data set of $(CaTiO_3)_n$, n=1-2, clusters we find significantly larger stabilities than for the other considered $(M_2O_3)_n$ clusters, namely $Al_2O_3$ and $MgSiO_3$.

Future investigations, in particular on titanates and on Ca-rich silicates, are required to draw a more thorough and complete picture of the condensation sequence at the (sub-)nanoscale.

## 1. Introduction

Asymptotic Giant Branch (AGB) stars with their highly dynamical atmospheres and circumstellar envelopes are unique astrochemical laboratories (Höfner and Olofsson 2018). Shell burning and mixing processes inside of AGB stars lead to a changing elemental composition at the atmosphere over time (Herwig 2005). In AGB circumstellar envelopes the presence of large-amplitude stellar pulsations, active dust formation from the gas-phase, and maser emission indicate departures from thermodynamic equilibrium conditions, which require non-equilibrium modelling (Gobrecht et al 2016).

In equilibrium, the chemistry is controlled by the C/O ratio, which is a consequence of the triple bonded CO molecule corresponding to the diatomic species with the highest bond energy (11.3 eV) known. As a result, an oxygen-rich chemistry with C/O < 1 and carbon-rich environments with C/O > 1 are expected, as the CO molecule locks the lesser abundant element (C or O). The observationally confirmed presence of HCN and CS in oxygen-dominated atmospheres (Schöier et al 2013, Decin et al 2010), as well as the detection of $H_2O$ vapour and SiO in Carbon-rich envelopes challenge this traditional dichotomy (Neufeld et al 2013, Fonfria et al. 2014). With regard to the dust populations the situation is different. To the best of our knowledge there is no confirmed active carbonaceous dust formation in oxygen-dominated AGB stellar atmospheres and no oxide dust synthesis in carbon-rich AGB envelopes.

Circumstellar dust species in oxygen-rich conditions comprise minerals and amorphous solids that are made of metal oxides with the stoichiometric formula $M_2O_3$ including corundum ($Al_2O_3$), enstatite ($MgSiO_3$), perovskite ($CaTiO_3$). This is also true for metal oxides with the generic formula $M_3O_4$ that includes $MgAl_2O_4$ (spinel), $CaAl_2O_4$ (krotite), and $Mg_2SiO_4$ (forsterite). Evidence for the presence of alumina and spinel is found in spectral features observed at around 13 microns (Sloan et al. 2003, Posch et al.1999), whereas the silicates, here enstatite and forsterite, show Si-O stretching modes at 10 micron and O-Si-O bending modes at around 18 microns (Woolf and Ney 1969).

Iron-bearing solid oxides like hematite ($Fe_2O_3$), ferrosilit ($FeSiO_3$) or fayalite ($Fe_2SiO_4$) fall also in the $M_2O_3$ and $M_3O_4$ stoichiometric families, but are not expected to act as seed particles for dust formation, owing to the too large opacities of the Fe-bearing seeds (Woitke 2006).

Traditionally, the thermodynamic stabilities of condensates are assessed based on a top-down approach starting with the macroscopic crystalline structure of a specific mineral. In classical nucleation theories (CNTs) the cohesive energy of a microscopic particle like a cluster is determined by an attractive volume term and a repulsive surface term resulting in a spherical particle as lowest-energy geometry. Depending on the material in question, top-down approaches like CNTs can predict reliable particle energies down to typically tens of nanometers, in rare cases even down to a few nanometers.

However, at the nanoscale and below, the lowest energy structures are significantly different from top-down derived geometries. Owing to finite size and quantum effects the properties of the small clusters including atomic coordination, bond lengths, formal charges, dipole moments are notably different from the macroscopic bulk phase. Notably, the most favourable cluster structures are non-crystalline. Therefore, the drawbacks of CNTs at the (sub-)nanoscale comprise the spherical cluster structures, fully coordinated atoms in the homogeneous interior of the cluster, growth by monomeric additions, bimolecular association reactions, and most prominently, unrealistic potential energies.

In this study we will focus on the thermochemistry of bottom-up generated, (sub-)nanometer sized clusters as precursors of dust grains in oxygen-rich circumstellar atmospheres and envelopes. Apart from $Al_2O_3$ the clusters presented in this study are ternary oxides comprising two different metal elements. In contrast, binary oxides contain just one type of metal and are therefore chemically and structurally less complex. Still, it is a challenging task to find the lowest-energy isomers referred to as global minima (GM) structures in the case of binary oxide clusters.

Smaller gas phase dust precursors in oxygen-rich environments include simple diatomic and triatomic molecules with well-defined spectroscopic constants, rotational and vibrational transitions. In the case of silicates, a very likely molecular precursor is SiO (Nuth and Donn 1981), whereas the aluminates including alumina are preceded by AlO and AlOH (Kaminski et al 2016, Decin et al 2017, Danilovich et al 2020). Titanantes originate from TiO and $TiO_2$ molecules showing relatively strong Ti-O bonds (Kaminski et al 2017, Danilovich et al 2020). Moreover, the hydroxyl radical, OH, and water vapour, $H_2O$, are considered to act as the most efficient oxidizer of the metal oxide molecules and clusters (Baudry et al 2022).

In this study we will focus on the thermochemistry of (sub-)nanometersized clusters with $M_2O_3$ and $M_3O_4$ stoichiometries as precursors of dust grains in oxygen-rich circumstellar atmospheres and envelopes. Moreover, this study aims at overcoming the drawbacks of classical nucleation theories by employing a bottom-up approach using lowest-energy cluster configuration with the most accurate yet affordable quantum calculations, instead of using top-down generated spherical particle structures.

This paper is organized as follows. In Section 2, we describe the methods used to compute the most favourable $(M_2O_3)_n$ and $(M_3O_4)_n$ cluster structures. The results are presented in Section 3. Section 4 discusses our results in the light of previous studies and summarizes our findings.

## 2. Methods

The energetically most favourable cluster isomers, or GM candidates, of the considered dust precursors of alumina, spinel, Ca aluminates, Mg-rich silicates and calcium titanates were recently investigated. For our purposes we consult the GM candidates for $(Al_2O_3)_n$, n=1-10 (Gobrecht et al 2022), $MgAl_2O_4$, $CaAl_2O_4$ (Gobrecht et al 2023), $(MgSiO_3)_n$, n=1-10, $(Mg_2SiO_4)_n$, n=1-7, (Macià Escatllar et al 2019) and $(CaTiO_3)_n$, n=1,2 (Plane 2013).
A comparison among the trioxides $Al_2O_3$, $MgSiO_3$, and $CaTiO_3$ is straightforward as these cluster contain the same number of atoms, metal atoms and oxygen atoms, respectively, per formula unit or cluster size n. The same reasoning is true for a comparison among clusters of $MgAl_2O_4$, $CaAl_2O_4$ and $Mg_2SiO_4$, showing seven atoms per formula unit of which three are metals and four are oxygens. For consistency we apply the B3LYP density functional in combination with a cc-pVTZ basis set including a vibrational frequency analysis for all calculations presented in this study (Becke 1993). This density functional basis set combination was chosen as a sensible compromise between computational cost and desired accuracy. Moreover, this combination has shown a reasonable agreement with experimental results for transition metal oxides such as titania and vanadia (Sindel et al 2022, Lecoq-Molinos 2025).
We use the RRHO (Rigid Rotor Harmonic Oscillator) approximation to compute the partition functions of the cluster leading the thermodynamic potential of interest, namely the enthalpy of formation, entropy and Gibbs Free energy of formation. For consistency, all presented quantum-chemical density functional calculations were computed with the Gaussian16 programme suite (Frisch et al 2016).

## 3. Results:

In Figure 1 the thermodynamic stabilities represented by the normalized Gibbs free energy $\Delta_f G(T)$ of formation of the $(M_2O_3)_n$ clusters are shown as a function of the cluster size n. The energies of $CaTiO_3$ clusters are colour-coded in orange, those of $Al_2O_3$ clusters in purple and those of $MgSiO_3$ clusters in green, respectively. Solid lines correspond to a temperature of T= 0 K, dash-dotted lines to a temperature of T= 1000 K, and dashed lines to a temperature of T=2000 K.
Clearly, the $(CaTiO_3)_n$ family is most favourable with regard to the studied $M_2O_3$ clusters for all considered temperatures. Alumina clusters have stabilities that are more than 200 kJ/mol less per formula unit than the calcium titanates, but are more favourable than the Mg-rich pyroxene clusters by about 100 - 150 kJ/mol, except for the monomer. The monomers $Al_2O_3$ and $MgSiO_3$ show almost identical Gibbs free energies at a temperature of 2000 K. Moreover, we note a comparatively lower stability for $(Al_2O_3)_n$ clusters at sizes n=8 and 10 and for $(MgSiO_3)_n$ at size n=5.

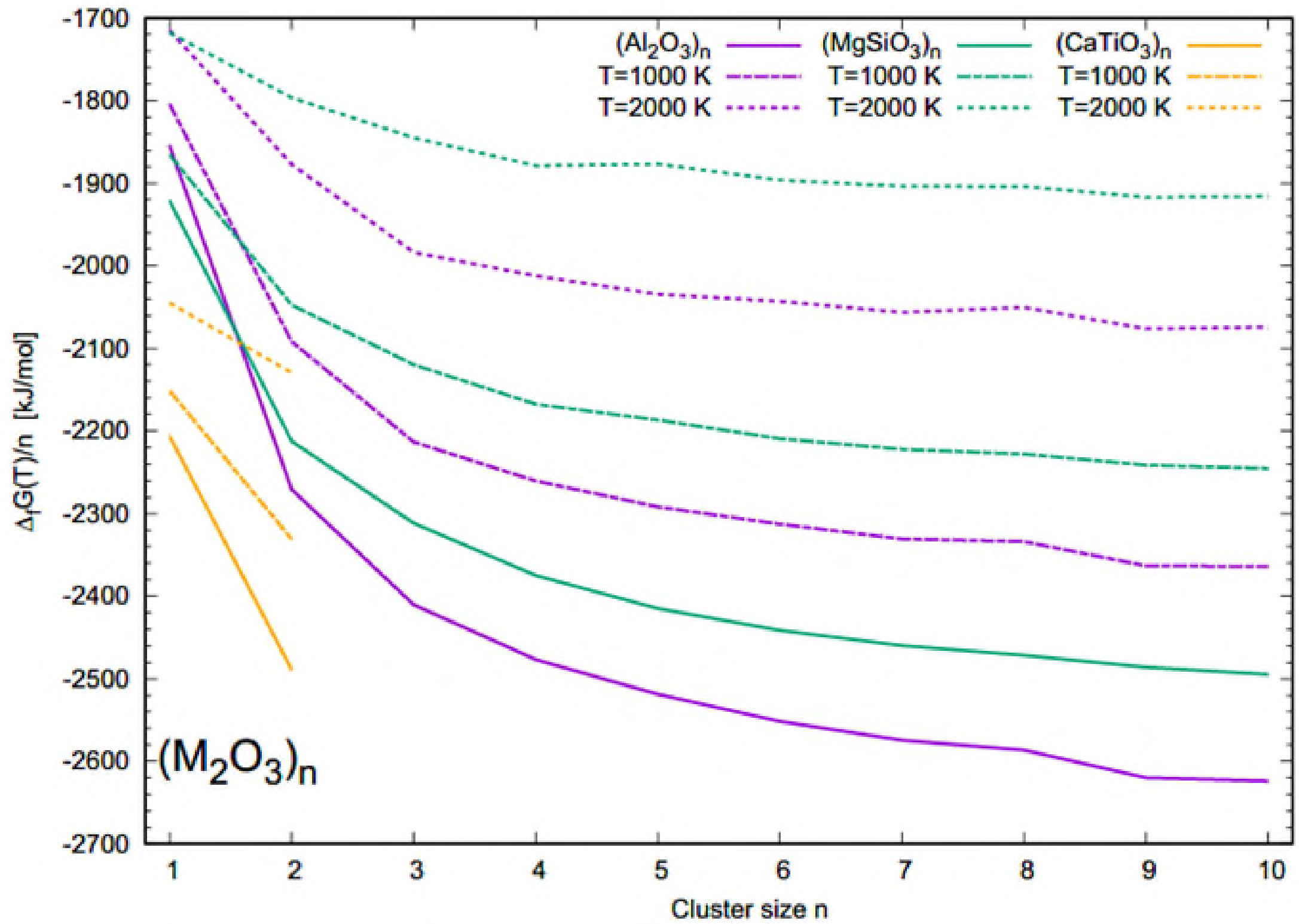

**Fig 1:** Gibbs free energies of formation $\Delta_f$ G(T) normalized to the cluster size n of the $(M_2O_3)_n$ clusters as a function of the cluster size n. $CaTiO_3$ clusters are shown in orange, $(Al_2O_3)_n$ clusters in purple, and $(MgSiO_3)_n$ clusters in green, respectively. Solid lines represent $\Delta_f$ G(0 K), dash-dotted lines $\Delta_f$ G(1000 K), and dashed lines $\Delta_f$ G(2000 K), respectively.

In Figure 2 the normalised Gibbs free energies $\Delta_f$ G(T) of formation of the $(M_3O_4)_n$ clusters are shown as a function of the cluster size n. The energies of $CaAl_2O_4$ clusters are colour-coded in red, those of $MgAl_2O_4$ clusters in purple and those of $Mg_2SiO_4$ clusters in green, respectively. As in Figure 1, solid lines correspond to a temperature of T= 0 K, dash-dotted lines to a temperature of T= 1000 K, and dashed lines to a temperature of T=2000 K. The calcium aluminate clusters constitute the most favourable clusters among the $M_3O_4$ set for all selected temperatures. The energy difference to their Mg-rich counterpart clusters, $(MgAl_2O_4)_n$, is in the range of ~150 – 200 kJ/mol. This energy difference is larger than the difference of ~100 – 150 kJ/mol between the spinel clusters, $(MgAl_2O_4)_n$, to the less favourable Mg-rich olivine silicate clusters. For cluster size n=5 we find a comparatively low stability of the aluminate clusters $(MgAl_2O_4)_5$ and $(CaAl_2O_4)_5$.

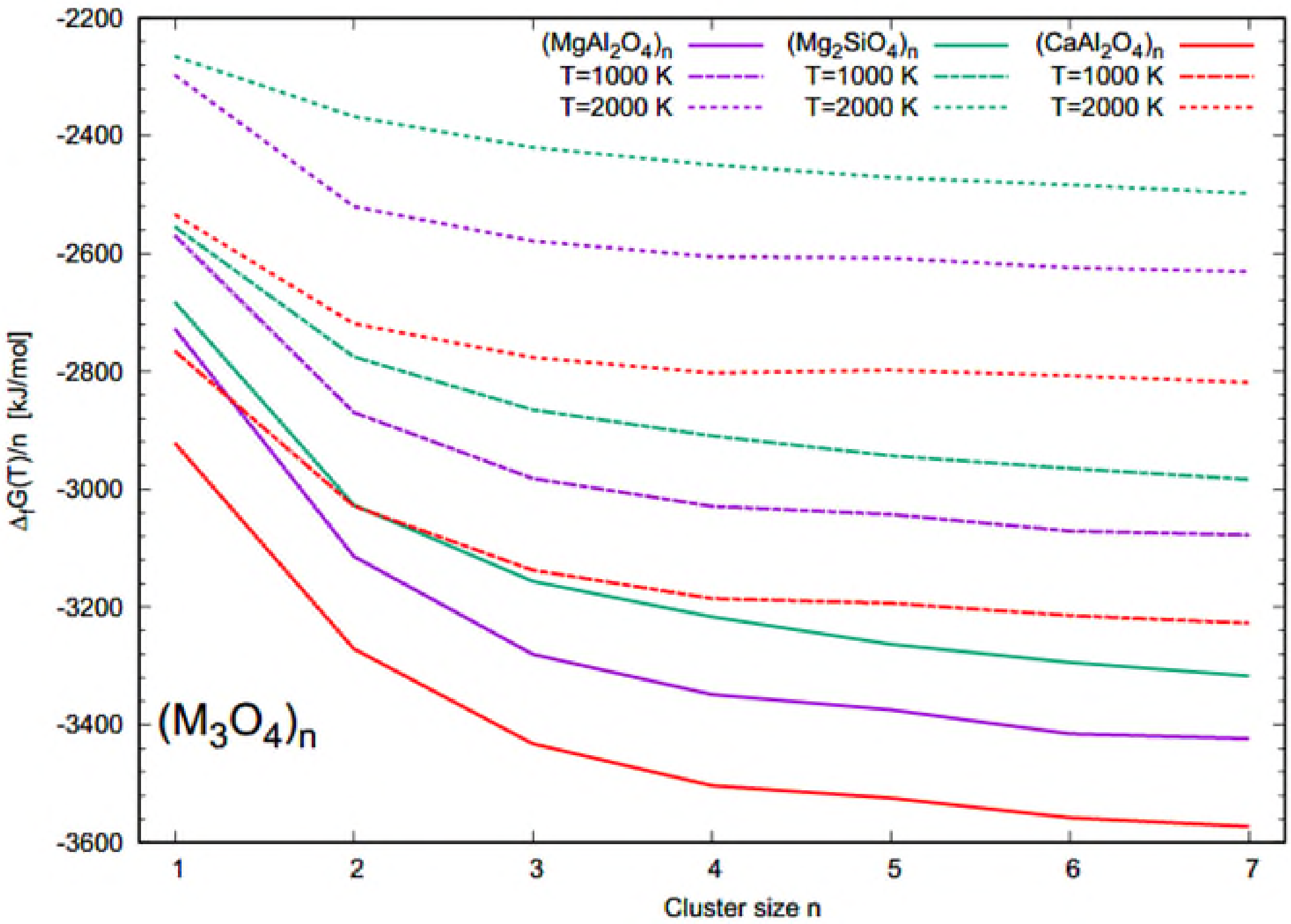

**Fig 2:** Gibbs free energies of formation $\Delta_f$ G(T) normalized to the cluster size n of the $(M_3O_4)_n$ clusters as a function of the cluster size n. $CaAl_2O_4$ clusters are shown in red, $(MgAl_2O_4)_n$ clusters in purple, and $(Mg_2SiO_4)_n$ clusters in green, respectively. Solid lines represent $\Delta_f$ G(0 K), dash-dotted lines $\Delta_f$ G(1000 K), and dashed lines $\Delta_f$ G(2000 K), respectively.

Note that the Gibbs free energies of formation in Figures 1und 2 are not scaled to the atomic heats of formation as in the JANAF tables (Chase et al 1998). In this way, the enthalpy of formation $\Delta_f$ H(0 K) and Gibbs free energy of formation $\Delta_f$ H(0 K) correspond to the unscaled total binding energy of the cluster at 0K. The corresponding thermochemical tables in the standard JANAF format can be found in the appendix A1. Note that the thermochemical tables of the cluster families $(MgAl_2O_4)_n$ and $(CaAl_2O_4)_n$ can be found in the supporting information of Gobrecht et al (2023).
An illustration of the applicability of the calculated Gibbs free energies of formation is shown in the form of chemical equilibrium abundances in the appendix A3. The corresponding computations are performed using the software GGchem (Woitke et al 2018) at a pressure of 1 bar using a solar elemental composition (Asplund et al 2009). The clusters of the $M_2O_3$ and $M_3O_4$ stoichiometries were treated separately in order to avoid interference between these families. In the $M_2O_3$ family, clusters larger than the monomer (n>1) are present in the form of $(CaTiO_3)_2$ for temperatures of T $\leq$ 1500 K, $(Al_2O_3)_{10}$ for T $\leq$ 1300K, and $(MgSiO_3)_{10}$ for T $\leq$ 1200 K. The $M_3O_4$ family shows no significant amount of $(MgAl_2O_4)_n$ clusters, which can be explained by the presence of the thermodynamically favoured, Al-bearing $(CaAl_2O_4)_n$ clusters at T $\leq$ 1400 K and the Mg-bearing silicate clusters $(Mg_2SiO_4)_n$ at T $\leq$ 1200 K. These equilibrium abundances reflect the results presented in Figures 1 and 2 weighted by elemental abundances and indicate trends in a potential sequence of condensation.
However, we emphasize that the equilibrium abundances shown here are not representative for the actual existence or presence of the corresponding clusters in stellar atmospheres, since chemical equilibrium cannot trace the complex chemical kinetics of cluster formation. For example, formation routes that are prohibited by energy barriers or spin cannot be identified or predicted in chemical equilibrium considerations.

## 4. Discussion and Summary

Alumina and ternary aluminium-bearing oxides are thermodynamically favoured over their silicate counterparts. Furthermore, calculations constrained to small cluster sizes indicate that titanium-bearing ternary oxides are even more favourable compared with aluminates and silicates.
The thermodynamic stabilities of the cluster families seem to follow an opposite trend with respect to the elemental abundances of their constituent metals, where Ti is about an order less abundant than Al and about two orders of magnitude less abundant than Si, according to solar abundances (Asplund 2009) and AGB stellar evolution models (Cristallo et al. 2015). This trend is also seen**,** when comparing Mg-containing clusters with the corresponding Ca-bearing counterparts, which are more favourable and Ca is about an order of magnitude less abundant than Mg.
Apart from alumina ($Al_2O_3$) the investigated $M_2O_3$ and $M_3O_4$ clusters are ternary metal oxides comprising two different metals. A comparison with binary oxide clusters such as titania ($TiO_2$), silica ($SiO_2$), and magnesia (MgO) is not included in our study, since their metal-to-oxygen ratio does not agree with a $M_2O_3$ or a $M_3O_4$ stoichiometry, respectively, making such comparisons not straight-forward. Moreover, the listed binary oxide clusters can be regarded as an integral part of the presented $M_2O_3$ and $M_3O_4$ clusters. $TiO_2$ is contained in $CaTiO_3$, whereas $SiO_2$ and MgO represent formal constituents of $MgSiO_3$ and $Mg_2SiO_4$. So, albeit clusters of $TiO_2$, SiO and MgO are tightly linked to the ternary oxide clusters of the presented study, they are not explicitly part of this study. However, we briefly discus their energetic and kinetic viabilities as well as limitations in AGB circumstellar envelopes with respect to homogeneous nucleation.
SiO is an abundant molecule in all chemical types of AGB stars (see e.g. Ramstedt et al 2009). However, the SiO nucleation via $(SiO)_n$ clusters was found to be inefficient and negligible under circumstellar conditions (Bromley et al 2016). Moreover, with increasing size the most favourable $(SiO)_n$ cluster exhibit segregated Si islands, further impeding the cluster growth by monomeric SiO additions. It should also be noted here that macroscopic solid SiO quickly segregates into islands of amorphous silica ($SiO_2$) and silicon (Friede 1996). Therefore, SiO cannot be regarded as a stable condensate. However, at the (sub-)nanoscale $(SiO)_n$ clusters can exist. SiO is a key molecule for the formation of silicates. Goumans and Bromley (2012) explored a mechanism, where the SiO molecule is dimerized to $Si_2O_2$, which subsequently oxidized and enriched with Mg atoms in an alternating manner to form $Mg_4Si_2O_8$, the dimer of Mg-rich olivine. The dimerisation of SiO represents the energetic bottleneck in their scheme at 1000 K, whereas the formation of the trimer, $Si_3O_3$, could proceed via $SiO_2$ and $Si_2O_3$ in terms of energy. However, the oxidation of SiO to $SiO_2$ is kinetically hampered and also the dimerization of $SiO_2$ to $Si_2O_4$ represents a bottleneck reaction in c**ir**cumstellar envelopes as has been confirmed by Anderson 2023 et al.
To our knowledge MgO does not exist in AGB circumstellar atmospheres, and Mg is predominantly expected to be in atomic form. Even if the MgO monomer**s were** present in circumstellar envelopes, its homogeneous nucleation is hampered by particularly favourable "magic" cluster sizes that act as bottlenecks for cluster growth (Köhler et al 1997, Bhatt and Ford 2007). For Mg-bearing titanates and aluminates the inclusion of magnesium constitutes a kinetic bottleneck (Plane 2013, Gobrecht et al 2023).  In these studies, it was also found that reactions with calcium atoms are significantly more efficient than those with Mg atoms and can, under certain cirumstances, lead to the formation of condensation seeds.
$TiO_2$ exists as a strongly bond gas phase molecule, i.e. as a monomer, but also in a macroscopic crystalline phase as anatase or rutile making it a prime candidate for nucleation. Its nucleation path via stoichiometric $(TiO_2)_n$ clusters follows indeed an energetically downhill process (Lamiel-Garcia et al 2017, Sindel et al 2022). However, a chemical-kinetic assessment of the dimerization of $TiO_2$ to $(TiO_2)_2$, representing the first reaction in this scenario, shows a slow and ineffective reaction rate with a negative temperature-dependence (Plane 2013). It implies that the rate is very slow at high temperatures and requires a third body to proceed. Moreover, homogeneous $TiO_2$ nucleation is limited by the availability of Ti having a low abundance.
Formation routes towards the alumina monomer, $Al_2O_3$, are thermodynamically and kinetically hampered. The synthesis of the triplet $Al_2O_3$ monomer synthesis is prohibited by an unfavourable, i.e. an endothermic oxidation of $Al_2O_2$ (Gobrecht et al 2022). Instead, a viable pathway leading to the formation of the alumina dimer, $(Al_2O_3)_2$, was identified. Alternative solutions to form $(Al_2O_3)_n$ clusters might involve $Al(OH)_3$ as suggested by Firth et al (2024).
In summary, SiO nucleation is hampered by increasing atomic segregation and slow growth rates, MgO nucleation is hindered by the absence of the MgO monomer and energy bottlenecks at magic cluster sizes, and homogeneous $TiO_2$ nucleation is viable, but constrained by the overall low Ti abundance. Given the chemical wealth of oxygen-rich AGB circumstellar envelopes, it is likely that ternary oxide clusters form. In particular, the above mentionned drawbacks of homogeneous nucleation of SiO, MgO and $TiO_2$, respectively, indicate that the formation of a ternary oxide involving an additional metal is required for subsequent cluster growth.
For a comparison between the binary oxide clusters of SiO, $TiO_2$, MgO and $Al_2O_3$ we refer to the study of Boulangier et al 2019. Although a (chemical-)kinetic treatment of the studied clusters would be desirable, an investigation of the presented clusters is computationally very expensive or even prohibitive for cluster sizes $n>3$. The reaction rates involving larger sized clusters are very expensive to compute at high level rate theories like statistical Ramsperger-Kassel-Marcus (RRKM) and transition state theories. For this reason, reaction systems comprising more than ~ 12 atoms the rates are typically assessed with theories of approximative nature including classical and kinetic nucleation, collisional rates, capture rates and detailed balance.

Spectral identification of the presented clusters is challenging. Owing to their large number of degrees of freedom the presented clusters exhibit many, partly overlapping and blurred spectral features in contrast to small, di- and tri-atomic molecules with sharp, discrete and well-defined transitions. In addition, the nucleating clusters might not have a longevity to be observed. However, recent studies have attempted to identify the spectral signatures of different cluster families in theoretical frameworks (Gobrecht 2022, Plane and Roberston 2022, Sindel 2023, Lecoq-Molinos 2024) as well as observationally (Decin et al. 2017, Baeyens et al. 2024).
Calcium silicate clusters, in particular the pyroxenes and olivines $(CaSiO_3)_n$ and $(Ca_2SiO_4)_n$, are entirely missing for both stoichiometries, $M_2O_3$ and $M_3O_4$, which constitutes one of the major limitations of the present study. Moreover, with the exception of the monomer and the dimer of perovskite ($CaTiO_3$), the calcium titanates clusters are not part of this study. We aim to investigate the cluster families of calcium silicates and titanates in a forthcoming study.

Therefore, no final conclusion can be drawn. However, our results indicate the following trends:

(i) alumina and aluminate clusters are energetically more favourable than their silicate counterparts
(ii) Owing to their thermodynamic stability calcium titanates are promising candidates as primary condensation seeds
(iii) Clusters comprising calcium are significantly more stable than their magnesium-containing counterparts
Other stoichiometric clusters of minerals like geikielite ($MgTiO_3$) and quandilite ($Mg_2TiO_4$) were not considered in this study. These condensates are rare and were, to the best of our knowledge, not found in meteorites. For consistency and completeness, however, these clusters as well as the aforementioned calcium silicates $CaSiO_3$ and $Ca_2SiO_4$ should be investigated in detail to draw firmer conclusions on the cluster condensation sequence among titanates, aluminates and silicates.

ACKNOWLEDGMENTS
I am deeply indebted to Tom Millar and Mauro Pirarba for enabling a publication in frontiers. The computations involved the Swedish National Infrastructure for Computing (SNIC) at Chalmers Centre for Computational Science and Engineering (C3SE) partially funded by the Swedish Research Council through grant no. 2018-05973. The computations were partially enabled by resources provided by the National Academic Infrastructure for Supercomputing in Sweden (NAISS) partially funded by the Swedish Research Council through grant agreement no. 2022-06725.

**APPENDIX A1: Thermo-chemical tables of the presented clusters in the standard JANF-NIST format**

**(Al2O3)1:**

| T (K) | S (J/mol.K) | Cp (J/mol.K) | ddH (kJ/mol) | dHf (kJ/mol) | dGf (kJ/mol) | log Kf |
|---|---|---|---|---|---|---|
| 0 | 0 | 0 | 0 | -459.077 | -459.077 | Inf |
| 100 | 249.94 | 46.257 | 3.797 | -460.602 | -458.203 | 239.338 |
| 200 | 287.071 | 62.794 | 9.248 | -463.122 | -454.833 | 118.789 |
| 298.15 | 314.914 | 76.906 | 16.139 | -465.041 | -450.325 | 78.894 |
| 300 | 315.39 | 77.128 | 16.282 | -465.069 | -450.233 | 78.392 |
| 400 | 339.01 | 86.817 | 24.515 | -466.308 | -445.086 | 58.122 |
| 500 | 359.094 | 92.959 | 33.527 | -467.151 | -439.677 | 45.932 |
| 600 | 376.417 | 96.915 | 43.034 | -467.854 | -434.116 | 37.793 |
| 700 | 391.568 | 99.556 | 52.866 | -468.6 | -428.438 | 31.97 |
| 800 | 404.988 | 101.385 | 62.919 | -469.511 | -422.636 | 27.595 |
| 900 | 417.01 | 102.696 | 73.126 | -470.695 | -416.711 | 24.185 |
| 1000 | 427.882 | 103.663 | 83.447 | -493.399 | -409.114 | 21.37 |
| 1100 | 437.798 | 104.395 | 93.851 | -494.609 | -400.628 | 19.024 |
| 1200 | 446.907 | 104.962 | 104.32 | -495.813 | -392.03 | 17.064 |
| 1300 | 455.327 | 105.409 | 114.839 | -497.019 | -383.334 | 15.402 |
| 1400 | 463.152 | 105.767 | 125.399 | -498.228 | -374.541 | 13.974 |
| 1500 | 470.459 | 106.059 | 135.991 | -499.449 | -365.666 | 12.733 |
| 1600 | 477.312 | 106.299 | 146.609 | -500.682 | -356.706 | 11.645 |
| 1700 | 483.763 | 106.499 | 157.249 | -501.93 | -347.673 | 10.683 |
| 1800 | 489.855 | 106.668 | 167.908 | -503.195 | -338.562 | 9.825 |
| 1900 | 495.626 | 106.811 | 178.582 | -504.481 | -329.379 | 9.055 |
| 2000 | 501.108 | 106.934 | 189.269 | -505.787 | -320.127 | 8.361 |
| 2100 | 506.328 | 107.04 | 199.968 | -507.117 | -310.812 | 7.731 |
| 2200 | 511.31 | 107.132 | 210.677 | -508.47 | -301.435 | 7.157 |
| 2300 | 516.074 | 107.212 | 221.394 | -509.85 | -291.995 | 6.631 |
| 2400 | 520.638 | 107.283 | 232.119 | -511.254 | -282.521 | 6.149 |
| 2500 | 525.019 | 107.345 | 242.85 | -512.686 | -272.946 | 5.703 |
| 2600 | 529.23 | 107.401 | 253.588 | -514.142 | -263.35 | 5.291 |
| 2700 | 533.284 | 107.451 | 264.33 | -515.625 | -253.667 | 4.907 |
| 2800 | 537.193 | 107.495 | -275.077 | 1104.937 | -241.976 | 4.514 |
| 2900 | 540.966 | 107.535 | -285.829 | 1104.281 | -211.165 | 3.803 |
| 3000 | 544.612 | 107.571 | -296.584 | 1103.649 | -180.408 | 3.141 |
| 3100 | 548.14 | 107.604 | -307.343 | 1103.044 | -149.658 | 2.522 |
| 3200 | 551.556 | 107.633 | -318.105 | 1102.465 | -118.916 | 1.941 |
| 3300 | 554.869 | 107.66 | -328.87 | 1101.907 | -88.188 | 1.396 |
| 3400 | 558.083 | 107.685 | -339.637 | 1101.374 | -57.419 | 0.882 |
| 3500 | 561.205 | 107.707 | -350.406 | 1100.863 | -26.746 | 0.399 |
| 3600 | 564.24 | 107.728 | -361.178 | 1100.377 | 3.955 | -0.057 |
| 3700 | 567.192 | 107.747 | -371.952 | 1099.912 | 34.619 | -0.489 |
| 3800 | 570.065 | 107.765 | -382.728 | 1099.468 | 65.236 | -0.897 |
| 3900 | 572.865 | 107.781 | -393.505 | 1099.048 | 95.863 | -1.284 |
| 4000 | 575.594 | 107.796 | -404.284 | 1098.651 | 126.505 | -1.652 |
| 4100 | 578.256 | 107.81 | -415.064 | 1098.279 | 157.154 | -2.002 |
| 4200 | 580.854 | 107.823 | -425.846 | 1097.928 | 187.732 | -2.335 |
| 4300 | 583.391 | 107.836 | -436.629 | 1097.601 | 218.395 | -2.653 |
| 4400 | 585.87 | 107.847 | -447.413 | 1097.3 | 248.966 | -2.956 |
| 4500 | 588.294 | 107.857 | -458.198 | 1097.022 | 279.528 | -3.245 |
| 4600 | 590.665 | 107.867 | -468.984 | 1096.774 | 310.173 | -3.522 |
| 4700 | 592.985 | 107.877 | -479.772 | 1096.555 | 340.705 | -3.786 |
| 4800 | 595.256 | 107.885 | -490.56 | 1096.368 | 371.307 | -4.041 |
| 4900 | 597.48 | 107.893 | -501.349 | 1096.211 | 401.893 | -4.284 |
| 5000 | 599.66 | 107.901 | -512.138 | 1096.093 | 432.442 | -4.518 |
| 5100 | 601.797 | 107.908 | -522.929 | 1096.009 | 463.063 | -4.743 |
| 5200 | 603.892 | 107.915 | -533.72 | 1095.968 | 493.638 | -4.959 |
| 5300 | 605.948 | 107.922 | -544.512 | 1095.971 | 524.173 | -5.166 |
| 5400 | 607.965 | 107.928 | -555.304 | 1096.021 | 554.764 | -5.366 |
| 5500 | 609.946 | 107.934 | -566.097 | 1096.119 | 585.3 | -5.559 |
| 5600 | 611.891 | 107.939 | -576.891 | 1096.27 | 615.902 | -5.745 |
| 5700 | 613.801 | 107.944 | -587.685 | 1096.479 | 646.436 | -5.924 |
| 5800 | 615.679 | 107.949 | -598.48 | 1096.744 | 677.018 | -6.097 |
| 5900 | 617.524 | 107.954 | -609.275 | 1097.076 | 707.649 | -6.265 |
| 6000 | 619.338 | 107.958 | -620.071 | 1097.434 | 738.251 | -6.427 |

**(Al2O3)2:**

| T (K) | S (J/mol.K) | Cp (J/mol.K) | ddH (kJ/mol) | dHf (kJ/mol) | dGf (kJ/mol) | log Kf |
|---|---|---|---|---|---|---|
| 0 | 0 | 0 | 0 | -1751.122 | -1751.122 | Inf |
| 100 | 252.697 | 59.879 | 4.052 | -1757.714 | -1728.197 | 902.707 |
| 200 | 310.467 | 111.531 | 12.685 | -1765.022 | -1695.71 | 442.869 |
| 298.15 | 362.819 | 150.949 | 25.683 | -1769.644 | -1660.604 | 290.927 |
| 300 | 363.755 | 151.553 | 25.963 | -1769.706 | -1659.927 | 289.015 |
| 400 | 411.207 | 177.544 | 42.52 | -1772.094 | -1622.924 | 211.93 |
| 500 | 452.704 | 193.732 | 61.146 | -1773.177 | -1585.488 | 165.633 |
| 600 | 489.005 | 204.055 | 81.073 | -1773.67 | -1547.898 | 134.755 |
| 700 | 521.009 | 210.902 | 101.843 | -1774.057 | -1510.243 | 112.694 |
| 800 | 549.497 | 215.626 | 123.183 | -1774.645 | -1472.513 | 96.144 |
| 900 | 575.1 | 219.001 | 144.923 | -1775.687 | -1434.691 | 83.266 |
| 1000 | 598.309 | 221.487 | 166.954 | -1819.706 | -1393.681 | 72.798 |
| 1100 | 619.511 | 223.366 | 189.2 | -1820.687 | -1351.032 | 64.154 |
| 1200 | 639.011 | 224.818 | 211.613 | -1821.621 | -1308.291 | 56.948 |
| 1300 | 657.054 | 225.963 | 234.154 | -1822.529 | -1265.479 | 50.847 |
| 1400 | 673.834 | 226.88 | 256.798 | -1823.424 | -1222.592 | 45.615 |
| 1500 | 689.513 | 227.625 | 279.524 | -1824.324 | -1179.65 | 41.079 |
| 1600 | 704.224 | 228.239 | 302.318 | -1825.231 | -1136.639 | 37.107 |
| 1700 | 718.077 | 228.751 | 325.169 | -1826.156 | -1093.58 | 33.601 |
| 1800 | 731.164 | 229.182 | 348.066 | -1827.108 | -1050.459 | 30.483 |
| 1900 | 743.566 | 229.548 | 371.003 | -1828.091 | -1007.284 | 27.692 |
| 2000 | 755.348 | 229.861 | 393.974 | -1829.106 | -964.05 | 25.178 |
| 2100 | 766.57 | 230.131 | 416.974 | -1830.164 | -920.774 | 22.903 |
| 2200 | 777.281 | 230.366 | 439.999 | -1831.263 | -877.447 | 20.833 |
| 2300 | 787.526 | 230.572 | 463.046 | -1832.409 | -834.069 | 18.942 |
| 2400 | 797.343 | 230.752 | 486.112 | -1833.602 | -790.697 | 17.209 |
| 2500 | 806.766 | 230.912 | 509.196 | -1834.843 | -747.183 | 15.611 |
| 2600 | 815.825 | 231.053 | 532.294 | -1836.133 | -703.698 | 14.137 |
| 2700 | 824.548 | 231.18 | 555.406 | -1837.472 | -660.102 | 12.77 |
| 2800 | 832.957 | 231.293 | 578.53 | -3014.466 | -612.542 | 11.427 |
| 2900 | 841.075 | 231.395 | 601.664 | -3011.523 | -526.806 | 9.489 |
| 3000 | 848.922 | 231.487 | 624.808 | -3008.626 | -441.238 | 7.683 |
| 3100 | 856.513 | 231.57 | 647.961 | -3005.781 | -355.731 | 5.994 |
| 3200 | 863.867 | 231.646 | 671.122 | -3002.986 | -270.305 | 4.412 |
| 3300 | 870.996 | 231.714 | 694.29 | -3000.231 | -184.945 | 2.927 |
| 3400 | 877.914 | 231.777 | 717.465 | -2997.524 | -99.558 | 1.53 |
| 3500 | 884.634 | 231.835 | 740.645 | -2994.861 | -14.412 | 0.215 |
| 3600 | 891.165 | 231.888 | 763.832 | -2992.246 | 70.752 | -1.027 |
| 3700 | 897.52 | 231.937 | 787.023 | -2989.673 | 155.785 | -2.199 |
| 3800 | 903.706 | 231.982 | 810.219 | -2987.141 | 240.677 | -3.308 |
| 3900 | 909.732 | 232.023 | 833.419 | -2984.655 | 325.559 | -4.36 |
| 4000 | 915.607 | 232.062 | 856.623 | -2982.215 | 410.421 | -5.359 |
| 4100 | 921.337 | 232.097 | 879.831 | -2979.822 | 495.26 | -6.31 |
| 4200 | 926.931 | 232.131 | 903.043 | -2977.473 | 579.91 | -7.212 |
| 4300 | 932.393 | 232.161 | 926.257 | -2975.17 | 664.694 | -8.074 |
| 4400 | 937.731 | 232.19 | 949.475 | -2972.918 | 749.253 | -8.895 |
| 4500 | 942.949 | 232.217 | 972.695 | -2970.712 | 833.763 | -9.678 |
| 4600 | 948.053 | 232.243 | 995.918 | -2968.566 | 918.401 | -10.429 |
| 4700 | 953.048 | 232.266 | 1019.144 | -2966.478 | 1002.775 | -11.144 |
| 4800 | 957.938 | 232.288 | 1042.371 | -2964.452 | 1087.252 | -11.832 |
| 4900 | 962.728 | 232.309 | 1065.601 | -2962.486 | 1171.658 | -12.49 |
| 5000 | 967.422 | 232.329 | 1088.833 | -2960.597 | 1255.963 | -13.121 |
| 5100 | 972.023 | 232.347 | 1112.067 | -2958.777 | 1340.38 | -13.728 |
| 5200 | 976.535 | 232.365 | 1135.303 | -2957.041 | 1424.666 | -14.311 |
| 5300 | 980.961 | 232.381 | 1158.54 | -2955.393 | 1508.85 | -14.87 |
| 5400 | 985.305 | 232.397 | 1181.779 | -2953.839 | 1593.106 | -15.41 |
| 5500 | 989.569 | 232.411 | 1205.019 | -2952.381 | 1677.233 | -15.929 |
| 5600 | 993.757 | 232.425 | 1228.261 | -2951.028 | 1761.456 | -16.43 |
| 5700 | 997.871 | 232.438 | 1251.504 | -2949.792 | 1845.504 | -16.912 |
| 5800 | 1001.914 | 232.451 | 1274.749 | -2948.667 | 1929.631 | -17.378 |
| 5900 | 1005.887 | 232.463 | 1297.994 | -2947.675 | 2013.823 | -17.829 |
| 6000 | 1009.794 | 232.474 | 1321.241 | -2946.736 | 2097.926 | -18.264 |

**(Al2O3)3:**

| T (K) | S (J/mol.K) | Cp (J/mol.K) | ddH (kJ/mol) | dHf (kJ/mol) | dGf (kJ/mol) | log Kf |
|---|---|---|---|---|---|---|
| 0 | 0 | 0 | 0 | -3045.892 | -3045.892 | Inf |
| 100 | 310.412 | 96.409 | 5.512 | -3056.346 | -3005.207 | 1569.740 |
| 200 | 403.757 | 178.46 | 19.452 | -3066.317 | -2949.959 | 770.441 |
| 298.15 | 486.552 | 236.379 | 39.994 | -3072.205 | -2891.449 | 506.564 |
| 300 | 488.017 | 237.259 | 40.432 | -3072.28 | -2890.326 | 503.244 |
| 400 | 561.891 | 275.199 | 66.202 | -3074.928 | -2829.204 | 369.452 |
| 500 | 626.06 | 299.007 | 95.003 | -3075.69 | -2767.658 | 289.132 |
| 600 | 682.026 | 314.306 | 125.723 | -3075.6 | -2706.052 | 235.58 |
| 700 | 731.293 | 324.512 | 157.696 | -3075.363 | -2644.488 | 197.332 |
| 800 | 775.115 | 331.584 | 190.522 | -3075.429 | -2582.925 | 168.646 |
| 900 | 814.478 | 336.652 | 223.947 | -3076.177 | -2521.327 | 146.332 |
| 1000 | 850.151 | 340.393 | 257.808 | -3141.391 | -2455.041 | 128.237 |
| 1100 | 882.732 | 343.225 | 291.995 | -3142.044 | -2386.374 | 113.318 |
| 1200 | 912.695 | 345.418 | 326.432 | -3142.628 | -2317.647 | 100.883 |
| 1300 | 940.414 | 347.147 | 361.063 | -3143.17 | -2248.878 | 90.36 |
| 1400 | 966.193 | 348.534 | 395.85 | -3143.692 | -2180.062 | 81.338 |
| 1500 | 990.279 | 349.662 | 430.761 | -3144.22 | -2111.222 | 73.518 |
| 1600 | 1012.876 | 350.593 | 465.776 | -3144.756 | -2042.332 | 66.675 |
| 1700 | 1034.155 | 351.368 | 500.875 | -3145.321 | -1973.425 | 60.635 |
| 1800 | 1054.257 | 352.021 | 536.045 | -3145.925 | -1904.47 | 55.266 |
| 1900 | 1073.306 | 352.575 | 571.276 | -3146.574 | -1835.481 | 50.46 |
| 2000 | 1091.403 | 353.051 | 606.557 | -3147.272 | -1766.45 | 46.134 |
| 2100 | 1108.638 | 353.461 | 641.884 | -3148.032 | -1697.391 | 42.22 |
| 2200 | 1125.09 | 353.817 | 677.248 | -3148.854 | -1628.3 | 38.66 |
| 2300 | 1140.824 | 354.129 | 712.645 | -3149.746 | -1559.166 | 35.409 |
| 2400 | 1155.902 | 354.403 | 748.072 | -3150.708 | -1490.079 | 32.43 |
| 2500 | 1170.374 | 354.645 | 783.525 | -3151.742 | -1420.815 | 29.686 |
| 2600 | 1184.288 | 354.86 | 819 | -3152.849 | -1351.628 | 27.154 |
| 2700 | 1197.684 | 355.052 | 854.496 | -3154.03 | -1282.302 | 24.807 |
| 2800 | 1210.6 | 355.224 | 890.01 | -4918.693 | -1207.067 | 22.518 |
| 2900 | 1223.068 | 355.379 | 925.541 | -4913.448 | -1074.594 | 19.355 |
| 3000 | 1235.118 | 355.518 | 961.086 | -4908.274 | -942.397 | 16.408 |
| 3100 | 1246.778 | 355.645 | 996.644 | -4903.178 | -810.328 | 13.654 |
| 3200 | 1258.071 | 355.76 | 1032.214 | -4898.157 | -678.4 | 11.074 |
| 3300 | 1269.02 | 355.864 | 1067.795 | -4893.195 | -546.602 | 8.652 |
| 3400 | 1279.645 | 355.96 | 1103.387 | -4888.305 | -414.788 | 6.372 |
| 3500 | 1289.964 | 356.048 | 1138.987 | -4883.481 | -283.352 | 4.229 |
| 3600 | 1299.996 | 356.128 | 1174.596 | -4878.73 | -151.926 | 2.204 |
| 3700 | 1309.754 | 356.202 | 1210.213 | -4874.04 | -20.705 | 0.292 |
| 3800 | 1319.255 | 356.27 | 1245.836 | -4869.413 | 110.271 | -1.516 |
| 3900 | 1328.51 | 356.333 | 1281.466 | -4864.854 | 241.212 | -3.231 |
| 4000 | 1337.532 | 356.392 | 1317.103 | -4860.363 | 372.105 | -4.859 |
| 4100 | 1346.333 | 356.446 | 1352.745 | -4855.943 | 502.937 | -6.407 |
| 4200 | 1354.923 | 356.497 | 1388.392 | -4851.591 | 633.473 | -7.878 |
| 4300 | 1363.312 | 356.544 | 1424.044 | -4847.305 | 764.184 | -9.283 |
| 4400 | 1371.509 | 356.587 | 1459.7 | -4843.098 | 894.544 | -10.619 |
| 4500 | 1379.523 | 356.628 | 1495.361 | -4838.958 | 1024.807 | -11.896 |
| 4600 | 1387.362 | 356.667 | 1531.026 | -4834.909 | 1155.243 | -13.118 |
| 4700 | 1395.033 | 356.703 | 1566.695 | -4830.947 | 1285.267 | -14.284 |
| 4800 | 1402.543 | 356.736 | 1602.367 | -4827.076 | 1415.427 | -15.403 |
| 4900 | 1409.899 | 356.768 | 1638.042 | -4823.297 | 1545.466 | -16.475 |
| 5000 | 1417.107 | 356.798 | 1673.72 | -4819.634 | 1675.336 | -17.502 |
| 5100 | 1424.173 | 356.826 | 1709.401 | -4816.074 | 1805.356 | -18.49 |
| 5200 | 1431.102 | 356.852 | 1745.085 | -4812.64 | 1935.164 | -19.439 |
| 5300 | 1437.9 | 356.877 | 1780.772 | -4809.336 | 2064.798 | -20.35 |
| 5400 | 1444.571 | 356.901 | 1816.461 | -4806.175 | 2194.531 | -21.228 |
| 5500 | 1451.12 | 356.923 | 1852.152 | -4803.157 | 2324.049 | -22.072 |
| 5600 | 1457.551 | 356.945 | 1887.845 | -4800.297 | 2453.702 | -22.887 |
| 5700 | 1463.869 | 356.965 | 1923.541 | -4797.612 | 2583.076 | -23.671 |
| 5800 | 1470.078 | 356.984 | 1959.238 | -4795.095 | 2712.553 | -24.429 |
| 5900 | 1476.18 | 357.002 | 1994.937 | -4792.775 | 2842.111 | -25.162 |
| 6000 | 1482.18 | 357.019 | 2030.638 | -4790.536 | 2971.523 | -25.869 |

**(Al2O3)4**

| T (K) | S (J/mol.K) | Cp (J/mol.K) | ddH (kJ/mol) | dHf (kJ/mol) | dGf (kJ/mol) | log Kf |
|---|---|---|---|---|---|---|
| 0 | 0 | 0 | 0 | -4326.559 | -4326.559 | Inf |
| 100 | 335.001 | 124.127 | 6.473 | -4341.374 | -4265.301 | 2227.938 |
| 200 | 458.142 | 238.045 | 24.895 | -4354.834 | -4183.65 | 1092.644 |
| 298.15 | 568.797 | 316.304 | 52.352 | -4362.617 | -4097.775 | 717.905 |
| 300 | 570.757 | 317.493 | 52.939 | -4362.714 | -4096.129 | 713.191 |
| 400 | 669.694 | 368.869 | 87.453 | -4366.09 | -4006.662 | 523.21 |
| 500 | 755.766 | 401.347 | 126.086 | -4366.875 | -3916.675 | 409.167 |
| 600 | 830.929 | 422.34 | 167.343 | -4366.458 | -3826.664 | 333.137 |
| 700 | 897.158 | 436.406 | 210.325 | -4365.79 | -3736.761 | 278.837 |
| 800 | 956.107 | 446.179 | 254.482 | -4365.489 | -3646.914 | 238.116 |
| 900 | 1009.086 | 453.198 | 299.469 | -4366.066 | -3557.071 | 206.445 |
| 1000 | 1057.117 | 458.387 | 345.061 | -4452.574 | -3461.023 | 180.783 |
| 1100 | 1100.999 | 462.321 | 391.105 | -4452.984 | -3361.849 | 159.639 |
| 1200 | 1141.362 | 465.369 | 437.496 | -4453.287 | -3262.635 | 142.017 |
| 1300 | 1178.71 | 467.775 | 484.158 | -4453.523 | -3163.406 | 127.106 |
| 1400 | 1213.449 | 469.705 | 531.035 | -4453.724 | -3064.153 | 114.324 |
| 1500 | 1245.911 | 471.277 | 578.087 | -4453.924 | -2964.903 | 103.246 |
| 1600 | 1276.369 | 472.572 | 625.281 | -4454.132 | -2865.622 | 93.552 |
| 1700 | 1305.052 | 473.653 | 672.594 | -4454.371 | -2766.346 | 84.998 |
| 1800 | 1332.152 | 474.562 | 720.006 | -4454.657 | -2667.041 | 77.395 |
| 1900 | 1357.831 | 475.336 | 767.502 | -4455.001 | -2567.714 | 70.591 |
| 2000 | 1382.23 | 475.998 | 815.07 | -4455.405 | -2468.361 | 64.466 |
| 2100 | 1405.468 | 476.57 | 862.699 | -4455.892 | -2369.001 | 58.925 |
| 2200 | 1427.65 | 477.068 | 910.381 | -4456.458 | -2269.619 | 53.887 |
| 2300 | 1448.867 | 477.502 | 958.11 | -4457.115 | -2170.209 | 49.286 |
| 2400 | 1469.197 | 477.885 | 1005.88 | -4457.863 | -2070.879 | 45.071 |
| 2500 | 1488.712 | 478.223 | 1053.686 | -4458.707 | -1971.337 | 41.188 |
| 2600 | 1507.475 | 478.523 | 1101.523 | -4459.646 | -1871.921 | 37.607 |
| 2700 | 1525.539 | 478.791 | 1149.389 | -4460.682 | -1772.338 | 34.288 |
| 2800 | 1542.956 | 479.031 | 1197.28 | -6813.027 | -1664.897 | 31.059 |
| 2900 | 1559.77 | 479.247 | 1245.194 | -6805.495 | -1481.159 | 26.678 |
| 3000 | 1576.02 | 479.442 | 1293.129 | -6798.054 | -1297.806 | 22.597 |
| 3100 | 1591.744 | 479.618 | 1341.082 | -6790.717 | -1114.642 | 18.781 |
| 3200 | 1606.974 | 479.778 | 1389.052 | -6783.479 | -931.684 | 15.208 |
| 3300 | 1621.74 | 479.924 | 1437.037 | -6776.32 | -748.917 | 11.854 |
| 3400 | 1636.069 | 480.058 | 1485.037 | -6769.256 | -566.143 | 8.698 |
| 3500 | 1649.986 | 480.18 | 1533.049 | -6762.278 | -383.892 | 5.729 |
| 3600 | 1663.515 | 480.292 | 1581.072 | -6755.399 | -201.668 | 2.926 |
| 3700 | 1676.676 | 480.396 | 1629.107 | -6748.6 | -19.736 | 0.279 |
| 3800 | 1689.489 | 480.491 | 1677.151 | -6741.884 | 161.861 | -2.225 |
| 3900 | 1701.971 | 480.579 | 1725.205 | -6735.258 | 343.394 | -4.599 |
| 4000 | 1714.139 | 480.661 | 1773.267 | -6728.724 | 524.848 | -6.854 |
| 4100 | 1726.009 | 480.737 | 1821.337 | -6722.284 | 706.206 | -8.997 |
| 4200 | 1737.594 | 480.807 | 1869.414 | -6715.933 | 887.159 | -11.033 |
| 4300 | 1748.909 | 480.873 | 1917.498 | -6709.671 | 1068.328 | -12.977 |
| 4400 | 1759.964 | 480.934 | 1965.588 | -6703.513 | 1249.02 | -14.828 |
| 4500 | 1770.773 | 480.991 | 2013.684 | -6697.445 | 1429.568 | -16.594 |
| 4600 | 1781.345 | 481.045 | 2061.786 | -6691.497 | 1610.339 | -18.286 |
| 4700 | 1791.691 | 481.095 | 2109.893 | -6685.666 | 1790.544 | -19.899 |
| 4800 | 1801.82 | 481.142 | 2158.005 | -6679.956 | 1970.92 | -21.448 |
| 4900 | 1811.742 | 481.186 | 2206.122 | -6674.367 | 2151.121 | -22.931 |
| 5000 | 1821.463 | 481.228 | 2254.242 | -6668.933 | 2331.092 | -24.352 |
| 5100 | 1830.993 | 481.267 | 2302.367 | -6663.636 | 2511.248 | -25.72 |
| 5200 | 1840.339 | 481.304 | 2350.496 | -6658.507 | 2691.108 | -27.032 |
| 5300 | 1849.507 | 481.339 | 2398.628 | -6653.553 | 2870.732 | -28.292 |
| 5400 | 1858.505 | 481.372 | 2446.763 | -6648.788 | 3050.47 | -29.507 |
| 5500 | 1867.338 | 481.403 | 2494.902 | -6644.213 | 3229.915 | -30.675 |
| 5600 | 1876.012 | 481.432 | 2543.044 | -6639.849 | 3409.53 | -31.802 |
| 5700 | 1884.534 | 481.46 | 2591.188 | -6635.719 | 3588.758 | -32.887 |
| 5800 | 1892.907 | 481.487 | 2639.336 | -6631.811 | 3768.128 | -33.935 |
| 5900 | 1901.138 | 481.512 | 2687.486 | -6628.167 | 3947.583 | -34.949 |
| 6000 | 1909.231 | 481.536 | 2735.638 | -6624.631 | 4126.835 | -35.927 |

**(Al2O3)5**

| T (K) | S (J/mol.K) | Cp (J/mol.K) | ddH (kJ/mol) | dHf (kJ/mol) | dGf (kJ/mol) | log Kf |
|---|---|---|---|---|---|---|
| 0 | 0 | 0 | 0 | -5618.124 | -5618.124 | Inf |
| 100 | 362.84 | 156.352 | 7.773 | -5636.961 | -5536.278 | 2891.820 |
| 200 | 518.251 | 300.168 | 31.023 | -5653.563 | -5428.698 | 1417.813 |
| 298.15 | 657.663 | 398.262 | 65.613 | -5663.023 | -5316.069 | 931.342 |
| 300 | 660.131 | 399.753 | 66.352 | -5663.139 | -5313.913 | 925.223 |
| 400 | 784.672 | 464.257 | 109.798 | -5667.056 | -5196.792 | 678.623 |
| 500 | 892.995 | 505.087 | 158.418 | -5667.708 | -5079.102 | 530.604 |
| 600 | 987.586 | 531.501 | 210.339 | -5666.837 | -4961.449 | 431.927 |
| 700 | 1070.933 | 549.207 | 264.43 | -5665.639 | -4843.991 | 361.459 |
| 800 | 1145.12 | 561.515 | 320.002 | -5664.887 | -4726.657 | 308.615 |
| 900 | 1211.795 | 570.356 | 376.618 | -5665.226 | -4609.375 | 267.518 |
| 1000 | 1272.243 | 576.894 | 433.997 | -5772.972 | -4484.38 | 234.237 |
| 1100 | 1327.47 | 581.851 | 491.945 | -5773.091 | -4355.516 | 206.824 |
| 1200 | 1378.269 | 585.692 | 550.33 | -5773.074 | -4226.639 | 183.979 |
| 1300 | 1425.273 | 588.723 | 609.056 | -5772.97 | -4097.775 | 164.649 |
| 1400 | 1468.995 | 591.156 | 668.055 | -5772.819 | -3968.912 | 148.08 |
| 1500 | 1509.85 | 593.137 | 727.272 | -5772.667 | -3840.082 | 133.722 |
| 1600 | 1548.184 | 594.77 | 786.67 | -5772.521 | -3711.24 | 121.158 |
| 1700 | 1584.284 | 596.132 | 846.217 | -5772.414 | -3582.43 | 110.073 |
| 1800 | 1618.391 | 597.278 | 905.89 | -5772.364 | -3453.606 | 100.22 |
| 1900 | 1650.711 | 598.253 | 965.667 | -5772.387 | -3324.78 | 91.403 |
| 2000 | 1681.419 | 599.089 | 1025.536 | -5772.483 | -3195.941 | 83.468 |
| 2100 | 1710.667 | 599.81 | 1085.481 | -5772.683 | -3067.115 | 76.289 |
| 2200 | 1738.585 | 600.436 | 1145.494 | -5772.98 | -2938.28 | 69.763 |
| 2300 | 1765.288 | 600.985 | 1205.566 | -5773.39 | -2809.427 | 63.803 |
| 2400 | 1790.876 | 601.467 | 1265.689 | -5773.915 | -2680.695 | 58.343 |
| 2500 | 1815.438 | 601.893 | 1325.857 | -5774.559 | -2551.717 | 53.315 |
| 2600 | 1839.052 | 602.271 | 1386.066 | -5775.32 | -2422.905 | 48.676 |
| 2700 | 1861.788 | 602.609 | 1446.31 | -5776.204 | -2293.907 | 44.378 |
| 2800 | 1883.709 | 602.911 | 1506.587 | -8716.222 | -2155.098 | 40.203 |
| 2900 | 1904.871 | 603.183 | 1566.892 | -8706.394 | -1920.934 | 34.599 |
| 3000 | 1925.324 | 603.429 | 1627.222 | -8696.682 | -1687.269 | 29.378 |
| 3100 | 1945.114 | 603.652 | 1687.577 | -8687.097 | -1453.848 | 24.497 |
| 3200 | 1964.282 | 603.854 | 1747.952 | -8677.637 | -1220.699 | 19.926 |
| 3300 | 1982.867 | 604.038 | 1808.347 | -8668.274 | -987.803 | 15.635 |
| 3400 | 2000.902 | 604.206 | 1868.759 | -8659.032 | -754.915 | 11.598 |
| 3500 | 2018.418 | 604.36 | 1929.187 | -8649.897 | -522.689 | 7.801 |
| 3600 | 2035.446 | 604.502 | 1989.631 | -8640.883 | -290.506 | 4.215 |
| 3700 | 2052.01 | 604.632 | 2050.087 | -8631.972 | -58.702 | 0.829 |
| 3800 | 2068.136 | 604.752 | 2110.557 | -8623.162 | 172.675 | -2.374 |
| 3900 | 2083.847 | 604.863 | 2171.038 | -8614.466 | 403.955 | -5.41 |
| 4000 | 2099.162 | 604.966 | 2231.529 | -8605.885 | 635.127 | -8.294 |
| 4100 | 2114.101 | 605.062 | 2292.031 | -8597.42 | 866.175 | -11.035 |
| 4200 | 2128.683 | 605.151 | 2352.541 | -8589.068 | 1096.697 | -13.639 |
| 4300 | 2142.923 | 605.233 | 2413.06 | -8580.826 | 1327.49 | -16.126 |
| 4400 | 2156.838 | 605.311 | 2473.588 | -8572.713 | 1557.669 | -18.492 |
| 4500 | 2170.442 | 605.383 | 2534.122 | -8564.714 | 1787.662 | -20.75 |
| 4600 | 2183.748 | 605.45 | 2594.664 | -8556.865 | 2017.923 | -22.914 |
| 4700 | 2196.77 | 605.513 | 2655.212 | -8549.162 | 2247.467 | -24.977 |
| 4800 | 2209.519 | 605.573 | 2715.767 | -8541.609 | 2477.216 | -26.957 |
| 4900 | 2222.006 | 605.628 | 2776.327 | -8534.209 | 2706.741 | -28.854 |
| 5000 | 2234.242 | 605.681 | 2836.892 | -8527.002 | 2935.963 | -30.671 |
| 5100 | 2246.236 | 605.73 | 2897.463 | -8519.966 | 3165.417 | -32.42 |
| 5200 | 2257.999 | 605.777 | 2958.038 | -8513.141 | 3394.487 | -34.098 |
| 5300 | 2269.538 | 605.821 | 3018.618 | -8506.533 | 3623.256 | -35.709 |
| 5400 | 2280.863 | 605.862 | 3079.202 | -8500.162 | 3852.16 | -37.262 |
| 5500 | 2291.98 | 605.902 | 3139.79 | -8494.029 | 4080.69 | -38.755 |
| 5600 | 2302.898 | 605.939 | 3200.383 | -8488.158 | 4309.421 | -40.196 |
| 5700 | 2313.623 | 605.974 | 3260.978 | -8482.581 | 4537.67 | -41.583 |
| 5800 | 2324.162 | 606.008 | 3321.577 | -8477.282 | 4766.079 | -42.923 |
| 5900 | 2334.522 | 606.04 | 3382.18 | -8472.311 | 4994.589 | -44.218 |
| 6000 | 2344.708 | 606.07 | 3442.785 | -8467.476 | 5222.841 | -45.468 |

**(Al2O3)6**

| T (K) | S (J/mol.K) | Cp (J/mol.K) | ddH (kJ/mol) | dHf (kJ/mol) | dGf (kJ/mol) | log Kf |
|---|---|---|---|---|---|---|
| 0 | 0 | 0 | 0 | -6944.302 | -6944.302 | Inf |
| 100 | 362.446 | 166.464 | 7.86 | -6968.374 | -6840.258 | 3572.941 |
| 200 | 537.216 | 350.445 | 34.137 | -6989.92 | -6703.144 | 1750.661 |
| 298.15 | 702.493 | 476.972 | 75.179 | -7001.738 | -6559.543 | 1149.191 |
| 300 | 705.449 | 478.864 | 76.063 | -7001.88 | -6556.796 | 1141.626 |
| 400 | 855.21 | 559.677 | 128.314 | -7006.464 | -6407.588 | 836.735 |
| 500 | 985.916 | 609.806 | 186.982 | -7006.923 | -6257.756 | 653.735 |
| 600 | 1100.135 | 641.83 | 249.678 | -7005.487 | -6108.04 | 531.746 |
| 700 | 1200.779 | 663.125 | 314.994 | -7003.642 | -5958.626 | 444.633 |
| 800 | 1290.345 | 677.849 | 382.086 | -7002.334 | -5809.419 | 379.312 |
| 900 | 1370.825 | 688.387 | 450.425 | -7002.341 | -5660.324 | 328.513 |
| 1000 | 1443.776 | 696.159 | 519.672 | -7131.244 | -5502.018 | 287.392 |
| 1100 | 1510.415 | 702.04 | 589.595 | -7131.002 | -5339.107 | 253.53 |
| 1200 | 1571.704 | 706.589 | 660.035 | -7130.603 | -5176.218 | 225.312 |
| 1300 | 1628.409 | 710.176 | 730.88 | -7130.105 | -5013.376 | 201.437 |
| 1400 | 1681.147 | 713.053 | 802.047 | -7129.555 | -4850.561 | 180.974 |
| 1500 | 1730.425 | 715.392 | 873.473 | -7129.007 | -4687.812 | 163.242 |
| 1600 | 1776.659 | 717.32 | 945.112 | -7128.471 | -4525.074 | 147.727 |
| 1700 | 1820.196 | 718.927 | 1016.926 | -7127.985 | -4362.398 | 134.039 |
| 1800 | 1861.328 | 720.279 | 1088.889 | -7127.569 | -4199.725 | 121.871 |
| 1900 | 1900.303 | 721.429 | 1160.976 | -7127.242 | -4037.068 | 110.985 |
| 2000 | 1937.333 | 722.413 | 1233.169 | -7127.007 | -3874.417 | 101.188 |
| 2100 | 1972.601 | 723.263 | 1305.454 | -7126.896 | -3711.796 | 92.325 |
| 2200 | 2006.265 | 724.001 | 1377.818 | -7126.904 | -3549.182 | 84.267 |
| 2300 | 2038.462 | 724.647 | 1450.251 | -7127.05 | -3386.562 | 76.91 |
| 2400 | 2069.315 | 725.215 | 1522.745 | -7127.333 | -3224.102 | 70.17 |
| 2500 | 2098.93 | 725.716 | 1595.292 | -7127.761 | -3061.361 | 63.963 |
| 2600 | 2127.402 | 726.162 | 1667.886 | -7128.331 | -2898.835 | 58.238 |
| 2700 | 2154.816 | 726.559 | 1740.522 | -7129.048 | -2736.102 | 52.932 |
| 2800 | 2181.245 | 726.916 | -1813.197 | 10656.727 | -2561.602 | 47.787 |
| 2900 | 2206.76 | 727.236 | -1885.904 | 10644.593 | -2272.693 | 40.935 |
| 3000 | 2231.419 | 727.525 | -1958.643 | 10632.595 | -1984.39 | 34.551 |
| 3100 | 2255.279 | 727.787 | -2031.408 | 10620.754 | -1696.396 | 28.584 |
| 3200 | 2278.389 | 728.024 | -2104.199 | 10609.061 | -1408.738 | 22.995 |
| 3300 | 2300.795 | 728.241 | -2177.013 | 10597.486 | -1121.391 | 17.75 |
| 3400 | 2322.538 | 728.439 | -2249.847 | 10586.056 | -834.064 | 12.814 |
| 3500 | 2343.656 | 728.62 | -2322.7 | 10574.754 | -547.544 | 8.172 |
| 3600 | 2364.184 | 728.787 | -2395.57 | 10563.6 | -261.084 | 3.788 |
| 3700 | 2384.154 | 728.94 | -2468.457 | 10552.567 | 24.912 | -0.352 |
| 3800 | 2403.596 | 729.082 | -2541.358 | 10541.658 | 310.382 | -4.266 |
| 3900 | 2422.536 | 729.212 | -2614.273 | 10530.885 | 595.733 | -7.979 |
| 4000 | 2441 | 729.333 | -2687.2 | 10520.25 | 880.942 | -11.504 |
| 4100 | 2459.01 | 729.446 | -2760.139 | 10509.756 | 1165.995 | -14.855 |
| 4200 | 2476.589 | 729.55 | -2833.089 | 10499.395 | 1450.412 | -18.038 |
| 4300 | 2493.757 | 729.648 | -2906.049 | 10489.168 | 1735.139 | -21.078 |
| 4400 | 2510.532 | 729.738 | -2979.018 | 10479.097 | 2019.125 | -23.97 |
| 4500 | 2526.933 | 729.823 | -3051.996 | 10469.161 | 2302.879 | -26.731 |
| 4600 | 2542.974 | 729.902 | -3124.983 | 10459.405 | 2586.95 | -29.375 |
| 4700 | 2558.672 | 729.977 | -3197.977 | 10449.825 | 2870.154 | -31.898 |
| 4800 | 2574.042 | 730.047 | -3270.978 | 10440.427 | 3153.591 | -34.318 |
| 4900 | 2589.095 | 730.112 | -3343.986 | 10431.211 | 3436.759 | -36.636 |
| 5000 | 2603.846 | 730.174 | -3417 | 10422.226 | 3719.554 | -38.857 |
| 5100 | 2618.306 | 730.232 | -3490.021 | 10413.447 | 4002.616 | -40.995 |
| 5200 | 2632.486 | 730.287 | -3563.047 | 10404.921 | 4285.22 | -43.045 |
| 5300 | 2646.398 | 730.339 | -3636.078 | 10396.657 | 4567.443 | -45.014 |
| 5400 | 2660.05 | 730.387 | -3709.114 | 10388.676 | 4849.833 | -46.912 |
| 5500 | 2673.452 | 730.434 | -3782.155 | 10380.981 | 5131.764 | -48.737 |
| 5600 | 2686.614 | 730.477 | -3855.201 | 10373.602 | 5413.929 | -50.498 |
| 5700 | 2699.543 | 730.519 | -3928.251 | 10366.573 | 5695.514 | -52.193 |
| 5800 | 2712.249 | 730.558 | -4001.305 | 10359.879 | 5977.277 | -53.831 |
| 5900 | 2724.737 | 730.596 | -4074.362 | 10353.581 | 6259.167 | -55.414 |
| 6000 | 2737.017 | 730.631 | -4147.424 | 10347.443 | 6540.733 | -56.941 |

**(Al2O3)7**

| T (K) | S (J/mol.K) | Cp (J/mol.K) | ddH (kJ/mol) | dHf (kJ/mol) | dGf (kJ/mol) | log Kf |
|---|---|---|---|---|---|---|
| 0 | 0 | 0 | 0 | -8255.737 | -8255.737 | Inf |
| 100 | 395.354 | 199.284 | 9.224 | -8283.767 | -8131.548 | 4247.434 |
| 200 | 602.384 | 411.785 | 40.319 | -8308.465 | -7969.021 | 2081.270 |
| 298.15 | 795.938 | 557.428 | 88.375 | -8322.079 | -7799.138 | 1366.360 |
| 300 | 799.393 | 559.617 | 89.408 | -8322.243 | -7795.89 | 1357.369 |
| 400 | 974.314 | 653.551 | 150.436 | -8327.523 | -7619.462 | 994.988 |
| 500 | 1126.955 | 712.246 | 218.951 | -8327.989 | -7442.322 | 777.485 |
| 600 | 1260.385 | 749.922 | 292.192 | -8326.218 | -7265.334 | 632.496 |
| 700 | 1377.998 | 775.055 | 368.521 | -8323.939 | -7088.717 | 528.96 |
| 800 | 1482.696 | 792.469 | 446.947 | -8322.261 | -6912.361 | 451.325 |
| 900 | 1576.795 | 804.951 | 526.851 | -8322.094 | -6736.157 | 390.952 |
| 1000 | 1662.105 | 814.167 | 607.83 | -8472.29 | -6549.226 | 342.092 |
| 1100 | 1740.046 | 821.147 | 689.611 | -8471.803 | -6356.945 | 301.862 |
| 1200 | 1811.736 | 826.549 | 772.007 | -8471.122 | -6164.704 | 268.339 |
| 1300 | 1878.07 | 830.811 | 854.883 | -8470.317 | -5972.538 | 239.977 |
| 1400 | 1939.77 | 834.23 | 938.141 | -8469.446 | -5780.424 | 215.668 |
| 1500 | 1997.423 | 837.011 | 1021.707 | -8468.571 | -5588.401 | 194.603 |
| 1600 | 2051.518 | 839.304 | 1105.527 | -8467.704 | -5396.407 | 176.172 |
| 1700 | 2102.46 | 841.215 | 1189.556 | -8466.891 | -5204.5 | 159.913 |
| 1800 | 2150.589 | 842.825 | 1273.76 | -8466.159 | -5012.612 | 145.46 |
| 1900 | 2196.196 | 844.192 | 1358.112 | -8465.527 | -4820.759 | 132.53 |
| 2000 | 2239.528 | 845.364 | 1442.592 | -8464.998 | -4628.922 | 120.894 |
| 2100 | 2280.799 | 846.375 | 1527.18 | -8464.613 | -4437.135 | 110.366 |
| 2200 | 2320.193 | 847.254 | 1611.862 | -8464.365 | -4245.368 | 100.797 |
| 2300 | 2357.872 | 848.022 | 1696.627 | -8464.275 | -4053.606 | 92.059 |
| 2400 | 2393.978 | 848.698 | 1781.464 | -8464.345 | -3862.041 | 84.054 |
| 2500 | 2428.636 | 849.296 | 1866.364 | -8464.582 | -3670.16 | 76.683 |
| 2600 | 2461.956 | 849.826 | 1951.321 | -8464.983 | -3478.538 | 69.884 |
| 2700 | 2494.038 | 850.299 | 2036.327 | -8465.556 | -3286.685 | 63.584 |
| 2800 | 2524.969 | 850.723 | -2121.379 | 12580.917 | -3081.118 | 57.478 |
| 2900 | 2554.829 | 851.105 | -2206.471 | 12566.493 | -2742.076 | 49.39 |
| 3000 | 2583.689 | 851.449 | -2291.599 | 12552.23 | -2403.758 | 41.853 |
| 3100 | 2611.613 | 851.761 | -2376.759 | 12538.148 | -2065.805 | 34.808 |
| 3200 | 2638.66 | 852.044 | -2461.95 | 12524.238 | -1728.254 | 28.211 |
| 3300 | 2664.883 | 852.302 | -2547.167 | 12510.466 | -1391.075 | 22.019 |
| 3400 | 2690.33 | 852.538 | -2632.409 | 12496.862 | -1053.926 | 16.191 |
| 3500 | 2715.046 | 852.754 | -2717.674 | 12483.407 | -717.728 | 10.711 |
| 3600 | 2739.072 | 852.952 | -2802.96 | 12470.123 | -381.607 | 5.537 |
| 3700 | 2762.444 | 853.135 | -2888.264 | 12456.982 | -46.035 | 0.65 |
| 3800 | 2785.198 | 853.303 | -2973.586 | 12443.984 | 288.919 | -3.971 |
| 3900 | 2807.365 | 853.459 | -3058.924 | 12431.145 | 623.725 | -8.354 |
| 4000 | 2828.975 | 853.603 | -3144.277 | 12418.466 | 958.358 | -12.515 |
| 4100 | 2850.054 | 853.737 | -3229.644 | 12405.951 | 1292.8 | -16.47 |
| 4200 | 2870.628 | 853.861 | -3315.024 | 12393.592 | 1626.497 | -20.228 |
| 4300 | 2890.722 | 853.977 | -3400.416 | 12381.388 | 1960.546 | -23.816 |
| 4400 | 2910.355 | 854.085 | -3485.82 | 12369.365 | 2293.729 | -27.23 |
| 4500 | 2929.55 | 854.186 | -3571.233 | 12357.501 | 2626.635 | -30.489 |
| 4600 | 2948.325 | 854.281 | -3656.657 | 12345.847 | 2959.899 | -33.61 |
| 4700 | 2966.699 | 854.369 | -3742.089 | 12334.398 | 3292.143 | -36.588 |
| 4800 | 2984.687 | 854.452 | -3827.53 | 12323.16 | 3624.664 | -39.444 |
| 4900 | 3002.306 | 854.531 | -3912.979 | 12312.135 | 3956.857 | -42.18 |
| 5000 | 3019.57 | 854.604 | -3998.436 | 12301.379 | 4288.616 | -44.802 |
| 5100 | 3036.495 | 854.673 | -4083.9 | 12290.864 | 4620.672 | -47.325 |
| 5200 | 3053.091 | 854.739 | -4169.371 | 12280.643 | 4952.196 | -49.745 |
| 5300 | 3069.373 | 854.8 | -4254.848 | 12270.727 | 5283.273 | -52.069 |
| 5400 | 3085.352 | 854.858 | -4340.331 | 12261.142 | 5614.532 | -54.309 |
| 5500 | 3101.038 | 854.913 | -4425.819 | 12251.891 | 5945.252 | -56.463 |
| 5600 | 3116.443 | 854.966 | -4511.313 | 12243.008 | 6276.242 | -58.542 |
| 5700 | 3131.576 | 855.015 | -4596.812 | 12234.534 | 6606.545 | -60.541 |
| 5800 | 3146.446 | 855.062 | -4682.316 | 12226.45 | 6937.063 | -62.474 |
| 5900 | 3161.064 | 855.107 | -4767.825 | 12218.826 | 7267.708 | -64.343 |
| 6000 | 3175.436 | 855.149 | -4853.337 | 12211.392 | 7597.983 | -66.146 |

**(Al2O3)8:**

| T (K) | S (J/mol.K) | Cp (J/mol.K) | ddH (kJ/mol) | dHf (kJ/mol) | dGf (kJ/mol) | log Kf |
|---|---|---|---|---|---|---|
| 0 | 0 | 0 | 0 | -9531.277 | -9531.277 | Inf |
| 100 | 393.453 | 203.432 | 8.955 | -9564.898 | -9385.095 | 4902.212 |
| 200 | 616.054 | 457.523 | 42.539 | -9595.078 | -9192.664 | 2400.849 |
| 298.15 | 833.999 | 633.306 | 96.689 | -9611.408 | -8991.206 | 1575.203 |
| 300 | 837.924 | 635.922 | 97.863 | -9611.602 | -8987.354 | 1564.819 |
| 400 | 1037.416 | 747.229 | 167.475 | -9617.77 | -8778.123 | 1146.292 |
| 500 | 1212.122 | 815.813 | 245.895 | -9618.186 | -8568.081 | 895.09 |
| 600 | 1365.004 | 859.444 | 329.814 | -9615.947 | -8358.246 | 727.641 |
| 700 | 1499.805 | 888.379 | 417.299 | -9613.09 | -8148.872 | 608.069 |
| 800 | 1619.813 | 908.348 | 507.194 | -9610.907 | -7939.835 | 518.412 |
| 900 | 1727.67 | 922.623 | 598.78 | -9610.449 | -7731.006 | 448.69 |
| 1000 | 1825.45 | 933.142 | 691.594 | -9781.835 | -7509.949 | 392.275 |
| 1100 | 1914.777 | 941.096 | 785.323 | -9781.014 | -7282.8 | 345.827 |
| 1200 | 1996.938 | 947.246 | 879.753 | -9779.972 | -7055.724 | 307.124 |
| 1300 | 2072.957 | 952.093 | 974.729 | -9778.792 | -6828.755 | 274.379 |
| 1400 | 2143.662 | 955.979 | 1070.14 | -9777.537 | -6601.863 | 246.316 |
| 1500 | 2209.729 | 959.138 | 1165.901 | -9776.28 | -6375.097 | 221.998 |
| 1600 | 2271.716 | 961.741 | 1261.949 | -9775.036 | -6148.38 | 200.722 |
| 1700 | 2330.088 | 963.911 | 1358.235 | -9773.854 | -5921.776 | 181.952 |
| 1800 | 2385.237 | 965.737 | 1454.72 | -9772.765 | -5695.212 | 165.269 |
| 1900 | 2437.495 | 967.288 | 1551.373 | -9771.792 | -5468.7 | 150.343 |
| 2000 | 2487.145 | 968.617 | 1648.17 | -9770.939 | -5242.221 | 136.911 |
| 2100 | 2534.432 | 969.763 | 1745.09 | -9770.251 | -5015.809 | 124.76 |
| 2200 | 2579.569 | 970.759 | 1842.117 | -9769.72 | -4789.432 | 113.714 |
| 2300 | 2622.741 | 971.631 | 1939.238 | -9769.371 | -4563.074 | 103.629 |
| 2400 | 2664.109 | 972.396 | 2036.44 | -9769.205 | -4336.951 | 94.39 |
| 2500 | 2703.818 | 973.073 | 2133.714 | -9769.231 | -4110.476 | 85.883 |
| 2600 | 2741.995 | 973.674 | 2231.052 | -9769.445 | -3884.311 | 78.036 |
| 2700 | 2778.752 | 974.211 | 2328.447 | -9769.854 | -3657.885 | 70.765 |
| 2800 | 2814.191 | 974.691 | -2425.892 | 14472.881 | -3415.801 | 63.722 |
| 2900 | 2848.401 | 975.123 | -2523.384 | 14456.153 | -3021.177 | 54.417 |
| 3000 | 2881.466 | 975.513 | -2620.916 | 14439.609 | -2627.391 | 45.746 |
| 3100 | 2913.459 | 975.866 | -2718.485 | 14423.272 | -2234.031 | 37.643 |
| 3200 | 2944.447 | 976.187 | -2816.088 | 14407.133 | -1841.139 | 30.053 |
| 3300 | 2974.49 | 976.479 | -2913.721 | 14391.152 | -1448.677 | 22.93 |
| 3400 | 3003.645 | 976.746 | -3011.383 | 14375.362 | -1056.259 | 16.227 |
| 3500 | 3031.962 | 976.99 | -3109.07 | 14359.743 | -664.936 | 9.924 |
| 3600 | 3059.488 | 977.215 | -3206.78 | 14344.321 | -273.706 | 3.971 |
| 3700 | 3086.265 | 977.422 | -3304.512 | 14329.061 | 116.89 | -1.65 |
| 3800 | 3112.334 | 977.612 | -3402.264 | 14313.965 | 506.773 | -6.966 |
| 3900 | 3137.73 | 977.789 | -3500.034 | 14299.051 | 896.48 | -12.007 |
| 4000 | 3162.488 | 977.952 | -3597.821 | 14284.32 | 1285.984 | -16.793 |
| 4100 | 3186.638 | 978.104 | -3695.624 | 14269.777 | 1675.263 | -21.343 |
| 4200 | 3210.21 | 978.245 | -3793.442 | 14255.411 | 2063.681 | -25.665 |
| 4300 | 3233.23 | 978.376 | -3891.273 | 14241.224 | 2452.503 | -29.792 |
| 4400 | 3255.724 | 978.498 | -3989.117 | 14227.244 | 2840.321 | -33.718 |
| 4500 | 3277.714 | 978.613 | -4086.972 | 14213.445 | 3227.826 | -37.467 |
| 4600 | 3299.225 | 978.72 | -4184.839 | 14199.886 | 3615.726 | -41.057 |
| 4700 | 3320.274 | 978.82 | -4282.716 | 14186.561 | 4002.468 | -44.482 |
| 4800 | 3340.883 | 978.914 | -4380.603 | 14173.478 | 4389.51 | -47.767 |
| 4900 | 3361.068 | 979.002 | -4478.498 | 14160.639 | 4776.176 | -50.914 |
| 5000 | 3380.847 | 979.086 | -4576.403 | 14148.106 | 5162.339 | -53.93 |
| 5100 | 3400.237 | 979.164 | -4674.315 | 14135.85 | 5548.84 | -56.831 |
| 5200 | 3419.251 | 979.238 | -4772.235 | 14123.93 | 5934.722 | -59.614 |
| 5300 | 3437.904 | 979.308 | -4870.163 | 14112.358 | 6320.096 | -62.288 |
| 5400 | 3456.21 | 979.373 | -4968.097 | 14101.164 | 6705.674 | -64.864 |
| 5500 | 3474.181 | 979.436 | -5066.037 | 14090.352 | 7090.627 | -67.34 |
| 5600 | 3491.83 | 979.495 | -5163.984 | 14079.961 | 7475.884 | -69.731 |
| 5700 | 3509.167 | 979.551 | -5261.936 | 14070.037 | 7860.354 | -72.031 |
| 5800 | 3526.204 | 979.604 | -5359.894 | 14060.559 | 8245.058 | -74.254 |
| 5900 | 3542.95 | 979.655 | -5457.857 | 14051.608 | 8629.916 | -76.403 |
| 6000 | 3559.415 | 979.702 | -5555.825 | 14042.872 | 9014.342 | -78.476 |

**(Al2O3)9:**

| T (K) | S (J/mol.K) | Cp (J/mol.K) | ddH (kJ/mol) | dHf (kJ/mol) | dGf (kJ/mol) | log Kf |
|---|---|---|---|---|---|---|
| 0 | 0 | 0 | 0 | -11022.189 | -11022.189 | Inf |
| 100 | 405.372 | 228.321 | 9.632 | -11060.455 | -10854.451 | 5669.72 |
| 200 | 657.124 | 517.789 | 47.626 | -11094.196 | -10634.293 | 2777.359 |
| 298.15 | 903.393 | 714.665 | 108.807 | -11112.305 | -10404.186 | 1822.748 |
| 300 | 907.823 | 717.591 | 110.132 | -11112.519 | -10399.788 | 1810.742 |
| 400 | 1132.782 | 842.204 | 188.628 | -11119.275 | -10160.949 | 1326.868 |
| 500 | 1329.654 | 919.198 | 276.997 | -11119.597 | -9921.237 | 1036.452 |
| 600 | 1501.902 | 968.298 | 371.547 | -11116.937 | -9681.787 | 842.864 |
| 700 | 1653.778 | 1000.92 | 470.113 | -11113.577 | -9442.881 | 704.628 |
| 800 | 1788.992 | 1023.464 | 571.398 | -11110.968 | -9204.375 | 600.977 |
| 900 | 1910.52 | 1039.595 | 674.594 | -11110.291 | -8966.12 | 520.374 |
| 1000 | 2020.699 | 1051.49 | 779.177 | -11302.933 | -8714.129 | 455.174 |
| 1100 | 2121.357 | 1060.489 | 884.796 | -11301.836 | -8455.302 | 401.504 |
| 1200 | 2213.943 | 1067.45 | 991.207 | -11300.486 | -8196.573 | 356.783 |
| 1300 | 2299.61 | 1072.939 | 1098.237 | -11298.977 | -7937.979 | 318.948 |
| 1400 | 2379.29 | 1077.339 | 1205.759 | -11297.38 | -7679.486 | 286.522 |
| 1500 | 2453.745 | 1080.918 | 1313.677 | -11295.779 | -7421.149 | 258.424 |
| 1600 | 2523.603 | 1083.868 | 1421.921 | -11294.19 | -7162.878 | 233.841 |
| 1700 | 2589.388 | 1086.326 | 1530.435 | -11292.668 | -6904.747 | 212.154 |
| 1800 | 2651.54 | 1088.395 | 1639.174 | -11291.249 | -6646.67 | 192.879 |
| 1900 | 2710.435 | 1090.153 | 1748.103 | -11289.96 | -6388.663 | 175.634 |
| 2000 | 2766.392 | 1091.66 | 1857.196 | -11288.804 | -6130.704 | 160.116 |
| 2100 | 2819.687 | 1092.959 | 1966.428 | -11287.833 | -5872.833 | 146.077 |
| 2200 | 2870.558 | 1094.089 | 2075.782 | -11287.037 | -5615.008 | 133.316 |
| 2300 | 2919.214 | 1095.077 | 2185.241 | -11286.447 | -5357.214 | 121.665 |
| 2400 | 2965.839 | 1095.945 | 2294.793 | -11286.06 | -5099.694 | 110.991 |
| 2500 | 3010.594 | 1096.712 | 2404.427 | -11285.889 | -4841.786 | 101.162 |
| 2600 | 3053.621 | 1097.394 | 2514.133 | -11285.929 | -4584.232 | 92.097 |
| 2700 | 3095.049 | 1098.002 | 2623.903 | -11286.188 | -4326.397 | 83.698 |
| 2800 | 3134.99 | 1098.547 | 2733.731 | -16576.891 | -4050.947 | 75.57 |
| 2900 | 3173.548 | 1099.037 | 2843.611 | -16557.871 | -3603.904 | 64.913 |
| 3000 | 3210.815 | 1099.479 | 2953.537 | -16539.056 | -3157.808 | 54.982 |
| 3100 | 3246.873 | 1099.879 | 3063.505 | -16520.474 | -2712.197 | 45.7 |
| 3200 | 3281.799 | 1100.243 | 3173.512 | -16502.114 | -2267.119 | 37.006 |
| 3300 | 3315.66 | 1100.574 | 3283.553 | -16483.932 | -1822.532 | 28.848 |
| 3400 | 3348.52 | 1100.877 | 3393.626 | -16465.965 | -1378.001 | 21.17 |
| 3500 | 3380.436 | 1101.155 | 3503.727 | -16448.19 | -934.708 | 13.95 |
| 3600 | 3411.46 | 1101.409 | 3613.856 | -16430.635 | -491.524 | 7.132 |
| 3700 | 3441.641 | 1101.644 | 3724.009 | -16413.263 | -49.062 | 0.693 |
| 3800 | 3471.023 | 1101.86 | 3834.184 | -16396.076 | 392.594 | -5.397 |
| 3900 | 3499.647 | 1102.06 | 3944.38 | -16379.093 | 834.046 | -11.171 |
| 4000 | 3527.551 | 1102.245 | 4054.595 | -16362.316 | 1275.268 | -16.653 |
| 4100 | 3554.77 | 1102.417 | 4164.829 | -16345.75 | 1716.23 | -21.865 |
| 4200 | 3581.338 | 1102.577 | 4275.078 | -16329.384 | 2156.216 | -26.816 |
| 4300 | 3607.284 | 1102.726 | 4385.344 | -16313.218 | 2596.654 | -31.543 |
| 4400 | 3632.637 | 1102.865 | 4495.623 | -16297.286 | 3035.956 | -36.041 |
| 4500 | 3657.423 | 1102.994 | 4605.916 | -16281.556 | 3474.898 | -40.335 |
| 4600 | 3681.666 | 1103.116 | 4716.222 | -16266.096 | 3914.292 | -44.448 |
| 4700 | 3705.392 | 1103.229 | 4826.539 | -16250.9 | 4352.363 | -48.371 |
| 4800 | 3728.619 | 1103.336 | 4936.868 | -16235.976 | 4790.782 | -52.134 |
| 4900 | 3751.37 | 1103.437 | 5047.206 | -16221.326 | 5228.765 | -55.739 |
| 5000 | 3773.664 | 1103.531 | 5157.555 | -16207.02 | 5666.175 | -59.193 |
| 5100 | 3795.518 | 1103.62 | 5267.912 | -16193.026 | 6103.967 | -62.517 |
| 5200 | 3816.949 | 1103.704 | 5378.278 | -16179.41 | 6541.056 | -65.705 |
| 5300 | 3837.973 | 1103.783 | 5488.653 | -16166.186 | 6977.571 | -68.767 |
| 5400 | 3858.606 | 1103.857 | 5599.035 | -16153.386 | 7414.309 | -71.718 |
| 5500 | 3878.861 | 1103.928 | 5709.424 | -16141.016 | 7850.344 | -74.555 |
| 5600 | 3898.753 | 1103.995 | 5819.82 | -16129.121 | 8286.717 | -77.294 |
| 5700 | 3918.294 | 1104.059 | 5930.223 | -16117.749 | 8722.198 | -79.929 |
| 5800 | 3937.496 | 1104.119 | 6040.632 | -16106.88 | 9157.943 | -82.475 |
| 5900 | 3956.37 | 1104.176 | 6151.047 | -16096.604 | 9593.858 | -84.937 |
| 6000 | 3974.929 | 1104.231 | 6261.467 | -16086.57 | 10029.273 | -87.312 |

**(Al2O3)10:**

| T (K) | S (J/mol.K) | Cp (J/mol.K) | ddH (kJ/mol) | dHf (kJ/mol) | dGf (kJ/mol) | log Kf |
|---|---|---|---|---|---|---|
| 0 | 0 | 0 | 0 | -12287.473 | -12287.473 | Inf |
| 100 | 424.733 | 248.801 | 10.608 | -12330.085 | -12098.624 | 6319.60 |
| 200 | 699.553 | 568.718 | 52.108 | -12368.29 | -11851.17 | 3095.171 |
| 298.15 | 971.344 | 791.551 | 119.648 | -12388.85 | -11592.384 | 2030.913 |
| 300 | 976.251 | 794.864 | 121.115 | -12389.093 | -11587.438 | 2017.528 |
| 400 | 1225.861 | 935.669 | 208.219 | -12396.714 | -11318.793 | 1478.065 |
| 500 | 1444.712 | 1022.277 | 306.456 | -12397.022 | -11049.171 | 1154.285 |
| 600 | 1636.32 | 1077.312 | 411.632 | -12393.946 | -10779.859 | 938.459 |
| 700 | 1805.31 | 1113.786 | 521.306 | -12390.057 | -10511.173 | 784.344 |
| 800 | 1955.776 | 1138.946 | 634.016 | -12386.987 | -10242.956 | 668.788 |
| 900 | 2091.019 | 1156.926 | 748.857 | -12386.056 | -9975.041 | 578.929 |
| 1000 | 2213.632 | 1170.172 | 865.244 | -12599.919 | -9691.881 | 506.246 |
| 1100 | 2325.653 | 1180.186 | 982.784 | -12598.514 | -9401.147 | 446.418 |
| 1200 | 2428.688 | 1187.928 | 1101.206 | -12596.827 | -9110.537 | 396.567 |
| 1300 | 2524.023 | 1194.03 | 1220.316 | -12594.962 | -8820.091 | 354.391 |
| 1400 | 2612.696 | 1198.921 | 1339.972 | -12593.001 | -8529.776 | 318.246 |
| 1500 | 2695.553 | 1202.897 | 1460.069 | -12591.034 | -8239.644 | 286.926 |
| 1600 | 2773.294 | 1206.173 | 1580.528 | -12589.08 | -7949.599 | 259.525 |
| 1700 | 2846.502 | 1208.903 | 1701.286 | -12587.202 | -7659.722 | 235.352 |
| 1800 | 2915.668 | 1211.201 | 1822.294 | -12585.439 | -7369.918 | 213.867 |
| 1900 | 2981.208 | 1213.153 | 1943.515 | -12583.818 | -7080.198 | 194.646 |
| 2000 | 3043.478 | 1214.825 | 2064.916 | -12582.347 | -6790.543 | 177.349 |
| 2100 | 3102.786 | 1216.268 | 2186.472 | -12581.081 | -6500.995 | 161.701 |
| 2200 | 3159.396 | 1217.522 | 2308.163 | -12580.01 | -6211.507 | 147.478 |
| 2300 | 3213.542 | 1218.618 | 2429.971 | -12579.167 | -5922.063 | 134.493 |
| 2400 | 3265.426 | 1219.581 | 2551.882 | -12578.551 | -5632.93 | 122.596 |
| 2500 | 3315.23 | 1220.433 | 2673.884 | -12578.174 | -5343.374 | 111.642 |
| 2600 | 3363.111 | 1221.19 | 2795.966 | -12578.032 | -5054.22 | 101.539 |
| 2700 | 3409.212 | 1221.864 | 2918.119 | -12578.134 | -4764.758 | 92.179 |
| 2800 | 3453.659 | 1222.469 | 3040.336 | -18456.507 | -4455.734 | 83.122 |
| 2900 | 3496.567 | 1223.012 | 3162.611 | -18435.187 | -3956.058 | 71.255 |
| 3000 | 3538.038 | 1223.503 | 3284.937 | -18414.096 | -3457.44 | 60.199 |
| 3100 | 3578.163 | 1223.947 | 3407.31 | -18393.263 | -2959.364 | 49.864 |
| 3200 | 3617.029 | 1224.351 | 3529.725 | -18372.678 | -2461.891 | 40.186 |
| 3300 | 3654.71 | 1224.718 | 3652.179 | -18352.289 | -1964.968 | 31.102 |
| 3400 | 3691.276 | 1225.054 | 3774.668 | -18332.14 | -1468.11 | 22.554 |
| 3500 | 3726.792 | 1225.362 | 3897.189 | -18312.204 | -972.634 | 14.516 |
| 3600 | 3761.315 | 1225.644 | 4019.739 | -18292.514 | -477.284 | 6.925 |
| 3700 | 3794.9 | 1225.905 | 4142.317 | -18273.026 | 17.258 | -0.244 |
| 3800 | 3827.596 | 1226.144 | 4264.919 | -18253.744 | 510.899 | -7.023 |
| 3900 | 3859.449 | 1226.366 | 4387.545 | -18234.688 | 1004.308 | -13.451 |
| 4000 | 3890.501 | 1226.572 | 4510.192 | -18215.861 | 1497.455 | -19.555 |
| 4100 | 3920.79 | 1226.762 | 4632.859 | -18197.269 | 1990.311 | -25.357 |
| 4200 | 3950.354 | 1226.94 | 4755.544 | -18178.899 | 2482.081 | -30.869 |
| 4300 | 3979.227 | 1227.105 | 4878.246 | -18160.752 | 2974.341 | -36.131 |
| 4400 | 4007.439 | 1227.259 | 5000.965 | -18142.863 | 3465.343 | -41.138 |
| 4500 | 4035.021 | 1227.403 | 5123.698 | -18125.2 | 3955.935 | -45.919 |
| 4600 | 4061.999 | 1227.537 | 5246.445 | -18107.838 | 4447.024 | -50.497 |
| 4700 | 4088.4 | 1227.663 | 5369.205 | -18090.768 | 4936.647 | -54.864 |
| 4800 | 4114.248 | 1227.782 | 5491.977 | -18074.001 | 5426.64 | -59.053 |
| 4900 | 4139.565 | 1227.893 | 5614.761 | -18057.537 | 5916.154 | -63.066 |
| 5000 | 4164.373 | 1227.998 | 5737.556 | -18041.457 | 6405.028 | -66.912 |
| 5100 | 4188.692 | 1228.096 | 5860.36 | -18025.723 | 6894.32 | -70.611 |
| 5200 | 4212.54 | 1228.189 | 5983.175 | -18010.408 | 7382.83 | -74.161 |
| 5300 | 4235.935 | 1228.277 | 6105.998 | -17995.53 | 7870.696 | -77.569 |
| 5400 | 4258.895 | 1228.36 | 6228.83 | -17981.123 | 8358.808 | -80.854 |
| 5500 | 4281.435 | 1228.439 | 6351.67 | -17967.193 | 8846.132 | -84.013 |
| 5600 | 4303.571 | 1228.513 | 6474.518 | -17953.79 | 9333.828 | -87.061 |
| 5700 | 4325.315 | 1228.583 | 6597.372 | -17940.971 | 9820.537 | -89.994 |
| 5800 | 4346.683 | 1228.65 | 6720.234 | -17928.709 | 10307.529 | -92.828 |
| 5900 | 4367.687 | 1228.714 | 6843.102 | -17917.106 | 10794.701 | -95.568 |
| 6000 | 4388.338 | 1228.774 | 6965.977 | -17905.771 | 11281.331 | -98.212 |

**(MgSiO3)1:**

| T (K) | S (J/mol.K) | Cp (J/mol.K) | ddH (kJ/mol) | dHf (kJ/mol) | dGf (kJ/mol) | log Kf |
|---|---|---|---|---|---|---|
| 0.00 | 0.000 | 0.000 | 0.000 | -589.172 | -589.172 | Inf |
| 100.00 | 242.269 | 41.997 | 3.597 | -590.839 | -587.736 | 306.998 |
| 200.00 | 276.547 | 58.961 | 8.641 | -593.347 | -583.650 | 152.432 |
| 298.15 | 302.871 | 73.108 | 15.158 | -595.254 | -578.456 | 101.342 |
| 300.00 | 303.324 | 73.332 | 15.293 | -595.283 | -578.352 | 100.699 |
| 400.00 | 325.876 | 83.264 | 23.157 | -596.553 | -572.501 | 74.760 |
| 500.00 | 345.216 | 89.874 | 31.836 | -597.406 | -566.383 | 59.169 |
| 600.00 | 362.021 | 94.326 | 41.060 | -598.065 | -560.114 | 48.762 |
| 700.00 | 376.807 | 97.404 | 50.655 | -598.673 | -553.743 | 41.320 |
| 800.00 | 389.964 | 99.593 | 60.511 | -599.318 | -547.279 | 35.733 |
| 900.00 | 401.791 | 101.193 | 70.555 | -600.064 | -540.731 | 31.383 |
| 1000.00 | 412.518 | 102.391 | 80.737 | -609.542 | -533.381 | 27.861 |
| 1100.00 | 422.322 | 103.309 | 91.024 | -610.605 | -525.714 | 24.964 |
| 1200.00 | 431.343 | 104.026 | 101.392 | -611.691 | -517.949 | 22.545 |
| 1300.00 | 439.693 | 104.595 | 111.824 | -612.804 | -510.093 | 20.496 |
| 1400.00 | 447.461 | 105.054 | 122.307 | -741.364 | -498.979 | 18.617 |
| 1500.00 | 454.723 | 105.430 | 132.832 | -741.205 | -481.675 | 16.773 |
| 1600.00 | 461.537 | 105.740 | 143.391 | -741.090 | -464.377 | 15.160 |
| 1700.00 | 467.956 | 106.000 | 153.978 | -791.205 | -446.637 | 13.723 |
| 1800.00 | 474.021 | 106.219 | 164.590 | -790.964 | -426.374 | 12.373 |
| 1900.00 | 479.769 | 106.406 | 175.221 | -790.741 | -406.126 | 11.165 |
| 2000.00 | 485.231 | 106.566 | 185.870 | -790.534 | -385.888 | 10.078 |
| 2100.00 | 490.434 | 106.705 | 196.533 | -790.348 | -365.662 | 9.095 |
| 2200.00 | 495.401 | 106.825 | 207.210 | -790.182 | -345.441 | 8.202 |
| 2300.00 | 500.152 | 106.931 | 217.898 | -790.040 | -325.232 | 7.386 |
| 2400.00 | 504.704 | 107.024 | 228.596 | -789.922 | -305.020 | 6.639 |
| 2500.00 | 509.075 | 107.106 | 239.302 | -789.829 | -284.824 | 5.951 |
| 2600.00 | 513.277 | 107.179 | 250.017 | -789.761 | -264.617 | 5.316 |
| 2700.00 | 517.323 | 107.244 | 260.738 | -789.722 | -244.423 | 4.729 |
| 2800.00 | 521.225 | 107.303 | 271.465 | -789.711 | -224.230 | 4.183 |
| 2900.00 | 524.991 | 107.355 | 282.198 | -789.725 | -204.033 | 3.675 |
| 3000.00 | 528.631 | 107.403 | 292.936 | -789.772 | -183.838 | 3.201 |
| 3100.00 | 532.154 | 107.446 | 303.679 | -789.846 | -163.636 | 2.757 |
| 3200.00 | 535.566 | 107.485 | 314.425 | -789.954 | -143.439 | 2.341 |
| 3300.00 | 538.874 | 107.521 | 325.176 | -790.091 | -123.229 | 1.951 |
| 3400.00 | 542.084 | 107.553 | 335.929 | -790.263 | -103.023 | 1.583 |
| 3500.00 | 545.202 | 107.583 | 346.686 | -790.467 | -82.804 | 1.236 |
| 3600.00 | 548.233 | 107.611 | 357.446 | -1174.860 | -52.124 | 0.756 |
| 3700.00 | 551.182 | 107.636 | 368.208 | -1174.723 | -20.937 | 0.296 |
| 3800.00 | 554.053 | 107.659 | 378.973 | -1174.625 | 10.249 | -0.141 |
| 3900.00 | 556.850 | 107.681 | 389.740 | -1174.570 | 41.429 | -0.555 |
| 4000.00 | 559.576 | 107.701 | 400.509 | -1174.558 | 72.610 | -0.948 |
| 4100.00 | 562.236 | 107.719 | 411.280 | -1174.589 | 103.787 | -1.322 |
| 4200.00 | 564.832 | 107.737 | 422.053 | -1174.667 | 134.967 | -1.679 |
| 4300.00 | 567.367 | 107.753 | 432.827 | -1174.792 | 166.152 | -2.018 |
| 4400.00 | 569.844 | 107.768 | 443.603 | -1174.968 | 197.341 | -2.343 |
| 4500.00 | 572.266 | 107.782 | 454.381 | -1175.195 | 228.526 | -2.653 |
| 4600.00 | 574.636 | 107.795 | 465.160 | -1175.478 | 259.722 | -2.949 |
| 4700.00 | 576.954 | 107.807 | 475.940 | -1175.815 | 290.928 | -3.233 |
| 4800.00 | 579.224 | 107.819 | 486.721 | -1176.212 | 322.134 | -3.505 |
| 4900.00 | 581.447 | 107.830 | 497.504 | -1176.670 | 353.357 | -3.767 |
| 5000.00 | 583.626 | 107.840 | 508.287 | -1177.194 | 384.586 | -4.018 |
| 5100.00 | 585.761 | 107.849 | 519.072 | -1177.782 | 415.833 | -4.259 |
| 5200.00 | 587.856 | 107.858 | 529.857 | -1178.439 | 447.084 | -4.491 |
| 5300.00 | 589.910 | 107.867 | 540.643 | -1179.169 | 478.350 | -4.714 |
| 5400.00 | 591.926 | 107.875 | 551.430 | -1179.976 | 509.636 | -4.930 |
| 5500.00 | 593.906 | 107.883 | 562.218 | -1180.857 | 540.921 | -5.137 |
| 5600.00 | 595.850 | 107.890 | 573.007 | -1181.819 | 572.246 | -5.338 |
| 5700.00 | 597.760 | 107.897 | 583.796 | -1182.866 | 603.574 | -5.531 |
| 5800.00 | 599.636 | 107.903 | 594.586 | -1183.999 | 634.928 | -5.718 |
| 5900.00 | 601.481 | 107.910 | 605.377 | -1185.220 | 666.297 | -5.899 |
| 6000.00 | 603.294 | 107.915 | 616.168 | -1186.535 | 697.690 | -6.074 |

**(MgSiO3)2:**

| T (K) | S (J/mol.K) | Cp (J/mol.K) | ddH (kJ/mol) | dHf (kJ/mol) | dGf (kJ/mol) | log Kf |
|---|---|---|---|---|---|---|
| 0 | 0 | 0 | 0 | -1761.005 | -1761.005 | Inf |
| 100 | 267.397 | 57.435 | 4.03 | -1767.503 | -1739.583 | 908.654 |
| 200 | 324.102 | 111.713 | 12.526 | -1774.112 | -1708.918 | 446.318 |
| 298.15 | 376.651 | 151.338 | 25.573 | -1777.913 | -1676.013 | 293.627 |
| 300 | 377.589 | 151.929 | 25.854 | -1777.96 | -1675.379 | 291.706 |
| 400 | 425.052 | 177.252 | 42.413 | -1779.668 | -1640.884 | 214.275 |
| 500 | 466.442 | 193.113 | 60.991 | -1780.155 | -1606.113 | 167.787 |
| 600 | 502.621 | 203.359 | 80.85 | -1780.062 | -1571.307 | 136.793 |
| 700 | 534.519 | 210.235 | 101.552 | -1779.765 | -1536.539 | 114.657 |
| 800 | 562.923 | 215.023 | 122.828 | -1779.491 | -1501.809 | 98.057 |
| 900 | 588.459 | 218.469 | 144.512 | -1779.387 | -1467.11 | 85.148 |
| 1000 | 611.615 | 221.02 | 166.492 | -1796.727 | -1430.984 | 74.746 |
| 1100 | 632.775 | 222.957 | 188.695 | -1797.225 | -1394.386 | 66.213 |
| 1200 | 652.242 | 224.46 | 211.069 | -1797.758 | -1357.742 | 59.1 |
| 1300 | 670.257 | 225.647 | 233.577 | -1798.341 | -1321.049 | 53.08 |
| 1400 | 687.016 | 226.6 | 256.191 | -2053.812 | -1277.974 | 47.681 |
| 1500 | 702.677 | 227.376 | 278.891 | -2051.844 | -1222.63 | 42.575 |
| 1600 | 717.372 | 228.017 | 301.662 | -2049.962 | -1167.411 | 38.112 |
| 1700 | 731.212 | 228.551 | 324.491 | -2148.537 | -1111.411 | 34.149 |
| 1800 | 744.289 | 229.001 | 347.369 | -2146.4 | -1050.464 | 30.483 |
| 1900 | 756.681 | 229.384 | 370.289 | -2144.296 | -989.637 | 27.207 |
| 2000 | 768.455 | 229.712 | 393.244 | -2142.225 | -928.919 | 24.261 |
| 2100 | 779.67 | 229.995 | 416.229 | -2140.194 | -868.305 | 21.598 |
| 2200 | 790.375 | 230.241 | 439.242 | -2138.203 | -807.781 | 19.179 |
| 2300 | 800.615 | 230.456 | 462.277 | -2136.261 | -747.36 | 16.973 |
| 2400 | 810.427 | 230.646 | 485.332 | -2134.365 | -687.006 | 14.952 |
| 2500 | 819.846 | 230.813 | 508.405 | -2132.519 | -626.749 | 13.095 |
| 2600 | 828.901 | 230.962 | 531.494 | -2130.724 | -566.538 | 11.382 |
| 2700 | 837.621 | 231.095 | 554.597 | -2128.984 | -506.419 | 9.797 |
| 2800 | 846.027 | 231.214 | 577.713 | -2127.3 | -446.353 | 8.327 |
| 2900 | 854.143 | 231.321 | 600.839 | -2125.669 | -386.35 | 6.959 |
| 3000 | 861.986 | 231.417 | 623.976 | -2124.101 | -326.405 | 5.683 |
| 3100 | 869.576 | 231.505 | 647.123 | -2122.588 | -266.5 | 4.49 |
| 3200 | 876.927 | 231.584 | 670.277 | -2121.142 | -206.655 | 3.373 |
| 3300 | 884.055 | 231.657 | 693.439 | -2119.757 | -146.845 | 2.324 |
| 3400 | 890.971 | 231.723 | 716.608 | -2118.438 | -87.087 | 1.338 |
| 3500 | 897.689 | 231.784 | 739.784 | -2117.183 | -27.354 | 0.408 |
| 3600 | 904.22 | 231.839 | 762.965 | -2884.308 | 53.249 | -0.773 |
| 3700 | 910.572 | 231.89 | 786.151 | -2882.372 | 134.83 | -1.903 |
| 3800 | 916.757 | 231.938 | 809.343 | -2880.514 | 216.361 | -2.974 |
| 3900 | 922.782 | 231.981 | 832.539 | -2878.742 | 297.836 | -3.989 |
| 4000 | 928.656 | 232.022 | 855.739 | -2877.056 | 379.264 | -4.953 |
| 4100 | 934.386 | 232.06 | 878.943 | -2875.457 | 460.648 | -5.869 |
| 4200 | 939.978 | 232.095 | 902.151 | -2873.95 | 541.999 | -6.741 |
| 4300 | 945.44 | 232.127 | 925.362 | -2872.538 | 623.315 | -7.572 |
| 4400 | 950.777 | 232.157 | 948.576 | -2871.228 | 704.6 | -8.365 |
| 4500 | 955.994 | 232.186 | 971.793 | -2870.021 | 785.842 | -9.122 |
| 4600 | 961.098 | 232.212 | 995.013 | -2868.924 | 867.077 | -9.846 |
| 4700 | 966.092 | 232.237 | 1018.236 | -2867.935 | 948.287 | -10.539 |
| 4800 | 970.982 | 232.261 | 1041.461 | -2867.067 | 1029.463 | -11.203 |
| 4900 | 975.771 | 232.283 | 1064.688 | -2866.322 | 1110.636 | -11.839 |
| 5000 | 980.464 | 232.303 | 1087.917 | -2865.706 | 1191.794 | -12.45 |
| 5100 | 985.064 | 232.323 | 1111.148 | -2865.221 | 1272.945 | -13.037 |
| 5200 | 989.576 | 232.341 | 1134.382 | -2864.871 | 1354.082 | -13.602 |
| 5300 | 994.002 | 232.358 | 1157.617 | -2864.669 | 1435.206 | -14.145 |
| 5400 | 998.345 | 232.375 | 1180.853 | -2864.62 | 1516.341 | -14.668 |
| 5500 | 1002.609 | 232.39 | 1204.091 | -2864.72 | 1597.452 | -15.171 |
| 5600 | 1006.797 | 232.405 | 1227.331 | -2864.983 | 1678.605 | -15.657 |
| 5700 | 1010.91 | 232.419 | 1250.572 | -2865.413 | 1759.744 | -16.126 |
| 5800 | 1014.953 | 232.432 | 1273.815 | -2866.016 | 1840.887 | -16.579 |
| 5900 | 1018.926 | 232.444 | 1297.059 | -2866.797 | 1922.051 | -17.016 |
| 6000 | 1022.833 | 232.456 | 1320.304 | -2867.764 | 2003.216 | -17.439 |

**(MgSiO3)3**

| T (K) | S (J/mol.K) | Cp (J/mol.K) | ddH (kJ/mol) | dHf (kJ/mol) | dGf (kJ/mol) | log Kf |
|---|---|---|---|---|---|---|
| 0 | 0 | 0 | 0 | -2938.927 | -2938.927 | Inf |
| 100 | 307.059 | 93.844 | 5.35 | -2949.369 | -2898.085 | 1513.786 |
| 200 | 399.416 | 178.112 | 19.162 | -2958.214 | -2843.076 | 742.526 |
| 298.15 | 481.927 | 234.998 | 39.631 | -2963.017 | -2785.407 | 487.986 |
| 300 | 483.383 | 235.853 | 40.066 | -2963.074 | -2784.304 | 484.785 |
| 400 | 556.707 | 272.889 | 65.643 | -2964.898 | -2724.374 | 355.762 |
| 500 | 620.336 | 296.562 | 94.202 | -2964.936 | -2664.211 | 278.325 |
| 600 | 675.871 | 312.064 | 124.685 | -2964.102 | -2604.134 | 226.707 |
| 700 | 724.814 | 322.561 | 156.449 | -2962.946 | -2544.232 | 189.851 |
| 800 | 768.394 | 329.914 | 189.093 | -2961.805 | -2484.49 | 162.219 |
| 900 | 807.575 | 335.227 | 222.364 | -2960.904 | -2424.887 | 140.735 |
| 1000 | 843.109 | 339.173 | 256.094 | -2986.154 | -2363.226 | 123.441 |
| 1100 | 875.583 | 342.175 | 290.167 | -2986.132 | -2300.937 | 109.261 |
| 1200 | 905.46 | 344.507 | 324.506 | -2986.154 | -2238.646 | 97.445 |
| 1300 | 933.111 | 346.352 | 359.053 | -2986.243 | -2176.35 | 87.446 |
| 1400 | 958.835 | 347.835 | 393.765 | -3368.659 | -2104.537 | 78.52 |
| 1500 | 982.875 | 349.044 | 428.611 | -3364.911 | -2014.38 | 70.146 |
| 1600 | 1005.435 | 350.042 | 463.567 | -3361.288 | -1924.466 | 62.827 |
| 1700 | 1026.682 | 350.875 | 498.614 | -3508.347 | -1833.427 | 56.334 |
| 1800 | 1046.758 | 351.577 | 533.737 | -3504.336 | -1735.017 | 50.348 |
| 1900 | 1065.783 | 352.174 | 568.926 | -3500.371 | -1636.83 | 44.999 |
| 2000 | 1083.86 | 352.686 | 604.169 | -3496.454 | -1538.85 | 40.19 |
| 2100 | 1101.079 | 353.128 | 639.46 | -3492.594 | -1441.067 | 35.844 |
| 2200 | 1117.516 | 353.513 | 674.793 | -3488.794 | -1343.459 | 31.897 |
| 2300 | 1133.238 | 353.849 | 710.161 | -3485.065 | -1246.039 | 28.298 |
| 2400 | 1148.304 | 354.145 | 745.561 | -3481.404 | -1148.759 | 25.002 |
| 2500 | 1162.766 | 354.407 | 780.989 | -3477.816 | -1051.654 | 21.973 |
| 2600 | 1176.671 | 354.639 | 816.442 | -3474.304 | -954.656 | 19.179 |
| 2700 | 1190.059 | 354.847 | 851.916 | -3470.875 | -857.822 | 16.595 |
| 2800 | 1202.967 | 355.033 | 887.411 | -3467.528 | -761.102 | 14.198 |
| 2900 | 1215.429 | 355.2 | 922.922 | -3464.259 | -664.504 | 11.969 |
| 3000 | 1227.473 | 355.351 | 958.45 | -3461.085 | -568.023 | 9.89 |
| 3100 | 1239.127 | 355.488 | 993.992 | -3457.994 | -471.627 | 7.947 |
| 3200 | 1250.416 | 355.612 | 1029.547 | -3455.001 | -375.353 | 6.127 |
| 3300 | 1261.36 | 355.725 | 1065.114 | -3452.099 | -279.148 | 4.418 |
| 3400 | 1271.981 | 355.829 | 1100.692 | -3449.296 | -183.054 | 2.812 |
| 3500 | 1282.297 | 355.924 | 1136.28 | -3446.59 | -87.019 | 1.299 |
| 3600 | 1292.325 | 356.011 | 1171.876 | -4596.453 | 40.3 | -0.585 |
| 3700 | 1302.08 | 356.091 | 1207.482 | -4592.722 | 169.06 | -2.387 |
| 3800 | 1311.578 | 356.165 | 1243.094 | -4589.111 | 297.72 | -4.092 |
| 3900 | 1320.83 | 356.233 | 1278.714 | -4585.627 | 426.277 | -5.709 |
| 4000 | 1329.85 | 356.297 | 1314.341 | -4582.271 | 554.745 | -7.244 |
| 4100 | 1338.649 | 356.355 | 1349.973 | -4579.046 | 683.123 | -8.703 |
| 4200 | 1347.237 | 356.41 | 1385.612 | -4575.959 | 811.43 | -10.091 |
| 4300 | 1355.624 | 356.461 | 1421.255 | -4573.014 | 939.67 | -11.415 |
| 4400 | 1363.819 | 356.509 | 1456.904 | -4570.221 | 1067.844 | -12.677 |
| 4500 | 1371.831 | 356.553 | 1492.557 | -4567.583 | 1195.931 | -13.882 |
| 4600 | 1379.669 | 356.594 | 1528.214 | -4565.111 | 1323.989 | -15.034 |
| 4700 | 1387.338 | 356.633 | 1563.876 | -4562.8 | 1451.992 | -16.137 |
| 4800 | 1394.847 | 356.67 | 1599.541 | -4560.67 | 1579.929 | -17.193 |
| 4900 | 1402.201 | 356.704 | 1635.21 | -4558.724 | 1707.844 | -18.206 |
| 5000 | 1409.408 | 356.737 | 1670.882 | -4556.972 | 1835.718 | -19.177 |
| 5100 | 1416.473 | 356.767 | 1706.557 | -4555.416 | 1963.56 | -20.111 |
| 5200 | 1423.401 | 356.796 | 1742.235 | -4554.064 | 2091.372 | -21.008 |
| 5300 | 1430.197 | 356.823 | 1777.916 | -4552.932 | 2219.151 | -21.871 |
| 5400 | 1436.867 | 356.848 | 1813.6 | -4552.029 | 2346.925 | -22.702 |
| 5500 | 1443.415 | 356.873 | 1849.286 | -4551.35 | 2474.65 | -23.502 |
| 5600 | 1449.846 | 356.896 | 1884.974 | -4550.916 | 2602.423 | -24.274 |
| 5700 | 1456.163 | 356.917 | 1920.665 | -4550.732 | 2730.155 | -25.019 |
| 5800 | 1462.371 | 356.938 | 1956.358 | -4550.808 | 2857.886 | -25.738 |
| 5900 | 1468.472 | 356.958 | 1992.052 | -4551.151 | 2985.63 | -26.432 |
| 6000 | 1474.472 | 356.976 | 2027.749 | -4551.772 | 3113.363 | -27.104 |

**(MgSiO3)4:**

| T (K) | S (J/mol.K) | Cp (J/mol.K) | ddH (kJ/mol) | dHf (kJ/mol) | dGf (kJ/mol) | log Kf |
|---|---|---|---|---|---|---|
| 0 | 0 | 0 | 0 | -4171.974 | -4171.974 | Inf |
| 100 | 343.693 | 129.914 | 6.937 | -4186.093 | -4111.143 | 2147.415 |
| 200 | 469.337 | 239.205 | 25.694 | -4197.546 | -4031.385 | 1052.877 |
| 298.15 | 579.839 | 314.494 | 53.103 | -4203.833 | -3948.317 | 691.72 |
| 300 | 581.788 | 315.646 | 53.686 | -4203.906 | -3946.729 | 687.178 |
| 400 | 680.007 | 365.933 | 87.948 | -4206.178 | -3860.572 | 504.133 |
| 500 | 765.408 | 398.378 | 126.28 | -4205.976 | -3774.155 | 394.279 |
| 600 | 840.056 | 419.691 | 167.256 | -4204.532 | -3687.912 | 321.057 |
| 700 | 905.905 | 434.137 | 209.992 | -4202.606 | -3601.962 | 268.778 |
| 800 | 964.575 | 444.258 | 253.94 | -4200.662 | -3516.282 | 229.587 |
| 900 | 1017.345 | 451.571 | 298.75 | -4199.012 | -3430.843 | 199.119 |
| 1000 | 1065.218 | 457.002 | 344.192 | -4232.21 | -3342.712 | 174.603 |
| 1100 | 1108.977 | 461.133 | 390.108 | -4231.696 | -3253.788 | 154.508 |
| 1200 | 1149.244 | 464.342 | 436.388 | -4231.23 | -3164.91 | 137.763 |
| 1300 | 1186.515 | 466.88 | 482.954 | -4230.846 | -3076.065 | 123.596 |
| 1400 | 1221.192 | 468.92 | 529.747 | -4740.223 | -2974.571 | 110.981 |
| 1500 | 1253.603 | 470.583 | 576.725 | -4734.709 | -2848.656 | 99.198 |
| 1600 | 1284.019 | 471.955 | 623.854 | -4729.358 | -2723.097 | 88.899 |
| 1700 | 1312.666 | 473.101 | 671.109 | -4924.911 | -2596.071 | 79.767 |
| 1800 | 1339.736 | 474.066 | 718.468 | -4919.034 | -2459.248 | 71.365 |
| 1900 | 1365.39 | 474.887 | 765.917 | -4913.217 | -2322.753 | 63.856 |
| 2000 | 1389.767 | 475.591 | 813.442 | -4907.46 | -2186.562 | 57.106 |
| 2100 | 1412.986 | 476.199 | 861.032 | -4901.778 | -2050.659 | 51.007 |
| 2200 | 1435.152 | 476.728 | 908.679 | -4896.175 | -1915.017 | 45.468 |
| 2300 | 1456.353 | 477.191 | 956.376 | -4890.664 | -1779.645 | 40.417 |
| 2400 | 1476.671 | 477.597 | 1004.116 | -4885.242 | -1644.486 | 35.791 |
| 2500 | 1496.175 | 477.957 | 1051.894 | -4879.918 | -1509.586 | 31.541 |
| 2600 | 1514.927 | 478.277 | 1099.706 | -4874.694 | -1374.847 | 27.621 |
| 2700 | 1532.983 | 478.562 | 1147.548 | -4869.578 | -1240.349 | 23.996 |
| 2800 | 1550.392 | 478.818 | 1195.417 | -4864.573 | -1106.027 | 20.633 |
| 2900 | 1567.198 | 479.048 | 1243.311 | -4859.669 | -971.877 | 17.505 |
| 3000 | 1583.442 | 479.255 | 1291.226 | -4854.892 | -837.91 | 14.589 |
| 3100 | 1599.16 | 479.443 | 1339.161 | -4850.225 | -704.074 | 11.863 |
| 3200 | 1614.385 | 479.614 | 1387.114 | -4845.688 | -570.415 | 9.311 |
| 3300 | 1629.146 | 479.77 | 1435.083 | -4841.273 | -436.869 | 6.915 |
| 3400 | 1643.47 | 479.912 | 1483.067 | -4836.989 | -303.484 | 4.662 |
| 3500 | 1657.384 | 480.042 | 1531.065 | -4832.833 | -170.196 | 2.54 |
| 3600 | 1670.909 | 480.162 | 1579.075 | -6365.435 | 4.79 | -0.07 |
| 3700 | 1684.066 | 480.272 | 1627.097 | -6359.913 | 181.679 | -2.565 |
| 3800 | 1696.875 | 480.374 | 1675.13 | -6354.548 | 358.429 | -4.927 |
| 3900 | 1709.355 | 480.468 | 1723.172 | -6349.354 | 535.016 | -7.166 |
| 4000 | 1721.52 | 480.555 | 1771.223 | -6344.331 | 711.477 | -9.291 |
| 4100 | 1733.387 | 480.636 | 1819.282 | -6339.482 | 887.805 | -11.311 |
| 4200 | 1744.97 | 480.711 | 1867.35 | -6334.816 | 1064.021 | -13.233 |
| 4300 | 1756.282 | 480.781 | 1915.424 | -6330.34 | 1240.137 | -15.064 |
| 4400 | 1767.336 | 480.846 | 1963.506 | -6326.066 | 1416.147 | -16.812 |
| 4500 | 1778.143 | 480.907 | 2011.593 | -6321.999 | 1592.029 | -18.48 |
| 4600 | 1788.713 | 480.964 | 2059.687 | -6318.151 | 1767.871 | -20.075 |
| 4700 | 1799.058 | 481.018 | 2107.786 | -6314.52 | 1943.615 | -21.601 |
| 4800 | 1809.185 | 481.068 | 2155.89 | -6311.13 | 2119.268 | -23.062 |
| 4900 | 1819.105 | 481.115 | 2204 | -6307.984 | 2294.872 | -24.463 |
| 5000 | 1828.825 | 481.159 | 2252.113 | -6305.097 | 2470.418 | -25.808 |
| 5100 | 1838.354 | 481.201 | 2300.231 | -6302.471 | 2645.907 | -27.099 |
| 5200 | 1847.698 | 481.241 | 2348.354 | -6300.116 | 2821.349 | -28.34 |
| 5300 | 1856.865 | 481.278 | 2396.48 | -6298.056 | 2996.729 | -29.534 |
| 5400 | 1865.862 | 481.313 | 2444.609 | -6296.301 | 3172.091 | -30.684 |
| 5500 | 1874.694 | 481.346 | 2492.742 | -6294.844 | 3347.382 | -31.79 |
| 5600 | 1883.367 | 481.378 | 2540.878 | -6293.714 | 3522.733 | -32.858 |
| 5700 | 1891.888 | 481.408 | 2589.018 | -6292.916 | 3698.01 | -33.888 |
| 5800 | 1900.26 | 481.436 | 2637.16 | -6292.466 | 3873.287 | -34.882 |
| 5900 | 1908.491 | 481.463 | 2685.305 | -6292.371 | 4048.553 | -35.843 |
| 6000 | 1916.583 | 481.489 | 2733.452 | -6292.648 | 4223.81 | -36.771 |

**(MgSiO3)5:**

| T (K) | S (J/mol.K) | Cp (J/mol.K) | ddH (kJ/mol) | dHf (kJ/mol) | dGf (kJ/mol) | log Kf |
|---|---|---|---|---|---|---|
| 0 | 0 | 0 | 0 | -5415.277 | -5415.277 | Inf |
| 100 | 336.727 | 130.345 | 6.273 | -5435.324 | -5332.348 | 2785.299 |
| 200 | 477.957 | 287.51 | 27.56 | -5451.8 | -5222.356 | 1363.923 |
| 298.15 | 613.739 | 391.832 | 61.278 | -5460.202 | -5107.694 | 894.836 |
| 300 | 616.168 | 393.379 | 62.005 | -5460.295 | -5105.503 | 888.936 |
| 400 | 739.17 | 459.723 | 104.92 | -5463.047 | -4986.704 | 651.189 |
| 500 | 846.588 | 501.483 | 153.136 | -5462.494 | -4867.632 | 508.512 |
| 600 | 940.583 | 528.572 | 204.732 | -5460.313 | -4748.845 | 413.419 |
| 700 | 1023.52 | 546.806 | 258.558 | -5457.499 | -4630.491 | 345.527 |
| 800 | 1097.415 | 559.529 | 313.911 | -5454.651 | -4512.533 | 294.635 |
| 900 | 1163.876 | 568.697 | 370.346 | -5452.166 | -4394.931 | 255.072 |
| 1000 | 1224.163 | 575.494 | 427.572 | -5493.24 | -4274.008 | 223.249 |
| 1100 | 1279.266 | 580.657 | 485.391 | -5492.174 | -4152.139 | 197.166 |
| 1200 | 1329.969 | 584.664 | 543.665 | -5491.167 | -4030.364 | 175.435 |
| 1300 | 1376.897 | 587.831 | 602.295 | -5490.265 | -3908.668 | 157.05 |
| 1400 | 1420.556 | 590.375 | 661.21 | -6126.562 | -3771.19 | 140.703 |
| 1500 | 1461.361 | 592.447 | 720.355 | -6119.247 | -3603.216 | 125.474 |
| 1600 | 1499.653 | 594.158 | 779.688 | -6112.137 | -3435.718 | 112.163 |
| 1700 | 1535.717 | 595.585 | 839.177 | -6356.158 | -3266.411 | 100.363 |
| 1800 | 1569.795 | 596.787 | 898.797 | -6348.39 | -3084.882 | 89.52 |
| 1900 | 1602.09 | 597.81 | 958.528 | -6340.699 | -2903.789 | 79.83 |
| 2000 | 1632.776 | 598.686 | 1018.354 | -6333.083 | -2723.095 | 71.119 |
| 2100 | 1662.005 | 599.443 | 1078.262 | -6325.56 | -2542.783 | 63.248 |
| 2200 | 1689.907 | 600.101 | 1138.239 | -6318.138 | -2362.818 | 56.1 |
| 2300 | 1716.595 | 600.676 | 1198.279 | -6310.831 | -2183.211 | 49.582 |
| 2400 | 1742.171 | 601.183 | 1258.372 | -6303.635 | -2003.888 | 43.613 |
| 2500 | 1766.721 | 601.63 | 1318.514 | -6296.561 | -1824.901 | 38.129 |
| 2600 | 1790.326 | 602.028 | 1378.697 | -6289.613 | -1646.139 | 33.071 |
| 2700 | 1813.053 | 602.383 | 1438.918 | -6282.799 | -1467.688 | 28.394 |
| 2800 | 1834.966 | 602.701 | 1499.172 | -6276.125 | -1289.475 | 24.055 |
| 2900 | 1856.121 | 602.987 | 1559.457 | -6269.578 | -1111.496 | 20.02 |
| 3000 | 1876.567 | 603.245 | 1619.769 | -6263.188 | -933.754 | 16.258 |
| 3100 | 1896.352 | 603.479 | 1680.105 | -6256.937 | -756.195 | 12.742 |
| 3200 | 1915.515 | 603.692 | 1740.464 | -6250.848 | -578.864 | 9.449 |
| 3300 | 1934.094 | 603.885 | 1800.843 | -6244.912 | -401.689 | 6.358 |
| 3400 | 1952.125 | 604.062 | 1861.24 | -6239.14 | -224.735 | 3.453 |
| 3500 | 1969.637 | 604.224 | 1921.655 | -6233.527 | -47.906 | 0.715 |
| 3600 | 1986.661 | 604.373 | 1982.085 | -8148.862 | 181.03 | -2.627 |
| 3700 | 2003.222 | 604.51 | 2042.529 | -8141.543 | 412.331 | -5.821 |
| 3800 | 2019.345 | 604.637 | 2102.986 | -8134.421 | 643.446 | -8.845 |
| 3900 | 2035.052 | 604.754 | 2163.456 | -8127.511 | 874.353 | -11.711 |
| 4000 | 2050.365 | 604.862 | 2223.937 | -8120.815 | 1105.085 | -14.431 |
| 4100 | 2065.302 | 604.962 | 2284.428 | -8114.337 | 1335.642 | -17.016 |
| 4200 | 2079.881 | 605.056 | 2344.929 | -8108.088 | 1566.051 | -19.476 |
| 4300 | 2094.119 | 605.143 | 2405.439 | -8102.076 | 1796.324 | -21.821 |
| 4400 | 2108.032 | 605.224 | 2465.957 | -8096.318 | 2026.456 | -24.057 |
| 4500 | 2121.634 | 605.3 | 2526.484 | -8090.816 | 2256.421 | -26.192 |
| 4600 | 2134.939 | 605.371 | 2587.017 | -8085.59 | 2486.318 | -28.233 |
| 4700 | 2147.959 | 605.437 | 2647.558 | -8080.634 | 2716.093 | -30.186 |
| 4800 | 2160.706 | 605.5 | 2708.105 | -8075.98 | 2945.739 | -32.056 |
| 4900 | 2173.191 | 605.559 | 2768.658 | -8071.632 | 3175.321 | -33.849 |
| 5000 | 2185.426 | 605.614 | 2829.216 | -8067.606 | 3404.814 | -35.569 |
| 5100 | 2197.419 | 605.666 | 2889.78 | -8063.907 | 3634.235 | -37.222 |
| 5200 | 2209.18 | 605.715 | 2950.349 | -8060.548 | 3863.585 | -38.81 |
| 5300 | 2220.719 | 605.761 | 3010.923 | -8057.557 | 4092.844 | -40.337 |
| 5400 | 2232.042 | 605.805 | 3071.501 | -8054.946 | 4322.086 | -41.807 |
| 5500 | 2243.158 | 605.846 | 3132.084 | -8052.708 | 4551.226 | -43.223 |
| 5600 | 2254.075 | 605.885 | 3192.671 | -8050.879 | 4780.429 | -44.589 |
| 5700 | 2264.799 | 605.923 | 3253.261 | -8049.466 | 5009.538 | -45.907 |
| 5800 | 2275.338 | 605.958 | 3313.855 | -8048.487 | 5238.628 | -47.178 |
| 5900 | 2285.697 | 605.991 | 3374.452 | -8047.953 | 5467.711 | -48.407 |
| 6000 | 2295.882 | 606.023 | 3435.053 | -8047.882 | 5696.771 | -49.594 |

**(MgSiO3)6:**

| T (K) | S (J/mol.K) | Cp (J/mol.K) | ddH (kJ/mol) | dHf (kJ/mol) | dGf (kJ/mol) | log Kf |
|---|---|---|---|---|---|---|
| 0 | 0 | 0 | 0 | -6657.299 | -6657.299 | Inf |
| 100 | 370.459 | 170.108 | 8.006 | -6680.877 | -6553.944 | 3423.388 |
| 200 | 548.058 | 353.297 | 34.688 | -6699.51 | -6419.079 | 1676.471 |
| 298.15 | 713.488 | 474.694 | 75.75 | -6708.992 | -6279.125 | 1100.063 |
| 300 | 716.43 | 476.508 | 76.63 | -6709.096 | -6276.454 | 1092.815 |
| 400 | 865.083 | 554.705 | 128.49 | -6712.037 | -6131.657 | 800.703 |
| 500 | 994.606 | 604.386 | 186.627 | -6711.095 | -5986.61 | 625.409 |
| 600 | 1107.866 | 636.82 | 248.797 | -6708.223 | -5841.961 | 508.582 |
| 700 | 1207.784 | 658.749 | 313.644 | -6704.591 | -5697.873 | 425.175 |
| 800 | 1296.808 | 674.097 | 380.33 | -6700.911 | -5554.297 | 362.654 |
| 900 | 1376.879 | 685.183 | 448.322 | -6697.659 | -5411.181 | 314.053 |
| 1000 | 1449.516 | 693.414 | 517.271 | -6746.67 | -5264.112 | 274.966 |
| 1100 | 1515.913 | 699.675 | 586.94 | -6745.104 | -5115.935 | 242.933 |
| 1200 | 1577.009 | 704.538 | 657.16 | -6743.605 | -4967.896 | 216.244 |
| 1300 | 1633.56 | 708.385 | 727.813 | -6742.225 | -4819.977 | 193.667 |
| 1400 | 1686.174 | 711.477 | 798.812 | -7505.481 | -4653.143 | 173.609 |
| 1500 | 1735.35 | 713.998 | 870.09 | -7496.399 | -4449.737 | 154.952 |
| 1600 | 1781.499 | 716.079 | 941.597 | -7487.559 | -4246.92 | 138.646 |
| 1700 | 1824.964 | 717.815 | 1013.294 | -7780.074 | -4041.954 | 124.193 |
| 1800 | 1866.036 | 719.279 | 1085.151 | -7770.44 | -3822.338 | 110.92 |
| 1900 | 1904.96 | 720.524 | 1157.143 | -7760.896 | -3603.263 | 99.059 |
| 2000 | 1941.946 | 721.591 | 1229.25 | -7751.441 | -3384.685 | 88.398 |
| 2100 | 1977.175 | 722.513 | 1301.456 | -7742.097 | -3166.579 | 78.763 |
| 2200 | 2010.805 | 723.315 | 1373.748 | -7732.871 | -2948.903 | 70.015 |
| 2300 | 2042.974 | 724.016 | 1446.116 | -7723.782 | -2731.676 | 62.038 |
| 2400 | 2073.801 | 724.633 | 1518.549 | -7714.826 | -2514.799 | 54.733 |
| 2500 | 2103.393 | 725.179 | 1591.04 | -7706.016 | -2298.343 | 48.021 |
| 2600 | 2131.845 | 725.663 | 1663.583 | -7697.355 | -2082.166 | 41.831 |
| 2700 | 2159.24 | 726.096 | 1736.171 | -7688.856 | -1866.379 | 36.107 |
| 2800 | 2185.653 | 726.484 | 1808.8 | -7680.523 | -1650.885 | 30.797 |
| 2900 | 2211.153 | 726.832 | 1881.466 | -7672.342 | -1435.686 | 25.859 |
| 3000 | 2235.799 | 727.147 | 1954.166 | -7664.349 | -1220.784 | 21.255 |
| 3100 | 2259.647 | 727.432 | 2026.895 | -7656.522 | -1006.107 | 16.953 |
| 3200 | 2282.746 | 727.691 | 2099.651 | -7648.89 | -791.719 | 12.923 |
| 3300 | 2305.142 | 727.927 | 2172.432 | -7641.44 | -577.529 | 9.141 |
| 3400 | 2326.876 | 728.143 | 2245.236 | -7634.186 | -363.609 | 5.586 |
| 3500 | 2347.986 | 728.341 | 2318.06 | -7627.125 | -149.854 | 2.236 |
| 3600 | 2368.506 | 728.522 | 2390.903 | -9925.2 | 126.425 | -1.834 |
| 3700 | 2388.47 | 728.689 | 2463.764 | -9916.089 | 405.527 | -5.725 |
| 3800 | 2407.904 | 728.844 | 2536.641 | -9907.214 | 684.405 | -9.408 |
| 3900 | 2426.838 | 728.986 | 2609.533 | -9898.594 | 963.019 | -12.898 |
| 4000 | 2445.296 | 729.118 | 2682.438 | -9890.231 | 1241.417 | -16.211 |
| 4100 | 2463.302 | 729.241 | 2755.356 | -9882.128 | 1519.595 | -19.36 |
| 4200 | 2480.876 | 729.355 | 2828.286 | -9874.301 | 1797.587 | -22.356 |
| 4300 | 2498.039 | 729.461 | 2901.227 | -9866.757 | 2075.41 | -25.211 |
| 4400 | 2514.811 | 729.56 | 2974.178 | -9859.518 | 2353.052 | -27.934 |
| 4500 | 2531.207 | 729.653 | 3047.138 | -9852.588 | 2630.489 | -30.534 |
| 4600 | 2547.245 | 729.739 | 3120.108 | -9845.987 | 2907.84 | -33.019 |
| 4700 | 2562.94 | 729.82 | 3193.086 | -9839.711 | 3185.033 | -35.397 |
| 4800 | 2578.306 | 729.897 | 3266.072 | -9833.796 | 3462.065 | -37.674 |
| 4900 | 2593.356 | 729.968 | 3339.065 | -9828.249 | 3739.013 | -39.858 |
| 5000 | 2608.104 | 730.035 | 3412.065 | -9823.088 | 4015.852 | -41.953 |
| 5100 | 2622.562 | 730.099 | 3485.072 | -9818.319 | 4292.591 | -43.965 |
| 5200 | 2636.739 | 730.159 | 3558.085 | -9813.958 | 4569.242 | -45.898 |
| 5300 | 2650.648 | 730.215 | 3631.104 | -9810.038 | 4845.782 | -47.757 |
| 5400 | 2664.298 | 730.269 | 3704.128 | -9806.575 | 5122.287 | -49.548 |
| 5500 | 2677.698 | 730.319 | 3777.157 | -9803.56 | 5398.666 | -51.272 |
| 5600 | 2690.858 | 730.367 | 3850.192 | -9801.034 | 5675.115 | -52.935 |
| 5700 | 2703.785 | 730.412 | 3923.231 | -9799.008 | 5951.449 | -54.538 |
| 5800 | 2716.489 | 730.455 | 3996.274 | -9797.503 | 6227.752 | -56.086 |
| 5900 | 2728.976 | 730.496 | 4069.322 | -9796.53 | 6504.043 | -57.582 |
| 6000 | 2741.254 | 730.535 | 4142.373 | -9796.115 | 6780.295 | -59.027 |

**(MgSiO3)7:**

| T (K) | S (J/mol.K) | Cp (J/mol.K) | ddH (kJ/mol) | dHf (kJ/mol) | dGf (kJ/mol) | log Kf |
|---|---|---|---|---|---|---|
| 0 | 0 | 0 | 0 | -7895.474 | -7895.474 | Inf |
| 100 | 393.387 | 196.054 | 9.115 | -7923.207 | -7771.237 | 4059.229 |
| 200 | 598.965 | 410.788 | 40.016 | -7945.174 | -7609.917 | 1987.483 |
| 298.15 | 791.792 | 554.342 | 87.886 | -7956.272 | -7442.652 | 1303.906 |
| 300 | 795.228 | 556.487 | 88.913 | -7956.393 | -7439.462 | 1295.310 |
| 400 | 968.996 | 648.887 | 149.537 | -7959.703 | -7266.486 | 948.894 |
| 500 | 1120.564 | 707.439 | 217.57 | -7958.398 | -7093.261 | 741.019 |
| 600 | 1253.152 | 745.563 | 290.35 | -7954.799 | -6920.545 | 602.48 |
| 700 | 1370.136 | 771.28 | 366.273 | -7950.293 | -6748.528 | 503.575 |
| 800 | 1474.368 | 789.248 | 444.35 | -7945.723 | -6577.148 | 429.439 |
| 900 | 1568.115 | 802.209 | 523.957 | -7941.646 | -6406.336 | 371.81 |
| 1000 | 1653.158 | 811.823 | 604.681 | -7998.542 | -6230.947 | 325.467 |
| 1100 | 1730.891 | 819.13 | 686.245 | -7996.432 | -6054.293 | 287.491 |
| 1200 | 1802.418 | 824.802 | 768.453 | -7994.398 | -5877.827 | 255.852 |
| 1300 | 1868.621 | 829.286 | 851.166 | -7992.504 | -5701.523 | 229.087 |
| 1400 | 1930.215 | 832.889 | 934.281 | -8882.686 | -5503.176 | 205.324 |
| 1500 | 1987.781 | 835.826 | 1017.721 | -8871.808 | -5262.178 | 183.243 |
| 1600 | 2041.804 | 838.249 | 1101.429 | -8861.212 | -5021.888 | 163.946 |
| 1700 | 2092.685 | 840.271 | 1185.358 | -9202.197 | -4779.109 | 146.842 |
| 1800 | 2140.763 | 841.975 | 1269.472 | -9190.676 | -4519.255 | 131.144 |
| 1900 | 2186.326 | 843.424 | 1353.744 | -9179.26 | -4260.046 | 117.115 |
| 2000 | 2229.621 | 844.666 | 1438.15 | -9167.948 | -4001.434 | 104.505 |
| 2100 | 2270.859 | 845.739 | 1522.672 | -9156.765 | -3743.386 | 93.111 |
| 2200 | 2310.224 | 846.671 | 1607.293 | -9145.721 | -3485.852 | 82.764 |
| 2300 | 2347.879 | 847.487 | 1692.002 | -9134.838 | -3228.855 | 73.329 |
| 2400 | 2383.963 | 848.205 | 1776.787 | -9124.109 | -2972.279 | 64.689 |
| 2500 | 2418.602 | 848.839 | 1861.64 | -9113.551 | -2716.208 | 56.751 |
| 2600 | 2451.905 | 849.403 | 1946.553 | -9103.167 | -2460.47 | 49.431 |
| 2700 | 2483.971 | 849.906 | 2031.519 | -9092.971 | -2205.197 | 42.662 |
| 2800 | 2514.889 | 850.357 | 2116.532 | -9082.97 | -1950.282 | 36.383 |
| 2900 | 2544.736 | 850.762 | 2201.589 | -9073.146 | -1695.714 | 30.543 |
| 3000 | 2573.584 | 851.128 | 2286.684 | -9063.542 | -1441.505 | 25.099 |
| 3100 | 2601.498 | 851.46 | 2371.813 | -9054.132 | -1187.569 | 20.01 |
| 3200 | 2628.536 | 851.761 | 2456.975 | -9044.948 | -933.979 | 15.245 |
| 3300 | 2654.75 | 852.036 | 2542.165 | -9035.978 | -680.627 | 10.773 |
| 3400 | 2680.19 | 852.287 | 2627.381 | -9027.237 | -427.601 | 6.569 |
| 3500 | 2704.899 | 852.516 | 2712.621 | -9018.72 | -174.775 | 2.608 |
| 3600 | 2728.918 | 852.728 | -2797.884 | 11699.528 | 150.988 | -2.191 |
| 3700 | 2752.284 | 852.922 | -2883.166 | 11688.621 | 480.042 | -6.777 |
| 3800 | 2775.033 | 853.101 | -2968.467 | 11677.989 | 808.815 | -11.118 |
| 3900 | 2797.195 | 853.267 | -3053.786 | 11667.654 | 1137.28 | -15.232 |
| 4000 | 2818.8 | 853.42 | -3139.12 | 11657.619 | 1465.485 | -19.137 |
| 4100 | 2839.874 | 853.563 | -3224.47 | 11647.887 | 1793.434 | -22.848 |
| 4200 | 2860.445 | 853.695 | -3309.833 | 11638.477 | 2121.149 | -26.38 |
| 4300 | 2880.534 | 853.819 | -3395.208 | 11629.399 | 2448.662 | -29.745 |
| 4400 | 2900.164 | 853.934 | -3480.596 | 11620.675 | 2775.965 | -32.954 |
| 4500 | 2919.356 | 854.042 | -3565.995 | 11612.311 | 3103.013 | -36.018 |
| 4600 | 2938.128 | 854.142 | -3651.404 | 11604.332 | 3429.958 | -38.948 |
| 4700 | 2956.498 | 854.237 | -3736.823 | 11596.732 | 3756.716 | -41.751 |
| 4800 | 2974.484 | 854.325 | -3822.251 | 11589.554 | 4083.274 | -44.435 |
| 4900 | 2992.1 | 854.408 | -3907.688 | 11582.804 | 4409.731 | -47.008 |
| 5000 | 3009.363 | 854.487 | -3993.133 | 11576.504 | 4736.051 | -49.477 |
| 5100 | 3026.284 | 854.56 | -4078.585 | 11570.663 | 5062.26 | -51.847 |
| 5200 | 3042.879 | 854.63 | -4164.045 | 11565.297 | 5388.349 | -54.126 |
| 5300 | 3059.159 | 854.696 | -4249.511 | 11560.447 | 5714.307 | -56.317 |
| 5400 | 3075.135 | 854.758 | -4334.984 | 11556.128 | 6040.225 | -58.427 |
| 5500 | 3090.82 | 854.816 | -4420.463 | 11552.332 | 6365.983 | -60.458 |
| 5600 | 3106.223 | 854.872 | -4505.947 | 11549.109 | 6691.822 | -62.418 |
| 5700 | 3121.354 | 854.925 | -4591.437 | 11546.467 | 7017.518 | -64.308 |
| 5800 | 3136.223 | 854.975 | -4676.932 | 11544.433 | 7343.18 | -66.132 |
| 5900 | 3150.839 | 855.022 | -4762.432 | 11543.021 | 7668.816 | -67.894 |
| 6000 | 3165.21 | 855.067 | -4847.936 | 11542.259 | 7994.404 | -69.597 |

**$(MgSiO_3)_8$:**

| T (K) | S (J/mol.K) | Cp (J/mol.K) | ddH (kJ/mol) | dHf (kJ/mol) | dGf (kJ/mol) | log Kf |
|---|---|---|---|---|---|---|
| 0 | 0 | 0 | 0 | -9116.984 | -9116.984 | Inf |
| 100 | 401.031 | 218.895 | 9.571 | -9149.525 | -8970.989 | 4685.908 |
| 200 | 634.689 | 470.303 | 44.736 | -9174.78 | -8781.661 | 2293.507 |
| 298.15 | 855.718 | 635.737 | 99.609 | -9187.299 | -8585.64 | 1504.150 |
| 300 | 859.658 | 638.199 | 100.788 | -9187.432 | -8581.902 | 1494.224 |
| 400 | 1058.939 | 744.102 | 170.313 | -9190.975 | -8379.333 | 1094.215 |
| 500 | 1232.729 | 811.077 | 248.32 | -9189.228 | -8176.542 | 854.187 |
| 600 | 1384.729 | 854.64 | 331.755 | -9184.857 | -7974.386 | 694.224 |
| 700 | 1518.819 | 884.01 | 418.779 | -9179.453 | -7773.071 | 580.027 |
| 800 | 1638.279 | 904.521 | 508.264 | -9173.976 | -7572.519 | 494.429 |
| 900 | 1745.716 | 919.312 | 599.494 | -9169.066 | -7372.651 | 427.892 |
| 1000 | 1843.17 | 930.282 | 692 | -9233.84 | -7167.578 | 374.391 |
| 1100 | 1932.244 | 938.617 | 785.463 | -9231.181 | -6961.084 | 330.55 |
| 1200 | 2014.203 | 945.087 | 879.661 | -9228.611 | -6754.829 | 294.027 |
| 1300 | 2090.06 | 950.202 | 974.435 | -9226.201 | -6548.778 | 263.13 |
| 1400 | 2160.633 | 954.312 | 1069.668 | -10243.308 | -6317.552 | 235.708 |
| 1500 | 2226.592 | 957.66 | 1165.272 | -10230.632 | -6037.604 | 210.246 |
| 1600 | 2288.489 | 960.423 | 1261.181 | -10218.279 | -5758.479 | 187.993 |
| 1700 | 2346.785 | 962.729 | 1357.342 | -10607.734 | -5476.523 | 168.271 |
| 1800 | 2401.87 | 964.672 | 1453.714 | -10594.326 | -5175.069 | 150.175 |
| 1900 | 2454.073 | 966.324 | 1550.266 | -10581.038 | -4874.367 | 134.004 |
| 2000 | 2503.675 | 967.74 | 1646.971 | -10567.869 | -4574.355 | 119.468 |
| 2100 | 2550.922 | 968.964 | 1743.808 | -10554.848 | -4275.004 | 106.334 |
| 2200 | 2596.023 | 970.027 | 1840.759 | -10541.985 | -3976.25 | 94.407 |
| 2300 | 2639.164 | 970.957 | 1937.809 | -10529.307 | -3678.122 | 83.532 |
| 2400 | 2680.505 | 971.775 | 2034.947 | -10516.805 | -3380.484 | 73.573 |
| 2500 | 2720.19 | 972.499 | 2132.161 | -10504.499 | -3083.434 | 64.424 |
| 2600 | 2758.344 | 973.142 | 2229.444 | -10492.392 | -2786.772 | 55.986 |
| 2700 | 2795.082 | 973.715 | 2326.787 | -10480.501 | -2490.655 | 48.184 |
| 2800 | 2830.503 | 974.229 | 2424.185 | -10468.831 | -2194.951 | 40.947 |
| 2900 | 2864.698 | 974.691 | 2521.631 | -10457.365 | -1899.656 | 34.216 |
| 3000 | 2897.749 | 975.109 | 2619.121 | -10446.151 | -1604.782 | 27.941 |
| 3100 | 2929.729 | 975.487 | 2716.652 | -10435.156 | -1310.222 | 22.077 |
| 3200 | 2960.705 | 975.83 | 2814.218 | -10424.422 | -1016.067 | 16.585 |
| 3300 | 2990.738 | 976.143 | 2911.817 | -10413.931 | -722.194 | 11.431 |
| 3400 | 3019.883 | 976.429 | 3009.445 | -10403.703 | -428.698 | 6.586 |
| 3500 | 3048.191 | 976.691 | 3107.102 | -10393.73 | -135.436 | 2.021 |
| 3600 | 3075.709 | 976.932 | 3204.783 | -13457.273 | 241.17 | -3.499 |
| 3700 | 3102.479 | 977.153 | 3302.487 | -13444.569 | 621.533 | -8.774 |
| 3800 | 3128.54 | 977.358 | 3400.213 | -13432.179 | 1001.574 | -13.767 |
| 3900 | 3153.93 | 977.547 | 3497.958 | -13420.13 | 1381.252 | -18.5 |
| 4000 | 3178.682 | 977.722 | 3595.722 | -13408.422 | 1760.626 | -22.991 |
| 4100 | 3202.826 | 977.884 | 3693.502 | -13397.062 | 2139.7 | -27.26 |
| 4200 | 3226.393 | 978.035 | 3791.298 | -13386.07 | 2518.503 | -31.322 |
| 4300 | 3249.408 | 978.176 | 3889.109 | -13375.455 | 2897.07 | -35.192 |
| 4400 | 3271.897 | 978.307 | 3986.933 | -13365.247 | 3275.39 | -38.883 |
| 4500 | 3293.884 | 978.43 | 4084.77 | -13355.45 | 3653.416 | -42.407 |
| 4600 | 3315.39 | 978.545 | 4182.619 | -13346.093 | 4031.318 | -45.777 |
| 4700 | 3336.436 | 978.652 | 4280.479 | -13337.169 | 4408.997 | -49 |
| 4800 | 3357.041 | 978.753 | 4378.349 | -13328.727 | 4786.449 | -52.087 |
| 4900 | 3377.223 | 978.848 | 4476.229 | -13320.775 | 5163.775 | -55.046 |
| 5000 | 3397 | 978.937 | 4574.119 | -13313.337 | 5540.943 | -57.885 |
| 5100 | 3416.386 | 979.021 | 4672.017 | -13306.423 | 5917.976 | -60.612 |
| 5200 | 3435.398 | 979.101 | 4769.923 | -13300.053 | 6294.868 | -63.232 |
| 5300 | 3454.048 | 979.175 | 4867.837 | -13294.271 | 6671.614 | -65.752 |
| 5400 | 3472.352 | 979.246 | 4965.758 | -13289.098 | 7048.296 | -68.178 |
| 5500 | 3490.321 | 979.313 | 5063.686 | -13284.522 | 7424.799 | -70.514 |
| 5600 | 3507.967 | 979.376 | 5161.62 | -13280.6 | 7801.39 | -72.767 |
| 5700 | 3525.302 | 979.437 | 5259.561 | -13277.343 | 8177.811 | -74.94 |
| 5800 | 3542.337 | 979.494 | 5357.507 | -13274.781 | 8554.186 | -77.038 |
| 5900 | 3559.081 | 979.548 | 5455.459 | -13272.929 | 8930.535 | -79.064 |
| 6000 | 3575.545 | 979.599 | 5553.417 | -13271.819 | 9306.823 | -81.022 |

**$(MgSiO_3)_9$:**

| T (K) | S (J/mol.K) | Cp (J/mol.K) | ddH (kJ/mol) | dHf (kJ/mol) | dGf (kJ/mol) | log Kf |
|---|---|---|---|---|---|---|
| 0 | 0 | 0 | 0 | -10384.645 | -10384.645 | Inf |
| 100 | 436.034 | 258.318 | 11.477 | -10420.544 | -10218.179 | 5337.26 |
| 200 | 705.607 | 535.798 | 51.973 | -10448.02 | -10004.078 | 2612.766 |
| 298.15 | 956.329 | 719.099 | 114.204 | -10461.605 | -9782.845 | 1713.893 |
| 300 | 960.786 | 721.838 | 115.537 | -10461.748 | -9778.628 | 1702.590 |
| 400 | 1185.921 | 839.935 | 194.078 | -10465.409 | -9550.157 | 1247.108 |
| 500 | 1382.014 | 914.875 | 282.095 | -10463.184 | -9321.51 | 973.8 |
| 600 | 1553.436 | 963.708 | 376.191 | -10458.035 | -9093.625 | 791.661 |
| 700 | 1704.626 | 996.667 | 474.312 | -10451.737 | -8866.725 | 661.635 |
| 800 | 1839.303 | 1019.703 | 575.196 | -10445.362 | -8640.714 | 564.174 |
| 900 | 1960.417 | 1036.322 | 678.04 | -10439.628 | -8415.499 | 488.417 |
| 1000 | 2070.274 | 1048.651 | 782.318 | -10512.29 | -8184.453 | 427.507 |
| 1100 | 2170.68 | 1058.023 | 887.672 | -10509.09 | -7951.828 | 377.596 |
| 1200 | 2263.065 | 1065.298 | 993.853 | -10505.991 | -7719.49 | 336.017 |
| 1300 | 2348.57 | 1071.051 | 1100.681 | -10503.072 | -7487.4 | 300.844 |
| 1400 | 2428.118 | 1075.673 | 1208.026 | -11647.11 | -7227.003 | 269.64 |
| 1500 | 2502.465 | 1079.44 | 1315.788 | -11632.642 | -6911.809 | 240.688 |
| 1600 | 2572.233 | 1082.548 | 1423.892 | -11618.538 | -6597.556 | 215.386 |
| 1700 | 2637.942 | 1085.142 | 1532.28 | -12056.468 | -6280.132 | 192.963 |
| 1800 | 2700.03 | 1087.328 | 1640.907 | -12041.176 | -5940.78 | 172.395 |
| 1900 | 2758.87 | 1089.187 | 1749.735 | -12026.02 | -5602.287 | 154.016 |
| 2000 | 2814.78 | 1090.781 | 1858.735 | -12010.998 | -5264.586 | 137.495 |
| 2100 | 2868.033 | 1092.157 | 1967.884 | -11996.142 | -4927.633 | 122.567 |
| 2200 | 2918.869 | 1093.354 | 2077.161 | -11981.464 | -4591.367 | 109.012 |
| 2300 | 2967.494 | 1094.401 | 2186.549 | -11966.994 | -4255.811 | 96.651 |
| 2400 | 3014.091 | 1095.321 | 2296.037 | -11952.722 | -3920.816 | 85.333 |
| 2500 | 3058.821 | 1096.135 | 2405.61 | -11938.67 | -3586.49 | 74.935 |
| 2600 | 3101.827 | 1096.859 | 2515.261 | -11924.842 | -3252.614 | 65.345 |
| 2700 | 3143.235 | 1097.504 | 2624.979 | -11911.258 | -2919.355 | 56.478 |
| 2800 | 3183.159 | 1098.083 | 2734.759 | -11897.922 | -2586.568 | 48.252 |
| 2900 | 3221.701 | 1098.603 | 2844.594 | -11884.814 | -2254.248 | 40.603 |
| 3000 | 3258.954 | 1099.073 | 2954.478 | -11871.991 | -1922.41 | 33.472 |
| 3100 | 3294.999 | 1099.498 | 3064.407 | -11859.415 | -1590.931 | 26.807 |
| 3200 | 3329.913 | 1099.885 | 3174.376 | -11847.132 | -1259.916 | 20.566 |
| 3300 | 3363.764 | 1100.237 | 3284.383 | -11835.121 | -929.224 | 14.708 |
| 3400 | 3396.614 | 1100.559 | 3394.423 | -11823.406 | -598.961 | 9.202 |
| 3500 | 3428.521 | 1100.854 | 3504.494 | -11811.98 | -268.971 | 4.014 |
| 3600 | 3459.537 | 1101.125 | 3614.593 | -15258.258 | 154.779 | -2.246 |
| 3700 | 3489.71 | 1101.374 | 3724.718 | -15243.758 | 582.748 | -8.227 |
| 3800 | 3519.085 | 1101.604 | 3834.867 | -15229.612 | 1010.346 | -13.888 |
| 3900 | 3547.702 | 1101.817 | 3945.038 | -15215.849 | 1437.535 | -19.253 |
| 4000 | 3575.6 | 1102.014 | 4055.23 | -15202.47 | 1864.378 | -24.346 |
| 4100 | 3602.814 | 1102.197 | 4165.44 | -15189.482 | 2290.872 | -29.186 |
| 4200 | 3629.376 | 1102.367 | 4275.669 | -15176.908 | 2717.064 | -33.791 |
| 4300 | 3655.318 | 1102.525 | 4385.913 | -15164.759 | 3142.974 | -38.179 |
| 4400 | 3680.666 | 1102.673 | 4496.173 | -15153.067 | 3568.609 | -42.364 |
| 4500 | 3705.448 | 1102.811 | 4606.448 | -15141.837 | 3993.909 | -46.36 |
| 4600 | 3729.688 | 1102.94 | 4716.735 | -15131.104 | 4419.062 | -50.179 |
| 4700 | 3753.409 | 1103.061 | 4827.035 | -15120.857 | 4843.963 | -53.834 |
| 4800 | 3776.633 | 1103.175 | 4937.347 | -15111.151 | 5268.605 | -57.333 |
| 4900 | 3799.381 | 1103.281 | 5047.67 | -15101.997 | 5693.096 | -60.688 |
| 5000 | 3821.671 | 1103.382 | 5158.003 | -15093.423 | 6117.412 | -63.907 |
| 5100 | 3843.522 | 1103.476 | 5268.346 | -15085.437 | 6541.565 | -66.998 |
| 5200 | 3864.95 | 1103.566 | 5378.698 | -15078.063 | 6965.561 | -69.969 |
| 5300 | 3885.972 | 1103.65 | 5489.059 | -15071.35 | 7389.379 | -72.826 |
| 5400 | 3906.603 | 1103.729 | 5599.428 | -15065.323 | 7813.127 | -75.576 |
| 5500 | 3926.856 | 1103.805 | 5709.805 | -15059.967 | 8236.672 | -78.224 |
| 5600 | 3946.745 | 1103.876 | 5820.189 | -15055.346 | 8660.312 | -80.779 |
| 5700 | 3966.284 | 1103.944 | 5930.58 | -15051.475 | 9083.753 | -83.242 |
| 5800 | 3985.484 | 1104.008 | 6040.977 | -15048.385 | 9507.145 | -85.62 |
| 5900 | 4004.357 | 1104.069 | 6151.381 | -15046.093 | 9930.498 | -87.917 |
| 6000 | 4022.914 | 1104.127 | 6261.791 | -15044.637 | 10353.78 | -90.137 |

**(MgSiO3)10:**

| T (K) | S (J/mol.K) | Cp (J/mol.K) | ddH (kJ/mol) | dHf (kJ/mol) | dGf (kJ/mol) | log Kf |
|---|---|---|---|---|---|---|
| 0 | 0 | 0 | 0 | -11626.666 | -11626.666 | Inf |
| 100 | 439.789 | 277.847 | 11.767 | -11667.539 | -11438.219 | 5974.64 |
| 200 | 734.583 | 591.169 | 56.109 | -11698.722 | -11195.568 | 2923.947 |
| 298.15 | 1012.145 | 797.908 | 125.014 | -11714.057 | -10944.839 | 1917.467 |
| 300 | 1017.09 | 800.99 | 126.493 | -11714.218 | -10940.061 | 1904.811 |
| 400 | 1267.159 | 933.624 | 213.736 | -11718.31 | -10681.152 | 1394.799 |
| 500 | 1485.199 | 1017.535 | 311.605 | -11715.766 | -10422.053 | 1088.771 |
| 600 | 1675.883 | 1072.115 | 416.275 | -11709.926 | -10163.821 | 884.829 |
| 700 | 1844.091 | 1108.91 | 525.441 | -11702.785 | -9906.705 | 739.239 |
| 800 | 1993.94 | 1134.607 | 637.69 | -11695.546 | -9650.598 | 630.112 |
| 900 | 2128.704 | 1153.136 | 752.125 | -11689.011 | -9395.396 | 545.288 |
| 1000 | 2250.945 | 1166.877 | 868.158 | -11769.578 | -9133.733 | 477.092 |
| 1100 | 2362.672 | 1177.319 | 985.391 | -11765.85 | -8870.333 | 421.212 |
| 1200 | 2465.473 | 1185.424 | 1103.544 | -11762.232 | -8607.268 | 374.66 |
| 1300 | 2560.621 | 1191.831 | 1222.419 | -11758.812 | -8344.493 | 335.282 |
| 1400 | 2649.14 | 1196.978 | 1341.869 | -13029.787 | -8050.281 | 300.356 |
| 1500 | 2731.871 | 1201.173 | 1461.783 | -13013.533 | -7695.195 | 267.967 |
| 1600 | 2809.507 | 1204.634 | 1582.079 | -12997.682 | -7341.165 | 239.662 |
| 1700 | 2882.626 | 1207.522 | 1702.691 | -13484.09 | -6983.623 | 214.578 |
| 1800 | 2951.717 | 1209.955 | 1823.568 | -13466.918 | -6601.731 | 191.575 |
| 1900 | 3017.193 | 1212.025 | 1944.67 | -13449.896 | -6220.801 | 171.02 |
| 2000 | 3079.408 | 1213.799 | 2065.963 | -13433.023 | -5840.759 | 152.543 |
| 2100 | 3138.667 | 1215.331 | 2187.422 | -13416.334 | -5461.559 | 135.847 |
| 2200 | 3195.236 | 1216.663 | 2309.023 | -13399.843 | -5083.13 | 120.687 |
| 2300 | 3249.345 | 1217.828 | 2430.749 | -13383.582 | -4705.498 | 106.864 |
| 2400 | 3301.197 | 1218.853 | 2552.584 | -13367.542 | -4328.499 | 94.206 |
| 2500 | 3350.972 | 1219.759 | 2674.515 | -13351.746 | -3952.251 | 82.577 |
| 2600 | 3398.828 | 1220.564 | 2796.532 | -13336.199 | -3576.509 | 71.852 |
| 2700 | 3444.906 | 1221.282 | 2918.625 | -13320.921 | -3201.459 | 61.935 |
| 2800 | 3489.333 | 1221.926 | 3040.786 | -13305.92 | -2826.942 | 52.737 |
| 2900 | 3532.222 | 1222.505 | 3163.008 | -13291.173 | -2452.951 | 44.182 |
| 3000 | 3573.676 | 1223.028 | 3285.285 | -13276.741 | -2079.499 | 36.207 |
| 3100 | 3613.787 | 1223.502 | 3407.612 | -13262.584 | -1706.457 | 28.753 |
| 3200 | 3652.638 | 1223.932 | 3529.984 | -13248.752 | -1333.93 | 21.774 |
| 3300 | 3690.307 | 1224.324 | 3652.397 | -13235.224 | -961.772 | 15.223 |
| 3400 | 3726.862 | 1224.682 | 3774.848 | -13222.023 | -590.095 | 9.066 |
| 3500 | 3762.367 | 1225.01 | 3897.333 | -13209.143 | -218.725 | 3.264 |
| 3600 | 3796.881 | 1225.311 | 4019.849 | -17038.157 | 256.815 | -3.726 |
| 3700 | 3830.457 | 1225.589 | 4142.394 | -17021.862 | 737.039 | -10.405 |
| 3800 | 3863.145 | 1225.845 | 4264.966 | -17005.96 | 1216.845 | -16.726 |
| 3900 | 3894.99 | 1226.082 | 4387.562 | -16990.484 | 1696.19 | -22.718 |
| 4000 | 3926.035 | 1226.301 | 4510.182 | -16975.434 | 2175.146 | -28.404 |
| 4100 | 3956.318 | 1226.504 | 4632.822 | -16960.819 | 2653.712 | -33.808 |
| 4200 | 3985.876 | 1226.694 | 4755.482 | -16946.664 | 3131.936 | -38.951 |
| 4300 | 4014.743 | 1226.87 | 4878.16 | -16932.981 | 3609.848 | -43.85 |
| 4400 | 4042.95 | 1227.034 | 5000.856 | -16919.805 | 4087.445 | -48.524 |
| 4500 | 4070.526 | 1227.188 | 5123.567 | -16907.144 | 4564.669 | -52.985 |
| 4600 | 4097.5 | 1227.332 | 5246.293 | -16895.033 | 5041.723 | -57.25 |
| 4700 | 4123.897 | 1227.466 | 5369.033 | -16883.463 | 5518.49 | -61.33 |
| 4800 | 4149.741 | 1227.593 | 5491.786 | -16872.495 | 5994.964 | -65.238 |
| 4900 | 4175.054 | 1227.712 | 5614.551 | -16862.14 | 6471.273 | -68.984 |
| 5000 | 4199.858 | 1227.823 | 5737.328 | -16852.428 | 6947.382 | -72.578 |
| 5100 | 4224.173 | 1227.929 | 5860.116 | -16843.37 | 7423.307 | -76.029 |
| 5200 | 4248.018 | 1228.028 | 5982.914 | -16834.992 | 7899.052 | -79.346 |
| 5300 | 4271.411 | 1228.122 | 6105.721 | -16827.35 | 8374.595 | -82.536 |
| 5400 | 4294.368 | 1228.21 | 6228.538 | -16820.468 | 8850.063 | -85.606 |
| 5500 | 4316.905 | 1228.294 | 6351.363 | -16814.333 | 9325.297 | -88.563 |
| 5600 | 4339.038 | 1228.373 | 6474.196 | -16809.015 | 9800.628 | -91.415 |
| 5700 | 4360.78 | 1228.449 | 6597.037 | -16804.529 | 10275.743 | -94.165 |
| 5800 | 4382.146 | 1228.52 | 6719.886 | -16800.91 | 10750.795 | -96.82 |
| 5900 | 4403.147 | 1228.588 | 6842.741 | -16798.18 | 11225.805 | -99.385 |
| 6000 | 4423.797 | 1228.652 | 6965.603 | -16796.378 | 11700.73 | |

**(CaTiO3)1:**

| T (K) | S (J/mol.K) | Cp (J/mol.K) | ddH (kJ/mol) | dHf (kJ/mol) | dGf (kJ/mol) | log Kf |
|---|---|---|---|---|---|---|
| 0 | 0 | 0 | 0 | -817.472 | -817.472 | Infinity |
| 100 | 254.982 | 47.437 | 3.845 | -819.533 | -816.612 | 426.549 |
| 200 | 294.261 | 67.352 | 9.627 | -822.296 | -812.56 | 212.216 |
| 298.15 | 323.886 | 80.954 | 16.954 | -824.108 | -807.358 | 141.444 |
| 300 | 324.387 | 81.156 | 17.104 | -824.127 | -807.246 | 140.552 |
| 400 | 349.012 | 89.755 | 25.684 | -825.259 | -801.443 | 104.656 |
| 500 | 369.658 | 95.081 | 34.946 | -826.09 | -795.389 | 83.093 |
| 600 | 387.317 | 98.487 | 44.637 | -826.902 | -789.169 | 68.702 |
| 700 | 402.68 | 100.756 | 54.606 | -827.844 | -782.811 | 58.413 |
| 800 | 416.243 | 102.327 | 64.765 | -829.803 | -776.193 | 50.68 |
| 900 | 428.364 | 103.452 | 75.057 | -831.146 | -769.426 | 44.656 |
| 1000 | 439.308 | 104.283 | 85.446 | -832.89 | -762.476 | 39.827 |
| 1100 | 449.278 | 104.912 | 95.907 | -835.125 | -755.326 | 35.867 |
| 1200 | 458.429 | 105.399 | 106.423 | -849.593 | -747.219 | 32.525 |
| 1300 | 466.881 | 105.783 | 116.983 | -850.887 | -738.634 | 29.678 |
| 1400 | 474.732 | 106.091 | 127.577 | -852.273 | -729.948 | 27.234 |
| 1500 | 482.06 | 106.342 | 138.199 | -853.764 | -721.159 | 25.113 |
| 1600 | 488.93 | 106.548 | 148.844 | -855.369 | -712.264 | 23.253 |
| 1700 | 495.395 | 106.721 | 159.508 | -857.113 | -703.265 | 21.608 |
| 1800 | 501.499 | 106.866 | 170.187 | -1007.691 | -691.955 | 20.08 |
| 1900 | 507.281 | 106.989 | 180.88 | -1008.334 | -674.392 | 18.54 |
| 2000 | 512.771 | 107.094 | 191.585 | -1023.864 | -656.346 | 17.142 |
| 2100 | 517.998 | 107.186 | 202.299 | -1025.649 | -637.921 | 15.867 |
| 2200 | 522.987 | 107.265 | 213.021 | -1027.47 | -619.418 | 14.707 |
| 2300 | 527.756 | 107.334 | 223.751 | -1029.341 | -600.832 | 13.645 |
| 2400 | 532.326 | 107.395 | 234.488 | -1031.249 | -582.155 | 12.67 |
| 2500 | 536.711 | 107.449 | 245.23 | -1033.218 | -563.411 | 11.772 |
| 2600 | 540.926 | 107.497 | 255.977 | -1035.235 | -544.573 | 10.94 |
| 2700 | 544.984 | 107.539 | 266.729 | -1037.316 | -525.659 | 10.169 |
| 2800 | 548.895 | 107.577 | 277.485 | -1039.478 | -506.676 | 9.452 |
| 2900 | 552.671 | 107.612 | 288.245 | -1041.706 | -487.601 | 8.783 |
| 3000 | 556.32 | 107.643 | 299.007 | -1044.023 | -468.455 | 8.156 |
| 3100 | 559.85 | 107.671 | 309.773 | -1046.441 | -449.238 | 7.57 |
| 3200 | 563.269 | 107.696 | 320.541 | -1048.951 | -429.93 | 7.018 |
| 3300 | 566.583 | 107.72 | 331.312 | -1051.576 | -410.546 | 6.498 |
| 3400 | 569.799 | 107.741 | 342.085 | -1054.308 | -391.079 | 6.008 |
| 3500 | 572.923 | 107.76 | 352.86 | -1057.161 | -371.53 | 5.545 |
| 3600 | 575.959 | 107.778 | 363.637 | -1060.155 | -351.905 | 5.106 |
| 3700 | 578.912 | 107.795 | 374.416 | -1472.372 | -324.399 | 4.58 |
| 3800 | 581.787 | 107.81 | 385.196 | -1474.4 | -293.341 | 4.032 |
| 3900 | 584.587 | 107.824 | 395.978 | -1476.623 | -262.227 | 3.512 |
| 4000 | 587.317 | 107.837 | 406.761 | -1479.057 | -231.057 | 3.017 |
| 4100 | 589.98 | 107.849 | 417.545 | -1481.694 | -199.829 | 2.546 |
| 4200 | 592.579 | 107.86 | 428.331 | -1484.546 | -168.533 | 2.096 |
| 4300 | 595.117 | 107.871 | 439.117 | -1487.596 | -137.162 | 1.666 |
| 4400 | 597.597 | 107.88 | 449.905 | -1490.853 | -105.722 | 1.255 |
| 4500 | 600.022 | 107.889 | 460.693 | -1494.294 | -74.207 | 0.861 |
| 4600 | 602.393 | 107.898 | 471.483 | -1497.951 | -42.603 | 0.484 |
| 4700 | 604.714 | 107.906 | 482.273 | -1501.807 | -10.929 | 0.121 |
| 4800 | 606.986 | 107.913 | 493.064 | -1505.856 | 20.829 | -0.227 |
| 4900 | 609.211 | 107.921 | 503.856 | -1510.097 | 52.677 | -0.562 |
| 5000 | 611.391 | 107.927 | 514.648 | -1514.516 | 84.624 | -0.884 |
| 5100 | 613.529 | 107.933 | 525.441 | -1519.134 | 116.745 | -1.196 |
| 5200 | 615.624 | 107.939 | 536.235 | -1523.885 | 148.776 | -1.494 |
| 5300 | 617.681 | 107.945 | 547.029 | -1528.545 | 181.439 | -1.788 |
| 5400 | 619.698 | 107.95 | 557.824 | -1533.91 | 213.298 | -2.063 |
| 5500 | 621.679 | 107.955 | 568.619 | -1538.816 | 246.311 | -2.339 |
| 5600 | 623.624 | 107.96 | 579.414 | -1544.536 | 278.208 | -2.595 |
| 5700 | 625.535 | 107.964 | 590.211 | -1540.493 | 309.377 | -2.835 |
| 5800 | 627.413 | 107.968 | 601.007 | -1555.719 | 343.491 | -3.093 |
| 5900 | 629.259 | 107.972 | 611.804 | -1562.032 | 375.725 | -3.326 |
| 6000 | 631.073 | 107.976 | 622.602 | -1567.388 | 409.183 | -3.562 |

**$(CaTiO_3)_2$:**

| T (K) | S (J/mol.K) | Cp (J/mol.K) | ddH (kJ/mol) | dHf (kJ/mol) | dGf (kJ/mol) | log Kf |
|---|---|---|---|---|---|---|
| 0 | 0 | 0 | 0 | -2200.939 | -2200.939 | Infinity |
| 100 | 293.233 | 80.087 | 4.993 | -2207.758 | -2180.244 | 1138.829 |
| 200 | 368.556 | 139.471 | 16.204 | -2213.638 | -2150.172 | 561.561 |
| 298.15 | 431.345 | 174.176 | 31.752 | -2216.368 | -2118.339 | 371.12 |
| 300 | 432.424 | 174.663 | 32.074 | -2216.384 | -2117.717 | 368.723 |
| 400 | 485.676 | 194.66 | 50.633 | -2217.248 | -2084.677 | 272.228 |
| 500 | 530.48 | 206.352 | 70.733 | -2217.335 | -2051.515 | 214.318 |
| 600 | 568.789 | 213.561 | 91.756 | -2217.318 | -2018.345 | 175.71 |
| 700 | 602.086 | 218.25 | 113.362 | -2217.533 | -1985.175 | 148.134 |
| 800 | 631.451 | 221.444 | 135.356 | -2219.775 | -1951.727 | 127.433 |
| 900 | 657.671 | 223.706 | 157.62 | -2220.781 | -1918.19 | 111.328 |
| 1000 | 681.331 | 225.363 | 180.078 | -2222.589 | -1884.476 | 98.434 |
| 1100 | 702.871 | 226.61 | 202.679 | -2225.381 | -1850.529 | 87.873 |
| 1200 | 722.632 | 227.57 | 225.39 | -2252.637 | -1814.817 | 78.996 |
| 1300 | 740.878 | 228.325 | 248.186 | -2253.55 | -1778.293 | 71.452 |
| 1400 | 757.822 | 228.929 | 271.05 | -2254.645 | -1741.696 | 64.983 |
| 1500 | 773.634 | 229.419 | 293.968 | -2255.953 | -1705.015 | 59.373 |
| 1600 | 788.453 | 229.822 | 316.931 | -2257.491 | -1668.229 | 54.461 |
| 1700 | 802.396 | 230.158 | 339.931 | -2259.307 | -1631.341 | 50.124 |
| 1800 | 815.56 | 230.44 | 362.961 | -2558.79 | -1589.929 | 46.138 |
| 1900 | 828.026 | 230.68 | 386.017 | -2558.406 | -1536.103 | 42.23 |
| 2000 | 839.864 | 230.885 | 409.096 | -2587.797 | -1481.405 | 38.69 |
| 2100 | 851.133 | 231.061 | 432.193 | -2589.698 | -1426.029 | 35.47 |
| 2200 | 861.885 | 231.215 | 455.307 | -2591.67 | -1370.571 | 32.541 |
| 2300 | 872.166 | 231.349 | 478.436 | -2593.744 | -1315.029 | 29.865 |
| 2400 | 882.015 | 231.467 | 501.576 | -2595.893 | -1259.376 | 27.409 |
| 2500 | 891.466 | 231.571 | 524.729 | -2598.163 | -1203.658 | 25.149 |
| 2600 | 900.55 | 231.664 | 547.89 | -2600.53 | -1147.819 | 23.06 |
| 2700 | 909.295 | 231.746 | 571.061 | -2603.024 | -1091.893 | 21.124 |
| 2800 | 917.725 | 231.82 | 594.239 | -2605.682 | -1035.896 | 19.325 |
| 2900 | 925.861 | 231.887 | 617.425 | -2608.473 | -979.766 | 17.647 |
| 3000 | 933.723 | 231.946 | 640.616 | -2611.439 | -923.552 | 16.08 |
| 3100 | 941.329 | 232.001 | 663.814 | -2614.609 | -867.254 | 14.613 |
| 3200 | 948.696 | 232.05 | 687.016 | -2617.963 | -810.827 | 13.235 |
| 3300 | 955.837 | 232.095 | 710.224 | -2621.548 | -754.302 | 11.939 |
| 3400 | 962.766 | 232.136 | 733.435 | -2625.347 | -697.659 | 10.718 |
| 3500 | 969.496 | 232.174 | 756.651 | -2629.386 | -640.9 | 9.565 |
| 3600 | 976.037 | 232.208 | 779.87 | -2633.709 | -584.038 | 8.474 |
| 3700 | 982.4 | 232.24 | 803.092 | -3456.479 | -511.464 | 7.22 |
| 3800 | 988.594 | 232.269 | 826.318 | -3458.869 | -431.827 | 5.936 |
| 3900 | 994.627 | 232.296 | 849.546 | -3461.651 | -352.126 | 4.716 |
| 4000 | 1000.509 | 232.321 | 872.777 | -3464.854 | -272.354 | 3.557 |
| 4100 | 1006.246 | 232.344 | 896.01 | -3468.464 | -192.507 | 2.453 |
| 4200 | 1011.845 | 232.366 | 919.246 | -3472.503 | -112.562 | 1.4 |
| 4300 | 1017.313 | 232.386 | 942.483 | -3476.939 | -32.51 | 0.395 |
| 4400 | 1022.655 | 232.405 | 965.723 | -3481.789 | 47.645 | -0.566 |
| 4500 | 1027.878 | 232.423 | 988.964 | -3487.006 | 127.916 | -1.485 |
| 4600 | 1032.987 | 232.439 | 1012.207 | -3492.656 | 208.315 | -2.365 |
| 4700 | 1037.986 | 232.454 | 1035.452 | -3498.703 | 288.83 | -3.21 |
| 4800 | 1042.88 | 232.469 | 1058.698 | -3505.138 | 369.475 | -4.021 |
| 4900 | 1047.674 | 232.482 | 1081.946 | -3511.956 | 450.257 | -4.8 |
| 5000 | 1052.371 | 232.495 | 1105.195 | -3519.128 | 531.207 | -5.549 |
| 5100 | 1056.975 | 232.507 | 1128.445 | -3526.7 | 612.48 | -6.273 |
| 5200 | 1061.49 | 232.518 | 1151.696 | -3534.539 | 693.524 | -6.966 |
| 5300 | 1065.919 | 232.529 | 1174.948 | -3542.196 | 775.82 | -7.646 |
| 5400 | 1070.265 | 232.539 | 1198.202 | -3551.261 | 856.462 | -8.285 |
| 5500 | 1074.532 | 232.549 | 1221.456 | -3559.409 | 939.387 | -8.921 |
| 5600 | 1078.723 | 232.558 | 1244.712 | -3569.184 | 1020.044 | -9.514 |
| 5700 | 1082.839 | 232.567 | 1267.968 | -3559.435 | 1099.221 | -10.073 |
| 5800 | 1086.884 | 232.575 | 1291.225 | -3588.222 | 1184.262 | -10.665 |
| 5900 | 1090.859 | 232.582 | 1314.483 | -3599.185 | 1265.518 | -11.204 |
| 6000 | 1094.769 | 232.59 | 1337.741 | -3608.235 | 1349.169 | -11.745 |

## M3O4 clusters:

## (Mg2SiO4)1:

| T (K) | S (J/mol.K) | Cp (J/mol.K) | ddH (kJ/mol) | dHf (kJ/mol) | dGf (kJ/mol) | log Kf |
|---|---|---|---|---|---|---|
| 0 | 0 | 0 | 0 | -957.838 | -957.838 | Inf |
| 100 | 270.151 | 64.453 | 4.687 | -960.509 | -950.579 | 496.525 |
| 200 | 323.868 | 92.155 | 12.597 | -963.614 | -939.404 | 245.344 |
| 298.15 | 364.477 | 111.387 | 22.642 | -965.776 | -927.023 | 162.409 |
| 300 | 365.167 | 111.686 | 22.848 | -965.807 | -926.782 | 161.365 |
| 400 | 399.239 | 124.928 | 34.724 | -967.105 | -913.558 | 119.297 |
| 500 | 428.129 | 133.745 | 47.687 | -967.871 | -900.075 | 94.029 |
| 600 | 453.073 | 139.686 | 61.377 | -968.411 | -886.463 | 77.172 |
| 700 | 474.933 | 143.789 | 75.563 | -968.927 | -872.766 | 65.126 |
| 800 | 494.334 | 146.705 | 90.096 | -969.551 | -858.985 | 56.085 |
| 900 | 511.743 | 148.833 | 104.878 | -970.381 | -845.119 | 49.049 |
| 1000 | 527.51 | 150.426 | 119.845 | -988.661 | -829.72 | 43.34 |
| 1100 | 541.907 | 151.645 | 134.951 | -990.091 | -813.758 | 38.642 |
| 1200 | 555.144 | 152.596 | 150.165 | -991.536 | -797.664 | 34.721 |
| 1300 | 567.389 | 153.351 | 165.464 | -993.005 | -781.448 | 31.399 |
| 1400 | 578.777 | 153.96 | 180.83 | -1249.328 | -758.782 | 28.31 |
| 1500 | 589.417 | 154.457 | 196.252 | -1248.172 | -723.787 | 25.204 |
| 1600 | 599.399 | 154.869 | 211.719 | -1247.061 | -688.866 | 22.489 |
| 1700 | 608.798 | 155.213 | 227.224 | -1296.183 | -653.561 | 20.081 |
| 1800 | 617.678 | 155.503 | 242.76 | -1294.953 | -615.793 | 17.87 |
| 1900 | 626.093 | 155.75 | 258.323 | -1293.747 | -578.097 | 15.893 |
| 2000 | 634.087 | 155.962 | 273.909 | -1292.563 | -540.461 | 14.115 |
| 2100 | 641.701 | 156.145 | 289.514 | -1291.407 | -502.887 | 12.508 |
| 2200 | 648.969 | 156.305 | 305.137 | -1290.278 | -465.362 | 11.049 |
| 2300 | 655.92 | 156.445 | 320.775 | -1289.182 | -427.894 | 9.718 |
| 2400 | 662.581 | 156.568 | 336.425 | -1288.119 | -390.466 | 8.498 |
| 2500 | 668.974 | 156.676 | 352.088 | -1287.088 | -353.091 | 7.377 |
| 2600 | 675.121 | 156.773 | 367.76 | -1286.095 | -315.741 | 6.343 |
| 2700 | 681.04 | 156.859 | 383.442 | -1285.139 | -278.444 | 5.387 |
| 2800 | 686.746 | 156.937 | 399.132 | -1284.223 | -241.176 | 4.499 |
| 2900 | 692.254 | 157.006 | 414.829 | -1283.345 | -203.939 | 3.673 |
| 3000 | 697.578 | 157.069 | 430.533 | -1282.513 | -166.738 | 2.903 |
| 3100 | 702.729 | 157.126 | 446.243 | -1281.722 | -129.551 | 2.183 |
| 3200 | 707.718 | 157.178 | 461.958 | -1280.981 | -92.402 | 1.508 |
| 3300 | 712.556 | 157.225 | 477.678 | -1280.287 | -55.268 | 0.875 |
| 3400 | 717.25 | 157.268 | 493.403 | -1279.645 | -18.16 | 0.279 |
| 3500 | 721.81 | 157.308 | 509.131 | -1279.055 | 18.934 | -0.283 |
| 3600 | 726.242 | 157.344 | 524.864 | -1662.677 | 66.468 | -0.964 |
| 3700 | 730.553 | 157.377 | 540.6 | -1661.791 | 114.494 | -1.616 |
| 3800 | 734.75 | 157.408 | 556.339 | -1660.971 | 162.497 | -2.234 |
| 3900 | 738.84 | 157.437 | 572.082 | -1660.221 | 210.472 | -2.819 |
| 4000 | 742.826 | 157.463 | 587.827 | -1659.544 | 258.432 | -3.375 |
| 4100 | 746.714 | 157.488 | 603.574 | -1658.943 | 306.371 | -3.903 |
| 4200 | 750.51 | 157.511 | 619.324 | -1658.422 | 354.298 | -4.406 |
| 4300 | 754.216 | 157.532 | 635.076 | -1657.985 | 402.218 | -4.886 |
| 4400 | 757.838 | 157.552 | 650.831 | -1657.636 | 450.127 | -5.344 |
| 4500 | 761.379 | 157.57 | 666.587 | -1657.381 | 498.015 | -5.781 |
| 4600 | 764.842 | 157.587 | 682.345 | -1657.224 | 545.918 | -6.199 |
| 4700 | 768.232 | 157.604 | 698.104 | -1657.167 | 593.809 | -6.599 |
| 4800 | 771.55 | 157.619 | 713.865 | -1657.217 | 641.7 | -6.983 |
| 4900 | 774.8 | 157.633 | 729.628 | -1657.379 | 689.598 | -7.351 |
| 5000 | 777.985 | 157.647 | 745.392 | -1657.657 | 737.498 | -7.704 |
| 5100 | 781.107 | 157.659 | 761.157 | -1658.055 | 785.406 | -8.044 |
| 5200 | 784.168 | 157.671 | 776.924 | -1658.576 | 833.326 | -8.371 |
| 5300 | 787.172 | 157.683 | 792.691 | -1659.23 | 881.245 | -8.685 |
| 5400 | 790.119 | 157.693 | 808.46 | -1660.02 | 929.193 | -8.988 |
| 5500 | 793.013 | 157.704 | 824.23 | -1660.946 | 977.134 | -9.28 |
| 5600 | 795.855 | 157.713 | 840.001 | -1662.018 | 1025.125 | -9.562 |
| 5700 | 798.646 | 157.722 | 855.773 | -1663.239 | 1073.12 | -9.834 |
| 5800 | 801.389 | 157.731 | 871.545 | -1664.615 | 1121.136 | -10.097 |
| 5900 | 804.086 | 157.739 | 887.319 | -1666.148 | 1169.179 | -10.351 |
| 6000 | 806.737 | 157.747 | 903.093 | -1667.846 | 1217.242 | -10.597 |

**(Mg2SiO4)2:**

| T (K) | S (J/mol.K) | Cp (J/mol.K) | ddH (kJ/mol) | dHf (kJ/mol) | dGf (kJ/mol) | log Kf |
|---|---|---|---|---|---|---|
| 0 | 0 | 0 | 0 | -2603.78 | -2603.78 | Inf |
| 100 | 304.603 | 88.043 | 5.234 | -2613.262 | -2569.831 | 1342.326 |
| 200 | 391.447 | 168.708 | 18.23 | -2622.296 | -2522.617 | 658.832 |
| 298.15 | 469.913 | 223.883 | 37.699 | -2627.241 | -2472.502 | 433.167 |
| 300 | 471.3 | 224.698 | 38.114 | -2627.3 | -2471.54 | 430.328 |
| 400 | 541.085 | 259.278 | 62.454 | -2629.308 | -2419.256 | 315.919 |
| 500 | 601.417 | 280.609 | 89.531 | -2629.689 | -2366.676 | 247.242 |
| 600 | 653.87 | 294.229 | 118.322 | -2629.358 | -2314.096 | 201.457 |
| 700 | 699.952 | 303.293 | 148.227 | -2628.857 | -2261.595 | 168.76 |
| 800 | 740.885 | 309.567 | 178.888 | -2628.51 | -2209.151 | 144.241 |
| 900 | 777.62 | 314.062 | 210.081 | -2628.541 | -2156.738 | 125.172 |
| 1000 | 810.89 | 317.38 | 241.661 | -2663.455 | -2101.443 | 109.767 |
| 1100 | 841.262 | 319.892 | 273.531 | -2664.657 | -2045.184 | 97.116 |
| 1200 | 869.184 | 321.837 | 305.621 | -2665.885 | -1988.817 | 86.57 |
| 1300 | 895.007 | 323.371 | 337.884 | -2667.158 | -1932.342 | 77.642 |
| 1400 | 919.018 | 324.602 | 370.285 | -3178.135 | -1863.093 | 69.512 |
| 1500 | 941.449 | 325.603 | 402.797 | -3174.155 | -1769.308 | 61.612 |
| 1600 | 962.49 | 326.428 | 435.4 | -3170.264 | -1675.781 | 54.708 |
| 1700 | 982.301 | 327.116 | 468.078 | -3266.84 | -1581.595 | 48.596 |
| 1800 | 1001.015 | 327.695 | 500.82 | -3262.71 | -1482.577 | 43.023 |
| 1900 | 1018.746 | 328.188 | 533.614 | -3258.63 | -1383.794 | 38.043 |
| 2000 | 1035.591 | 328.609 | 566.455 | -3254.593 | -1285.223 | 33.566 |
| 2100 | 1051.633 | 328.973 | 599.334 | -3250.612 | -1186.856 | 29.521 |
| 2200 | 1066.944 | 329.29 | 632.248 | -3246.686 | -1088.667 | 25.848 |
| 2300 | 1081.588 | 329.566 | 665.191 | -3242.827 | -990.672 | 22.499 |
| 2400 | 1095.619 | 329.81 | 698.16 | -3239.032 | -892.824 | 19.432 |
| 2500 | 1109.087 | 330.025 | 731.152 | -3235.304 | -795.157 | 16.614 |
| 2600 | 1122.035 | 330.216 | 764.164 | -3231.65 | -697.605 | 14.015 |
| 2700 | 1134.501 | 330.386 | 797.194 | -3228.072 | -600.219 | 11.612 |
| 2800 | 1146.519 | 330.539 | 830.241 | -3224.573 | -502.954 | 9.383 |
| 2900 | 1158.12 | 330.676 | 863.302 | -3221.15 | -405.813 | 7.309 |
| 3000 | 1169.333 | 330.8 | 896.375 | -3217.821 | -308.802 | 5.377 |
| 3100 | 1180.181 | 330.912 | 929.461 | -3214.573 | -211.873 | 3.57 |
| 3200 | 1190.689 | 331.014 | 962.557 | -3211.425 | -115.077 | 1.878 |
| 3300 | 1200.876 | 331.107 | 995.664 | -3208.37 | -18.353 | 0.29 |
| 3400 | 1210.762 | 331.192 | 1028.779 | -3205.421 | 78.258 | -1.202 |
| 3500 | 1220.364 | 331.27 | 1061.902 | -3202.574 | 174.8 | -2.609 |
| 3600 | 1229.697 | 331.341 | 1095.032 | -3968.154 | 292.168 | -4.239 |
| 3700 | 1238.776 | 331.407 | 1128.17 | -3964.716 | 410.474 | -5.795 |
| 3800 | 1247.615 | 331.467 | 1161.313 | -3961.411 | 528.688 | -7.267 |
| 3900 | 1256.226 | 331.523 | 1194.463 | -3958.247 | 646.81 | -8.663 |
| 4000 | 1264.62 | 331.575 | 1227.618 | -3955.228 | 764.852 | -9.988 |
| 4100 | 1272.808 | 331.623 | 1260.778 | -3952.36 | 882.811 | -11.247 |
| 4200 | 1280.8 | 331.668 | 1293.943 | -3949.653 | 1000.71 | -12.446 |
| 4300 | 1288.605 | 331.71 | 1327.112 | -3947.114 | 1118.548 | -13.588 |
| 4400 | 1296.231 | 331.749 | 1360.284 | -3944.754 | 1236.329 | -14.677 |
| 4500 | 1303.687 | 331.785 | 1393.461 | -3942.579 | 1354.033 | -15.717 |
| 4600 | 1310.979 | 331.819 | 1426.641 | -3940.601 | 1471.726 | -16.712 |
| 4700 | 1318.116 | 331.851 | 1459.825 | -3938.821 | 1589.366 | -17.664 |
| 4800 | 1325.103 | 331.881 | 1493.012 | -3937.256 | 1706.963 | -18.575 |
| 4900 | 1331.946 | 331.909 | 1526.201 | -3935.917 | 1824.542 | -19.45 |
| 5000 | 1338.652 | 331.936 | 1559.393 | -3934.809 | 1942.091 | -20.289 |
| 5100 | 1345.226 | 331.961 | 1592.588 | -3933.94 | 2059.62 | -21.095 |
| 5200 | 1351.672 | 331.984 | 1625.785 | -3933.319 | 2177.138 | -21.869 |
| 5300 | 1357.996 | 332.006 | 1658.985 | -3932.961 | 2294.634 | -22.615 |
| 5400 | 1364.202 | 332.027 | 1692.187 | -3932.877 | 2412.144 | -23.333 |
| 5500 | 1370.294 | 332.047 | 1725.39 | -3933.066 | 2529.621 | -24.024 |
| 5600 | 1376.278 | 332.066 | 1758.596 | -3933.546 | 2647.159 | -24.691 |
| 5700 | 1382.155 | 332.084 | 1791.804 | -3934.324 | 2764.675 | -25.335 |
| 5800 | 1387.931 | 332.101 | 1825.013 | -3935.411 | 2882.204 | -25.957 |
| 5900 | 1393.608 | 332.117 | 1858.224 | -3936.814 | 2999.768 | -26.558 |
| 6000 | 1399.19 | 332.132 | 1891.436 | -3938.546 | 3117.334 | -27.138 |

**$(Mg_2SiO_4)_3$:**

| T (K) | S (J/mol.K) | Cp (J/mol.K) | ddH (kJ/mol) | dHf (kJ/mol) | dGf (kJ/mol) | log Kf |
|---|---|---|---|---|---|---|
| 0 | 0 | 0 | 0 | -4294.432 | -4294.432 | Inf |
| 100 | 331.756 | 124.499 | 6.318 | -4310.188 | -4232.526 | 2210.818 |
| 200 | 460.874 | 256.73 | 25.713 | -4323.838 | -4149.06 | 1083.610 |
| 298.15 | 580.752 | 342.534 | 55.461 | -4330.711 | -4061.596 | 711.566 |
| 300 | 582.875 | 343.779 | 56.096 | -4330.787 | -4059.923 | 706.887 |
| 400 | 689.603 | 396.328 | 93.321 | -4333.084 | -3969.195 | 518.317 |
| 500 | 781.783 | 428.552 | 134.691 | -4332.901 | -3878.209 | 405.149 |
| 600 | 861.868 | 449.113 | 178.647 | -4331.635 | -3787.378 | 329.717 |
| 700 | 932.195 | 462.807 | 224.287 | -4330.101 | -3696.795 | 275.855 |
| 800 | 994.65 | 472.294 | 271.069 | -4328.79 | -3606.409 | 235.472 |
| 900 | 1050.692 | 479.096 | 318.657 | -4328.038 | -3516.168 | 204.071 |
| 1000 | 1101.443 | 484.12 | 366.83 | -4379.606 | -3421.696 | 178.729 |
| 1100 | 1147.771 | 487.927 | 415.44 | -4380.604 | -3325.859 | 157.93 |
| 1200 | 1190.358 | 490.875 | 464.387 | -4381.634 | -3229.929 | 140.593 |
| 1300 | 1229.744 | 493.202 | 513.595 | -4382.73 | -3133.909 | 125.921 |
| 1400 | 1266.365 | 495.068 | 563.012 | -5148.38 | -3018.789 | 112.631 |
| 1500 | 1300.575 | 496.587 | 612.597 | -5141.593 | -2866.924 | 99.834 |
| 1600 | 1332.665 | 497.84 | 662.32 | -5134.938 | -2715.501 | 88.651 |
| 1700 | 1362.879 | 498.884 | 712.158 | -5278.981 | -2563.139 | 78.755 |
| 1800 | 1391.42 | 499.763 | 762.092 | -5271.965 | -2403.58 | 69.749 |
| 1900 | 1418.461 | 500.511 | 812.106 | -5265.022 | -2244.417 | 61.703 |
| 2000 | 1444.151 | 501.151 | 862.19 | -5258.144 | -2085.618 | 54.47 |
| 2100 | 1468.615 | 501.704 | 912.334 | -5251.347 | -1927.16 | 47.935 |
| 2200 | 1491.966 | 502.185 | 962.529 | -5244.634 | -1769.014 | 42.001 |
| 2300 | 1514.299 | 502.605 | 1012.769 | -5238.02 | -1611.196 | 36.591 |
| 2400 | 1535.697 | 502.974 | 1063.048 | -5231.502 | -1453.633 | 31.637 |
| 2500 | 1556.236 | 503.301 | 1113.362 | -5225.084 | -1296.376 | 27.086 |
| 2600 | 1575.982 | 503.591 | 1163.707 | -5218.776 | -1139.324 | 22.889 |
| 2700 | 1594.993 | 503.85 | 1214.079 | -5212.582 | -982.554 | 19.008 |
| 2800 | 1613.321 | 504.082 | 1264.476 | -5206.507 | -825.996 | 15.409 |
| 2900 | 1631.013 | 504.291 | 1314.895 | -5200.545 | -669.654 | 12.062 |
| 3000 | 1648.113 | 504.479 | 1365.333 | -5194.723 | -513.535 | 8.941 |
| 3100 | 1664.657 | 504.649 | 1415.79 | -5189.023 | -357.567 | 6.025 |
| 3200 | 1680.682 | 504.804 | 1466.263 | -5183.472 | -201.824 | 3.294 |
| 3300 | 1696.218 | 504.946 | 1516.75 | -5178.063 | -46.219 | 0.732 |
| 3400 | 1711.294 | 505.075 | 1567.252 | -5172.81 | 109.196 | -1.678 |
| 3500 | 1725.936 | 505.193 | 1617.765 | -5167.711 | 264.485 | -3.947 |
| 3600 | 1740.17 | 505.301 | 1668.29 | -6315.251 | 450.985 | -6.544 |
| 3700 | 1754.016 | 505.401 | 1718.825 | -6309.266 | 638.868 | -9.019 |
| 3800 | 1767.495 | 505.493 | 1769.37 | -6303.478 | 826.595 | -11.362 |
| 3900 | 1780.627 | 505.578 | 1819.923 | -6297.904 | 1014.16 | -13.583 |
| 4000 | 1793.428 | 505.657 | 1870.485 | -6292.546 | 1201.582 | -15.691 |
| 4100 | 1805.915 | 505.731 | 1921.055 | -6287.414 | 1388.861 | -17.694 |
| 4200 | 1818.102 | 505.799 | 1971.631 | -6282.525 | 1576.032 | -19.601 |
| 4300 | 1830.005 | 505.862 | 2022.214 | -6277.887 | 1763.088 | -21.417 |
| 4400 | 1841.635 | 505.921 | 2072.803 | -6273.516 | 1950.04 | -23.15 |
| 4500 | 1853.005 | 505.976 | 2123.398 | -6269.424 | 2136.86 | -24.804 |
| 4600 | 1864.127 | 506.028 | 2173.999 | -6265.626 | 2323.637 | -26.385 |
| 4700 | 1875.01 | 506.077 | 2224.604 | -6262.127 | 2510.325 | -27.899 |
| 4800 | 1885.665 | 506.122 | 2275.214 | -6258.95 | 2696.929 | -29.348 |
| 4900 | 1896.101 | 506.165 | 2325.828 | -6256.111 | 2883.487 | -30.738 |
| 5000 | 1906.328 | 506.205 | 2376.447 | -6253.618 | 3069.982 | -32.072 |
| 5100 | 1916.352 | 506.243 | 2427.069 | -6251.485 | 3256.44 | -33.352 |
| 5200 | 1926.183 | 506.279 | 2477.695 | -6249.723 | 3442.854 | -34.583 |
| 5300 | 1935.827 | 506.312 | 2528.325 | -6248.356 | 3629.223 | -35.768 |
| 5400 | 1945.291 | 506.344 | 2578.958 | -6247.4 | 3815.598 | -36.908 |
| 5500 | 1954.582 | 506.374 | 2629.594 | -6246.852 | 4001.903 | -38.006 |
| 5600 | 1963.707 | 506.403 | 2680.232 | -6246.743 | 4188.292 | -39.066 |
| 5700 | 1972.67 | 506.43 | 2730.874 | -6247.08 | 4374.625 | -40.088 |
| 5800 | 1981.478 | 506.456 | 2781.518 | -6247.88 | 4560.971 | -41.075 |
| 5900 | 1990.136 | 506.48 | 2832.165 | -6249.154 | 4747.349 | -42.029 |
| 6000 | 1998.649 | 506.503 | 2882.814 | -6250.921 | 4933.715 | -42.951 |

**(Mg2SiO4)4:**

| T (K) | S (J/mol.K) | Cp (J/mol.K) | ddH (kJ/mol) | dHf (kJ/mol) | dGf (kJ/mol) | log Kf |
|---|---|---|---|---|---|---|
| 0 | 0 | 0 | 0 | -5968.417 | -5968.417 | Inf |
| 100 | 370.874 | 168.585 | 8.111 | -5989.738 | -5879.043 | 3070.860 |
| 200 | 545.671 | 346.302 | 34.358 | -6007.551 | -5760.749 | 1504.535 |
| 298.15 | 707.061 | 460.666 | 74.404 | -6016.333 | -5637.455 | 987.647 |
| 300 | 709.916 | 462.334 | 75.258 | -6016.427 | -5635.101 | 981.146 |
| 400 | 853.427 | 532.902 | 125.313 | -6019.068 | -5507.466 | 719.193 |
| 500 | 977.377 | 576.267 | 180.94 | -6018.357 | -5379.602 | 561.996 |
| 600 | 1085.067 | 603.927 | 240.048 | -6016.169 | -5252.04 | 457.225 |
| 700 | 1179.636 | 622.33 | 301.42 | -6013.605 | -5124.894 | 382.42 |
| 800 | 1263.617 | 635.067 | 364.327 | -6011.326 | -4998.085 | 326.338 |
| 900 | 1338.973 | 644.193 | 428.314 | -6009.787 | -4871.54 | 282.733 |
| 1000 | 1407.211 | 650.928 | 493.086 | -6078.003 | -4739.41 | 247.558 |
| 1100 | 1469.501 | 656.029 | 558.445 | -6078.788 | -4605.515 | 218.695 |
| 1200 | 1526.759 | 659.977 | 624.254 | -6079.615 | -4471.547 | 194.639 |
| 1300 | 1579.713 | 663.091 | 690.413 | -6080.528 | -4337.505 | 174.281 |
| 1400 | 1628.948 | 665.59 | 756.852 | -7100.845 | -4178.037 | 155.883 |
| 1500 | 1674.941 | 667.622 | 823.516 | -7091.245 | -3969.615 | 138.233 |
| 1600 | 1718.083 | 669.298 | 890.364 | -7081.821 | -3761.82 | 122.809 |
| 1700 | 1758.702 | 670.695 | 957.366 | -7273.327 | -3552.806 | 109.163 |
| 1800 | 1797.073 | 671.871 | 1024.496 | -7263.421 | -3334.232 | 96.756 |
| 1900 | 1833.426 | 672.87 | 1091.734 | -7253.611 | -3116.213 | 85.67 |
| 2000 | 1867.962 | 673.727 | 1159.065 | -7243.888 | -2898.708 | 75.706 |
| 2100 | 1900.852 | 674.466 | 1226.476 | -7234.273 | -2681.692 | 66.703 |
| 2200 | 1932.243 | 675.108 | 1293.955 | -7224.77 | -2465.112 | 58.528 |
| 2300 | 1962.266 | 675.67 | 1361.494 | -7215.399 | -2248.995 | 51.076 |
| 2400 | 1991.032 | 676.164 | 1429.087 | -7206.154 | -2033.242 | 44.252 |
| 2500 | 2018.644 | 676.6 | 1496.725 | -7197.044 | -1817.924 | 37.983 |
| 2600 | 2045.188 | 676.988 | 1564.405 | -7188.08 | -1602.895 | 32.202 |
| 2700 | 2070.745 | 677.334 | 1632.122 | -7179.267 | -1388.267 | 26.857 |
| 2800 | 2095.383 | 677.644 | 1699.871 | -7170.614 | -1173.941 | 21.9 |
| 2900 | 2119.168 | 677.923 | 1767.649 | -7162.112 | -959.929 | 17.29 |
| 3000 | 2142.155 | 678.174 | 1835.454 | -7153.795 | -746.224 | 12.993 |
| 3100 | 2164.396 | 678.402 | 1903.283 | -7145.642 | -532.747 | 8.977 |
| 3200 | 2185.937 | 678.609 | 1971.134 | -7137.687 | -319.578 | 5.217 |
| 3300 | 2206.822 | 678.798 | 2039.005 | -7129.92 | -106.616 | 1.688 |
| 3400 | 2227.089 | 678.97 | 2106.893 | -7122.364 | 106.073 | -1.63 |
| 3500 | 2246.773 | 679.128 | 2174.798 | -7115.011 | 318.579 | -4.754 |
| 3600 | 2265.907 | 679.273 | 2242.718 | -8644.511 | 572.688 | -8.309 |
| 3700 | 2284.52 | 679.406 | 2310.652 | -8635.977 | 828.623 | -11.698 |
| 3800 | 2302.64 | 679.53 | 2378.599 | -8627.706 | 1084.334 | -14.905 |
| 3900 | 2320.293 | 679.643 | 2446.558 | -8619.719 | 1339.816 | -17.945 |
| 4000 | 2337.501 | 679.749 | 2514.528 | -8612.021 | 1595.095 | -20.83 |
| 4100 | 2354.287 | 679.847 | 2582.508 | -8604.625 | 1850.166 | -23.571 |
| 4200 | 2370.671 | 679.938 | 2650.497 | -8597.552 | 2105.077 | -26.18 |
| 4300 | 2386.671 | 680.022 | 2718.495 | -8590.814 | 2359.828 | -28.666 |
| 4400 | 2402.306 | 680.101 | 2786.501 | -8584.432 | 2614.421 | -31.037 |
| 4500 | 2417.59 | 680.175 | 2854.515 | -8578.422 | 2868.831 | -33.3 |
| 4600 | 2432.54 | 680.244 | 2922.536 | -8572.805 | 3123.173 | -35.464 |
| 4700 | 2447.171 | 680.309 | 2990.564 | -8567.585 | 3377.376 | -37.535 |
| 4800 | 2461.494 | 680.37 | 3058.598 | -8562.795 | 3631.461 | -39.518 |
| 4900 | 2475.523 | 680.427 | 3126.638 | -8558.455 | 3885.472 | -41.419 |
| 5000 | 2489.27 | 680.481 | 3194.683 | -8554.578 | 4139.392 | -43.243 |
| 5100 | 2502.746 | 680.531 | 3262.734 | -8551.179 | 4393.243 | -44.995 |
| 5200 | 2515.961 | 680.579 | 3330.789 | -8548.276 | 4647.031 | -46.679 |
| 5300 | 2528.926 | 680.624 | 3398.849 | -8545.9 | 4900.74 | -48.299 |
| 5400 | 2541.648 | 680.667 | 3466.914 | -8544.071 | 5154.454 | -49.859 |
| 5500 | 2554.138 | 680.707 | 3534.982 | -8542.787 | 5408.062 | -51.361 |
| 5600 | 2566.404 | 680.745 | 3603.055 | -8542.086 | 5661.776 | -52.81 |
| 5700 | 2578.453 | 680.781 | 3671.131 | -8541.982 | 5915.401 | -54.208 |
| 5800 | 2590.293 | 680.816 | 3739.211 | -8542.494 | 6169.038 | -55.558 |
| 5900 | 2601.932 | 680.848 | 3807.294 | -8543.639 | 6422.702 | -56.862 |
| 6000 | 2613.375 | 680.879 | 3875.381 | -8545.44 | 6676.35 | -58.122 |

**(Mg2SiO4)5:**

| T (K) | S (J/mol.K) | Cp (J/mol.K) | ddH (kJ/mol) | dHf (kJ/mol) | dGf (kJ/mol) | log Kf |
|---|---|---|---|---|---|---|
| 0 | 0 | 0 | 0 | -7692.4 | -7692.4 | Inf |
| 100 | 397.523 | 207.731 | 9.458 | -7719.732 | -7574.756 | 3956.599 |
| 200 | 615.3 | 433.765 | 42.186 | -7742.079 | -7420.218 | 1937.939 |
| 298.15 | 817.778 | 578.523 | 92.432 | -7752.868 | -7259.578 | 1271.833 |
| 300 | 821.364 | 580.629 | 93.504 | -7752.981 | -7256.514 | 1263.456 |
| 400 | 1001.649 | 669.567 | 156.385 | -7755.97 | -7090.414 | 925.902 |
| 500 | 1157.39 | 724.076 | 226.281 | -7754.719 | -6924.109 | 723.348 |
| 600 | 1292.699 | 758.796 | 300.548 | -7751.602 | -6758.259 | 588.352 |
| 700 | 1411.516 | 781.879 | 377.656 | -7748.004 | -6592.994 | 491.969 |
| 800 | 1517.026 | 797.849 | 456.689 | -7744.756 | -6428.209 | 419.714 |
| 900 | 1611.695 | 809.287 | 537.076 | -7742.429 | -6263.801 | 363.537 |
| 1000 | 1697.42 | 817.728 | 618.447 | -7827.293 | -6092.458 | 318.234 |
| 1100 | 1775.671 | 824.12 | 700.553 | -7827.867 | -5918.95 | 281.064 |
| 1200 | 1847.6 | 829.067 | 783.223 | -7828.492 | -5745.388 | 250.087 |
| 1300 | 1914.121 | 832.969 | 866.332 | -7829.223 | -5571.768 | 223.874 |
| 1400 | 1975.969 | 836.099 | 949.791 | -9104.209 | -5366.397 | 200.22 |
| 1500 | 2033.744 | 838.646 | 1033.533 | -9091.797 | -5099.861 | 177.591 |
| 1600 | 2087.938 | 840.745 | 1117.505 | -9079.605 | -4834.138 | 157.817 |
| 1700 | 2138.962 | 842.495 | 1201.67 | -9318.575 | -4566.917 | 140.323 |
| 1800 | 2187.161 | 843.969 | 1285.995 | -9305.78 | -4287.769 | 124.426 |
| 1900 | 2232.826 | 845.221 | 1370.456 | -9293.104 | -4009.339 | 110.223 |
| 2000 | 2276.208 | 846.294 | 1455.033 | -9280.537 | -3731.573 | 97.458 |
| 2100 | 2317.522 | 847.22 | 1539.71 | -9268.105 | -3454.438 | 85.923 |
| 2200 | 2356.954 | 848.024 | 1624.473 | -9255.812 | -3177.87 | 75.451 |
| 2300 | 2394.666 | 848.728 | 1709.312 | -9243.683 | -2901.895 | 65.903 |
| 2400 | 2430.801 | 849.347 | 1794.216 | -9231.714 | -2626.4 | 57.161 |
| 2500 | 2465.484 | 849.894 | 1879.179 | -9219.911 | -2351.459 | 49.13 |
| 2600 | 2498.827 | 850.379 | 1964.193 | -9208.292 | -2076.9 | 41.725 |
| 2700 | 2530.929 | 850.813 | 2049.253 | -9196.862 | -1802.855 | 34.878 |
| 2800 | 2561.878 | 851.201 | 2134.354 | -9185.631 | -1529.207 | 28.527 |
| 2900 | 2591.754 | 851.55 | 2219.492 | -9174.588 | -1255.961 | 22.622 |
| 3000 | 2620.628 | 851.866 | 2304.663 | -9163.777 | -983.116 | 17.117 |
| 3100 | 2648.566 | 852.151 | 2389.864 | -9153.171 | -710.572 | 11.973 |
| 3200 | 2675.624 | 852.41 | 2475.092 | -9142.813 | -438.426 | 7.156 |
| 3300 | 2701.858 | 852.647 | 2560.345 | -9132.69 | -166.55 | 2.636 |
| 3400 | 2727.315 | 852.862 | 2645.621 | -9122.829 | 104.975 | -1.613 |
| 3500 | 2752.041 | 853.06 | 2730.917 | -9113.223 | 376.253 | -5.615 |
| 3600 | 2776.075 | 853.242 | -2816.232 | 11024.683 | 699.527 | -10.15 |
| 3700 | 2799.455 | 853.409 | -2901.565 | 11013.6 | 1025.071 | -14.471 |
| 3800 | 2822.216 | 853.563 | -2986.914 | 11002.846 | 1350.323 | -18.561 |
| 3900 | 2844.39 | 853.706 | -3072.277 | 10992.448 | 1675.278 | -22.438 |
| 4000 | 2866.005 | 853.838 | -3157.654 | 10982.411 | 1999.969 | -26.117 |
| 4100 | 2887.09 | 853.96 | -3243.044 | 10972.751 | 2324.389 | -29.613 |
| 4200 | 2907.67 | 854.074 | -3328.446 | 10963.494 | 2648.601 | -32.94 |
| 4300 | 2927.768 | 854.18 | -3413.859 | 10954.656 | 2972.601 | -36.11 |
| 4400 | 2947.406 | 854.279 | -3499.282 | 10946.263 | 3296.401 | -39.133 |
| 4500 | 2966.605 | 854.372 | -3584.715 | 10938.335 | 3619.952 | -42.019 |
| 4600 | 2985.385 | 854.459 | -3670.156 | 10930.899 | 3943.408 | -44.778 |
| 4700 | 3003.762 | 854.54 | -3755.606 | 10923.959 | 4266.69 | -47.418 |
| 4800 | 3021.753 | 854.616 | -3841.064 | 10917.556 | 4589.814 | -49.947 |
| 4900 | 3039.376 | 854.687 | -3926.529 | 10911.716 | 4912.829 | -52.371 |
| 5000 | 3056.643 | 854.755 | -4012.001 | 10906.454 | 5235.731 | -54.697 |
| 5100 | 3073.57 | 854.818 | -4097.48 | 10901.79 | 5558.536 | -56.93 |
| 5200 | 3090.17 | 854.878 | -4182.965 | 10897.745 | 5881.251 | -59.077 |
| 5300 | 3106.454 | 854.934 | -4268.455 | 10894.36 | 6203.869 | -61.142 |
| 5400 | 3122.435 | 854.987 | -4353.951 | 10891.659 | 6526.473 | -63.13 |
| 5500 | 3138.124 | 855.038 | -4439.453 | 10889.637 | 6848.941 | -65.045 |
| 5600 | 3153.531 | 855.086 | -4524.959 | 10888.346 | 7171.536 | -66.893 |
| 5700 | 3168.666 | 855.131 | -4610.47 | 10887.8 | 7494.01 | -68.674 |
| 5800 | 3183.539 | 855.174 | -4695.985 | 10888.025 | 7816.488 | -70.394 |
| 5900 | 3198.158 | 855.215 | -4781.505 | 10889.04 | 8139.003 | -72.056 |
| 6000 | 3212.532 | 855.254 | -4867.028 | 10890.877 | 8461.481 | -73.663 |

**(Mg2SiO4)6:**

| T (K) | S (J/mol.K) | Cp (J/mol.K) | ddH (kJ/mol) | dHf (kJ/mol) | dGf (kJ/mol) | log Kf |
|---|---|---|---|---|---|---|
| 0 | 0 | 0 | 0 | -9415.742 | -9415.742 | Inf |
| 100 | 421.148 | 243.664 | 10.728 | -9449.162 | -9269.603 | 4841.885 |
| 200 | 679.688 | 518.668 | 49.621 | -9476.359 | -9078.391 | 2371.004 |
| 298.15 | 922.482 | 695.07 | 109.88 | -9489.342 | -8879.848 | 1555.694 |
| 300 | 926.789 | 697.632 | 111.169 | -9489.475 | -8876.06 | 1545.441 |
| 400 | 1143.588 | 805.667 | 186.788 | -9492.9 | -8670.876 | 1132.287 |
| 500 | 1331.038 | 871.67 | 270.915 | -9491.147 | -8465.5 | 884.374 |
| 600 | 1493.943 | 913.609 | 360.329 | -9487.113 | -8260.724 | 719.151 |
| 700 | 1637.005 | 941.443 | 453.171 | -9482.483 | -8056.701 | 601.191 |
| 800 | 1764.047 | 960.674 | 548.333 | -9478.263 | -7853.299 | 512.762 |
| 900 | 1878.036 | 974.436 | 645.125 | -9475.143 | -7650.391 | 444.012 |
| 1000 | 1981.254 | 984.586 | 743.1 | -9576.65 | -7439.198 | 388.579 |
| 1100 | 2075.472 | 992.267 | 841.96 | -9577.006 | -7225.439 | 343.103 |
| 1200 | 2162.076 | 998.21 | 941.496 | -9577.424 | -7011.646 | 305.205 |
| 1300 | 2242.168 | 1002.896 | 1041.56 | -9577.968 | -6797.811 | 273.136 |
| 1400 | 2316.633 | 1006.654 | 1142.045 | -11107.617 | -6545.9 | 244.228 |
| 1500 | 2386.192 | 1009.711 | 1242.868 | -11092.39 | -6220.615 | 216.618 |
| 1600 | 2451.44 | 1012.23 | 1343.969 | -11077.425 | -5896.327 | 192.493 |
| 1700 | 2512.871 | 1014.33 | 1445.3 | -11363.856 | -5570.264 | 171.151 |
| 1800 | 2570.9 | 1016.098 | 1546.824 | -11348.168 | -5229.907 | 151.766 |
| 1900 | 2625.879 | 1017.6 | 1648.511 | -11332.623 | -4890.432 | 134.446 |
| 2000 | 2678.109 | 1018.887 | 1750.337 | -11317.209 | -4551.771 | 118.879 |
| 2100 | 2727.848 | 1019.998 | 1852.282 | -11301.958 | -4213.883 | 104.813 |
| 2200 | 2775.321 | 1020.963 | 1954.331 | -11286.873 | -3876.69 | 92.043 |
| 2300 | 2820.724 | 1021.807 | 2056.471 | -11271.985 | -3540.227 | 80.4 |
| 2400 | 2864.228 | 1022.549 | 2158.689 | -11257.289 | -3204.353 | 69.74 |
| 2500 | 2905.984 | 1023.205 | 2260.978 | -11242.792 | -2869.157 | 59.947 |
| 2600 | 2946.126 | 1023.788 | 2363.328 | -11228.516 | -2534.433 | 50.917 |
| 2700 | 2984.774 | 1024.307 | 2465.733 | -11214.467 | -2200.339 | 42.568 |
| 2800 | 3022.034 | 1024.773 | 2568.188 | -11200.656 | -1866.733 | 34.824 |
| 2900 | 3058.002 | 1025.192 | 2670.686 | -11187.072 | -1533.622 | 27.623 |
| 3000 | 3092.764 | 1025.57 | 2773.225 | -11173.765 | -1201.003 | 20.911 |
| 3100 | 3126.398 | 1025.912 | 2875.799 | -11160.705 | -868.755 | 14.638 |
| 3200 | 3158.975 | 1026.223 | 2978.406 | -11147.942 | -537.001 | 8.766 |
| 3300 | 3190.558 | 1026.506 | 3081.043 | -11135.461 | -205.577 | 3.254 |
| 3400 | 3221.206 | 1026.765 | 3183.706 | -11123.296 | 125.414 | -1.927 |
| 3500 | 3250.973 | 1027.002 | 3286.395 | -11111.435 | 456.104 | -6.807 |
| 3600 | 3279.907 | 1027.22 | 3389.106 | -13404.854 | 849.177 | -12.321 |
| 3700 | 3308.055 | 1027.42 | 3491.838 | -13391.222 | 1244.961 | -17.575 |
| 3800 | 3335.457 | 1027.605 | 3594.59 | -13377.984 | 1640.387 | -22.548 |
| 3900 | 3362.152 | 1027.776 | 3697.359 | -13365.173 | 2035.451 | -27.261 |
| 4000 | 3388.175 | 1027.934 | 3800.144 | -13352.796 | 2430.184 | -31.735 |
| 4100 | 3413.559 | 1028.081 | 3902.945 | -13340.871 | 2824.589 | -35.985 |
| 4200 | 3438.335 | 1028.218 | 4005.76 | -13329.43 | 3218.734 | -40.03 |
| 4300 | 3462.531 | 1028.345 | 4108.588 | -13318.492 | 3612.616 | -43.884 |
| 4400 | 3486.173 | 1028.464 | 4211.429 | -13308.087 | 4006.252 | -47.56 |
| 4500 | 3509.287 | 1028.575 | 4314.281 | -13298.241 | 4399.58 | -51.068 |
| 4600 | 3531.895 | 1028.678 | 4417.144 | -13288.984 | 4792.793 | -54.423 |
| 4700 | 3554.019 | 1028.776 | 4520.016 | -13280.324 | 5185.783 | -57.633 |
| 4800 | 3575.679 | 1028.867 | 4622.899 | -13272.307 | 5578.575 | -60.707 |
| 4900 | 3596.895 | 1028.953 | 4725.79 | -13264.966 | 5971.234 | -63.653 |
| 5000 | 3617.683 | 1029.033 | 4828.689 | -13258.319 | 6363.746 | -66.481 |
| 5100 | 3638.061 | 1029.109 | 4931.596 | -13252.39 | 6756.139 | -69.196 |
| 5200 | 3658.046 | 1029.181 | 5034.511 | -13247.203 | 7148.414 | -71.806 |
| 5300 | 3677.65 | 1029.249 | 5137.432 | -13242.808 | 7540.57 | -74.316 |
| 5400 | 3696.89 | 1029.313 | 5240.36 | -13239.234 | 7932.697 | -76.733 |
| 5500 | 3715.777 | 1029.373 | 5343.295 | -13236.475 | 8324.664 | -79.06 |
| 5600 | 3734.326 | 1029.43 | 5446.235 | -13234.593 | 8716.769 | -81.306 |
| 5700 | 3752.546 | 1029.485 | 5549.181 | -13233.605 | 9108.731 | -83.471 |
| 5800 | 3770.451 | 1029.536 | 5652.132 | -13233.542 | 9500.689 | -85.562 |
| 5900 | 3788.051 | 1029.585 | 5755.088 | -13234.428 | 9892.681 | -87.582 |
| 6000 | 3805.356 | 1029.632 | 5858.049 | -13236.299 | 10284.625 | -89.535 |

**(Mg2SiO4)7:**

| T (K) | S (J/mol.K) | Cp (J/mol.K) | ddH (kJ/mol) | dHf (kJ/mol) | dGf (kJ/mol) | log Kf |
|---|---|---|---|---|---|---|
| 0 | 0 | 0 | 0 | -11146.776 | -11146.776 | Inf |
| 100 | 455.922 | 289.09 | 12.579 | -11185.703 | -10972.675 | 5731.47 |
| 200 | 760.895 | 609.143 | 58.432 | -11216.955 | -10746.244 | 2806.597 |
| 298.15 | 1045.382 | 813.075 | 129.03 | -11231.806 | -10511.532 | 1841.555 |
| 300 | 1050.421 | 816.04 | 130.537 | -11231.958 | -10507.057 | 1829.419 |
| 400 | 1303.846 | 941.372 | 218.929 | -11235.784 | -10264.62 | 1340.406 |
| 500 | 1522.843 | 1018.316 | 317.214 | -11233.602 | -10021.996 | 1046.978 |
| 600 | 1713.16 | 1067.389 | 421.673 | -11228.753 | -9780.102 | 851.423 |
| 700 | 1880.312 | 1100.043 | 530.15 | -11223.19 | -9539.109 | 711.808 |
| 800 | 2028.766 | 1122.647 | 641.35 | -11218.089 | -9298.866 | 607.146 |
| 900 | 2161.98 | 1138.844 | 754.467 | -11214.256 | -9059.223 | 525.777 |
| 1000 | 2282.618 | 1150.801 | 868.978 | -11332.474 | -8809.935 | 460.178 |
| 1100 | 2392.745 | 1159.856 | 984.531 | -11332.673 | -8557.675 | 406.365 |
| 1200 | 2493.979 | 1166.865 | 1100.881 | -11332.936 | -8305.397 | 361.52 |
| 1300 | 2587.605 | 1172.396 | 1217.855 | -11333.338 | -8053.087 | 323.573 |
| 1400 | 2674.657 | 1176.831 | 1335.324 | -13117.692 | -7756.375 | 289.39 |
| 1500 | 2755.977 | 1180.441 | 1453.194 | -13099.684 | -7374.076 | 256.785 |
| 1600 | 2832.259 | 1183.417 | 1571.391 | -13081.979 | -6992.958 | 228.294 |
| 1700 | 2904.08 | 1185.897 | 1689.861 | -13415.898 | -6609.783 | 203.091 |
| 1800 | 2971.924 | 1187.986 | 1808.558 | -13397.343 | -6209.944 | 180.206 |
| 1900 | 3036.204 | 1189.761 | 1927.447 | -13378.953 | -5811.152 | 159.758 |
| 2000 | 3097.271 | 1191.282 | 2046.501 | -13360.713 | -5413.323 | 141.38 |
| 2100 | 3155.426 | 1192.595 | 2165.697 | -13342.66 | -5016.405 | 124.775 |
| 2200 | 3210.932 | 1193.736 | 2285.015 | -13324.8 | -4620.313 | 109.699 |
| 2300 | 3264.019 | 1194.733 | 2404.439 | -13307.17 | -4225.086 | 95.954 |
| 2400 | 3314.885 | 1195.61 | 2523.957 | -13289.761 | -3830.554 | 83.369 |
| 2500 | 3363.708 | 1196.386 | 2643.558 | -13272.584 | -3436.826 | 71.808 |
| 2600 | 3410.645 | 1197.075 | 2763.232 | -13255.663 | -3043.661 | 61.147 |
| 2700 | 3455.835 | 1197.689 | 2882.97 | -13239.007 | -2651.24 | 51.291 |
| 2800 | 3499.402 | 1198.24 | 3002.767 | -13222.628 | -2259.398 | 42.149 |
| 2900 | 3541.459 | 1198.735 | 3122.616 | -13206.512 | -1868.144 | 33.649 |
| 3000 | 3582.105 | 1199.182 | 3242.513 | -13190.719 | -1477.471 | 25.725 |
| 3100 | 3621.433 | 1199.587 | 3362.451 | -13175.214 | -1087.241 | 18.32 |
| 3200 | 3659.524 | 1199.954 | 3482.429 | -13160.054 | -697.593 | 11.387 |
| 3300 | 3696.454 | 1200.289 | 3602.441 | -13145.224 | -308.342 | 4.881 |
| 3400 | 3732.291 | 1200.595 | 3722.486 | -13130.76 | 80.396 | -1.235 |
| 3500 | 3767.097 | 1200.876 | 3842.559 | -13116.653 | 468.776 | -6.996 |
| 3600 | 3800.93 | 1201.133 | 3962.66 | -15792.037 | 929.927 | -13.493 |
| 3700 | 3833.844 | 1201.37 | 4082.785 | -15775.862 | 1394.233 | -19.683 |
| 3800 | 3865.885 | 1201.589 | 4202.933 | -15760.147 | 1858.116 | -25.541 |
| 3900 | 3897.099 | 1201.791 | 4323.103 | -15744.928 | 2321.573 | -31.094 |
| 4000 | 3927.529 | 1201.978 | 4443.291 | -15730.216 | 2784.628 | -36.363 |
| 4100 | 3957.211 | 1202.152 | 4563.498 | -15716.031 | 3247.298 | -41.371 |
| 4200 | 3986.182 | 1202.314 | 4683.721 | -15702.411 | 3709.658 | -46.136 |
| 4300 | 4014.474 | 1202.464 | 4803.96 | -15689.377 | 4171.708 | -50.676 |
| 4400 | 4042.12 | 1202.605 | 4924.214 | -15676.965 | 4633.457 | -55.005 |
| 4500 | 4069.148 | 1202.736 | 5044.481 | -15665.205 | 5094.843 | -59.139 |
| 4600 | 4095.584 | 1202.859 | 5164.761 | -15654.132 | 5556.091 | -63.091 |
| 4700 | 4121.454 | 1202.974 | 5285.052 | -15643.755 | 6017.074 | -66.871 |
| 4800 | 4146.782 | 1203.082 | 5405.355 | -15634.129 | 6477.815 | -70.492 |
| 4900 | 4171.59 | 1203.183 | 5525.668 | -15625.291 | 6938.401 | -73.963 |
| 5000 | 4195.898 | 1203.278 | 5645.992 | -15617.261 | 7398.809 | -77.294 |
| 5100 | 4219.727 | 1203.368 | 5766.324 | -15610.07 | 7859.07 | -80.492 |
| 5200 | 4243.095 | 1203.453 | 5886.665 | -15603.745 | 8319.193 | -83.566 |
| 5300 | 4266.019 | 1203.533 | 6007.014 | -15598.343 | 8779.166 | -86.523 |
| 5400 | 4288.517 | 1203.609 | 6127.372 | -15593.898 | 9239.104 | -89.37 |
| 5500 | 4310.602 | 1203.68 | 6247.736 | -15590.406 | 9698.847 | -92.111 |
| 5600 | 4332.292 | 1203.748 | 6368.107 | -15587.936 | 10158.748 | -94.756 |
| 5700 | 4353.598 | 1203.812 | 6488.486 | -15586.508 | 10618.473 | -97.306 |
| 5800 | 4374.535 | 1203.873 | 6608.87 | -15586.16 | 11078.192 | -99.769 |
| 5900 | 4395.115 | 1203.931 | 6729.26 | -15586.919 | 11537.948 | -102.148 |
| 6000 | 4415.35 | 1203.986 | 6849.656 | -15588.827 | 11997.643 | -104.448 |

**APPENDIX A2:**

Fitting parameters for the logarithm of the equilibrium constant according to the parametrization of Woitke et al (2018):

| | | | | | |
|---|---|---|---|---|---|
| $(Al_2O_3)_1$ | 2.237420e+05 | -4.718610e+00 | -2.907880e+01 | 2.514060e-03 | -1.700610E-07 |
| $(Al_2O_3)_2$ | 5.481830e+05 | -1.033280e+01 | -8.471570e+01 | 6.433890e-03 | -4.345780E-07 |
| $(Al_2O_3)_3$ | 8.728800e+05 | -1.430190e+01 | -1.420050e+02 | 9.534330e-03 | -6.434270E-07 |
| $(Al_2O_3)_4$ | 1.195890e+06 | -1.905360e+01 | -1.991840e+02 | 1.299580e-02 | -8.759270E-07 |
| $(Al_2O_3)_5$ | 1.520180e+06 | -2.351910e+01 | -2.574180e+02 | 1.633540e-02 | -1.100730E-06 |
| $(Al_2O_3)_6$ | 1.848210e+06 | -2.847360e+01 | -3.165670e+02 | 1.998260e-02 | -1.348560E-06 |
| $(Al_2O_3)_7$ | 2.175490e+06 | -3.323350e+01 | -3.734430e+02 | 2.347250e-02 | -1.583700E-06 |
| $(Al_2O_3)_8$ | 2.498090e+06 | -3.874440e+01 | -4.308460e+02 | 2.742840e-02 | -1.853430E-06 |
| $(Al_2O_3)_9$ | 2.846460e+06 | -4.330300e+01 | -4.908090e+02 | 3.080760e-02 | -2.080710E-06 |
| $(Al_2O_3)_{10}$ | 3.167630e+06 | -4.852940e+01 | -5.465370e+02 | 3.457990e-02 | -2.336610E-06 |
| | | | | | |
| $(MgSiO_3)$ | 2.317600e+05 | -5.280170e+00 | -2.539090e+01 | 2.734370e-03 | -1.882740E-07 |
| $(MgSiO_3)_2$ | 5.341740e+05 | -1.039970e+01 | -7.961090e+01 | 6.449180e-03 | -4.451560E-07 |
| $(MgSiO_3)_3$ | 8.371960e+05 | -1.460620e+01 | -1.361070e+02 | 9.627140e-03 | -6.630400E-07 |
| $(MgSiO_3)_4$ | 1.146770e+06 | -1.931760e+01 | -1.902050e+02 | 1.305200e-02 | -8.974830E-07 |
| $(MgSiO_3)_5$ | 1.457940e+06 | -2.476560e+01 | -2.470150e+02 | 1.694840e-02 | -1.167400E-06 |
| $(MgSiO_3)_6$ | 1.768600e+06 | -2.903760e+01 | -3.037640e+02 | 2.014550e-02 | -1.386180E-06 |
| $(MgSiO_3)_7$ | 2.078910e+06 | -3.384020e+01 | -3.595630e+02 | 2.366740e-02 | -1.628870E-06 |
| $(MgSiO_3)_8$ | 2.387300e+06 | -3.846980e+01 | -4.181980e+02 | 2.709870e-02 | -1.865260E-06 |
| $(MgSiO_3)_9$ | 2.701040e+06 | -4.268300e+01 | -4.752110e+02 | 3.029640e-02 | -2.084950E-06 |
| $(MgSiO_3)_{10}$ | 3.011900e+06 | -4.769220e+01 | -5.323370e+02 | 3.392080e-02 | -2.334620E-06 |
| | | | | | |
| $(CaTiO_3)_1$ | 2.660150e+05 | -4.619310e+00 | -2.947260e+01 | 2.631050e-03 | -2.149990E-07 |
| $(CaTiO_3)_2$ | 6.006720e+05 | -8.075710e+00 | -9.280340e+01 | 5.740500e-03 | -4.644780E-07 |
| | | | | | |
| $(MgAl_2O_4)_1$ | 3.293490e+05 | -6.697340e+00 | -5.077280e+01 | 4.002720e-03 | -2.737570E-07 |
| $(MgAl_2O_4)_2$ | 7.518310e+05 | -1.212740e+01 | -1.314040e+02 | 8.287160e-03 | -5.640660E-07 |
| $(MgAl_2O_4)_3$ | 1.188240e+06 | -1.855500e+01 | -2.153940e+02 | 1.319850e-02 | -9.011320E-07 |
| $(MgAl_2O_4)_4$ | 1.617310e+06 | -2.464120e+01 | -2.984810e+02 | 1.788730e-02 | -1.221370e-06 |
| $(MgAl_2O_4)_5$ | 2.037620e+06 | -3.059080e+01 | -3.823830e+02 | 2.254310e-02 | -1.540540e-06 |
| $(MgAl_2O_4)_6$ | 2.474570e+06 | -3.714980e+01 | -4.652310e+02 | 2.754760e-02 | -1.884720e-06 |
| $(MgAl_2O_4)_7$ | 2.893780e+06 | -4.204060e+01 | -5.524310e+02 | 3.166310e-02 | -2.166670e-06 |
| | | | | | |
| $(Mg_2SiO_4)_1$ | 3.237250E+05 | -6.182100e+00 | -4.952900e+01 | 3.655940e-03 | -2.524420e-07 |
| $(Mg_2SiO_4)_2$ | 7.309130E+05 | -1.283890e+01 | -1.272780e+02 | 8.662100e-03 | -6.002540e-07 |
| $(Mg_2SiO_4)_3$ | 1.143430E+06 | -1.862940e+01 | -2.100780e+02 | 1.318370e-02 | -9.135890e-07 |
| $(Mg_2SiO_4)_4$ | 1.553840e+06 | -2.433600e+01 | -2.916120e+02 | 1.765640e-02 | -1.223310e-06 |
| $(Mg_2SiO_4)_5$ | 1.970320e+06 | -3.021030e+01 | -3.738920e+02 | 2.222970e-02 | -1.540490e-06 |
| $(Mg_2SiO_4)_6$ | 2.386730e+06 | -3.622730e+01 | -4.559380e+02 | 2.687200e-02 | -1.862120e-06 |
| $(Mg_2SiO_4)_7$ | 2.803980e+06 | -4.198270e+01 | -5.375930e+02 | 3.133660e-02 | -2.170410e-06 |
| $(CaAl_2O_4)_1$ | 3.526520e+05 | -6.380090e+00 | -5.233940e+01 | 3.953820e-03 | -2.851190e-07 |
| $(CaAl_2O_4)_2$ | 7.895310e+05 | -1.160540e+01 | -1.339440e+02 | 8.283860e-03 | -5.944380e-07 |
| $(CaAl_2O_4)_3$ | 1.242990e+06 | -1.729710e+01 | -2.216080e+02 | 1.292700e-02 | -9.275820e-07 |
| $(CaAl_2O_4)_4$ | 1.691680e+06 | -2.321660e+01 | -3.059610e+02 | 1.767010e-02 | -1.267140e-06 |
| $(CaAl_2O_4)_5$ | 2.127440e+06 | -2.880810e+01 | -3.922330e+02 | 2.226990e-02 | -1.597540e-06 |
| $(CaAl_2O_4)_6$ | 2.577340e+06 | -3.477370e+01 | -4.786330e+02 | 2.709240e-02 | -1.944160e-06 |
| $(CaAl_2O_4)_7$ | 3.019480e+06 | -3.965580e+01 | -5.663200e+02 | 3.134330e-02 | -2.251010e-06 |

**APPENDIX A3:**

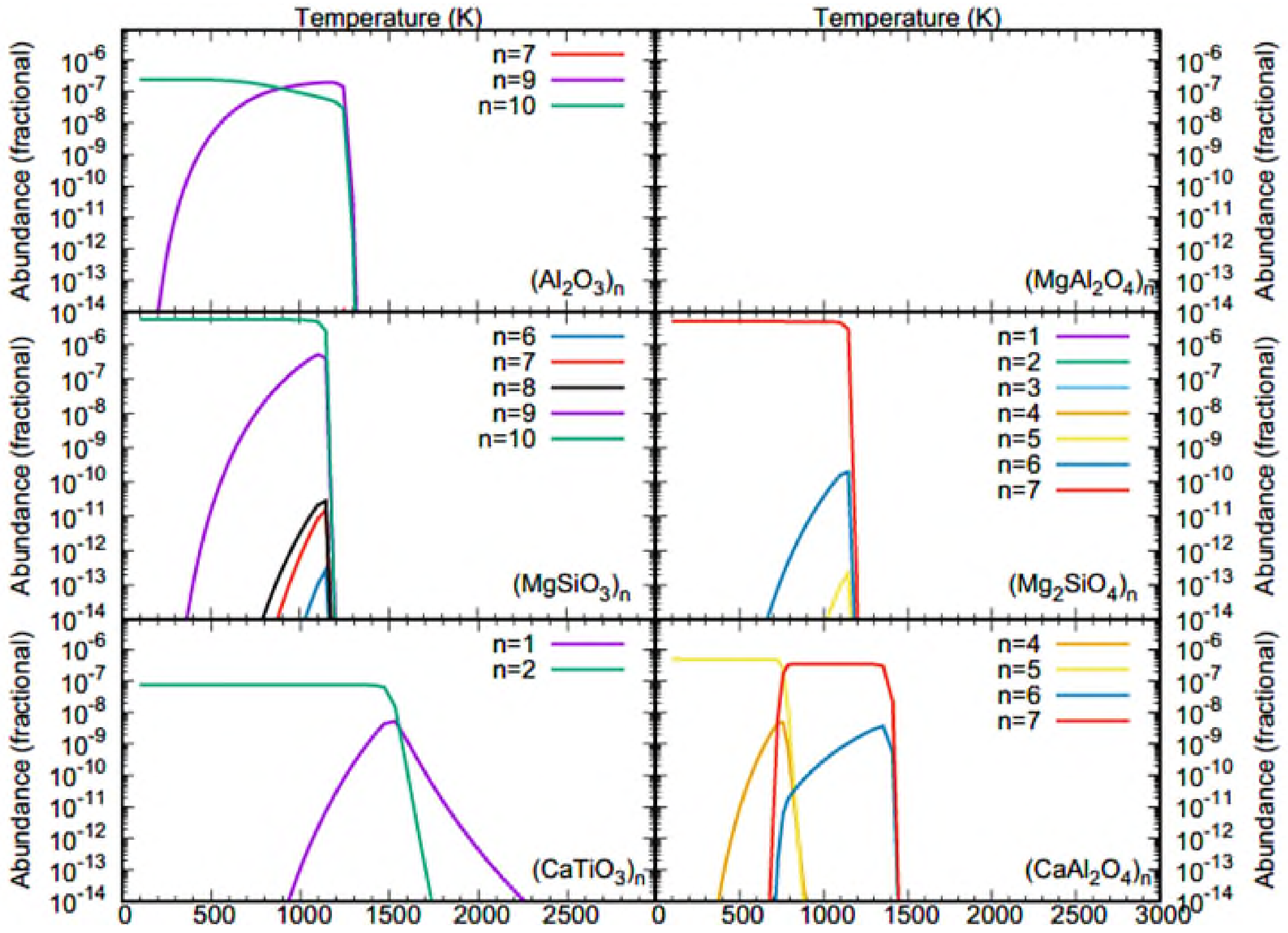

Figure A3: Chemical equilibrium abundances (with respect to the total gas) of the considered cluster families as a function of the temperature T. *Left panel:* $(M_2O_3)_n$ clusters. *Right panel*: $(M_3O_4)_n$ clusters.

Top left panel: $(Al_2O_3)_n$, n=1-10; middle left panel: $(MgSiO_3)$n, n=1-10; bottom left panel: $(CaTiO_3)_n$, n=1-2; top right panel: $(MgAl_2O_4)_1$ n=1-7; middle right panel: $(Mg_2SiO_4)_1$ n=1-7; bottom right panel: $(CaAlO_4)_n$, n=1-2;